\documentclass{jfm}
\usepackage{graphicx}

\usepackage{epstopdf, epsfig}
\usepackage{color} 
\graphicspath{{./figures/}} 

\usepackage{amssymb}
\usepackage{amsmath}
\usepackage{natbib}
\usepackage{subfloat}
\usepackage{subcaption}
\usepackage{bm}
\usepackage{lscape}
\usepackage [autostyle, english = american]{csquotes}
\MakeOuterQuote{"}

\newsavebox{\largestimage}

\newcommand{\changes}[1]{\textcolor{black}{#1}}    
\newcommand{\revii}[1]{\textcolor{black}{#1}} 
\newcommand{\revi}[1]{\textcolor{black}{#1}} 
\def\vec#1{\mbox{\boldmath $#1$}}
\newcommand{\bu}{\mathbf{u}}

\def\Otf{\Omega^\mathrm{f}(t)}
\def\bz{\mathbf z}
\newcommand{\xx}{\mbox{$\mathbf{x}$}}
\def\strain{\vec \epsilon}
\def\stress{{\vec \sigma}}
\def\div{\vec \nabla}

\usepackage{environ}
\NewEnviron{myequation}{%
\begin{equation}
\scalebox{0.7}{$\BODY$}
\end{equation}
}

\def\vec#1{\mbox{\boldmath $#1$}}

\usepackage[english]{babel}
\usepackage{algorithm}

\def\dd{\partial}
\def\bI{{\bf I}}
\def\bx{{\vec x}}
\def\bu{{\vec u}}
\def\force{{\vec f}}
\def\bT{{\vec T}}
\def\G{\Gamma}
\def\strain{\vec \epsilon}
\def\stress{{\vec \sigma}}
\def\Otf{\Omega^\mathrm{f}(t)}

\def\Otnm1f{\Omega^\mathrm{f}(t^{\mathrm{n}-1})}
\def\Otnm12f{\Omega^\mathrm{f}(t^{\mathrm{n}-\frac{1}{2}})}

\def\Otnm1s{\Omega^\mathrm{s}(t^{\mathrm{n}-1})}
\def\Otnm12s{\Omega^\mathrm{s}(t^{\mathrm{n}-\frac{1}{2}})}

\def\shalf{\frac{1}{2}}

\def\div{\vec \nabla}

\def\bz{\vec z}

\def\testf{\vec \phi^\mathrm{f}}
\def\testfbar{\bar{\vec \phi}^\mathrm{f}}
\def\testfdash{(\vec \phi')^\mathrm{f}}
\def\tests{\vec \phi^\mathrm{s}}
\def\testp{ q}
\def\testpbar{\bar{q}}
\def\testpdash{{q}'}

\def\dO{\mathrm{d}\vec{\Omega}}
\def\dG{\mathrm{d}\vec{\Gamma}}

\def\vphi{\vec{\varphi}^\mathrm{s}}
\def\uf{{\vec u}^\mathrm{f}}
\def\us{{\vec u}^\mathrm{s}}

\def\usnn{{\vec u}^\mathrm{s,n-1}}
\def\usnnn{{\vec u}^\mathrm{s,n-2}}

\def\vphi{\vec{\varphi}^\mathrm{s}}
\def\vphin{\vec{\varphi}^\mathrm{s,n}}
\def\vphinn{\vec{\varphi}^\mathrm{s,n-1}}

\def\grad{\bm \nabla}
\def\Os{\Omega^\mathrm{s}}
\def\Of{\Omega^\mathrm{f}}

\shorttitle{Large Amplitude Flapping of Inverted Foil}
\shortauthor{P. S. Gurugubelli and R. K. Jaiman}

\title{On the Mechanism of Large Amplitude Flapping of Inverted Foil in a Uniform Flow}

\author{P. S. Gurugubelli
  \and  R. K. Jaiman
   \corresp{\email{mperkj@nus.edu.sg}}
  }
 
\affiliation{Department of Mechanical Engineering, National University Singapore, Singapore 119077}

\begin{document}
\maketitle

\begin{abstract}
An elastic foil interacting with a uniform flow with its trailing edge clamped, also known as the inverted foil, exhibits a wide range of complex self-induced flapping regimes such as {large amplitude flapping (LAF)}, deformed and flipped flapping. In particular, the {LAF}  has a profound impact on the development of energy harvesting devices. Here, we perform three-dimensional numerical experiments to examine the role of vortex structures, the vortex-vortex interaction and shedding process on the LAF response at Reynolds number $Re=30,000$. We first show that the LAF can produce a secondary vortex in addition to the dominant counter-rotating vortex pair from both leading edge and trailing edge over the one-half cycle. To understand the role of the counter-rotating periodic vortices generated from the edges of the foil and the interaction between them on the LAF, we investigate the dynamics of the inverted foil for a novel configuration wherein we introduce a fixed splitter plate at the trailing edge to suppress the vortex shedding from trailing edge and inhibit the interaction between the counter-rotating vortices. Unlike the vortex-induced vibration of a circular cylinder, we find that the inhibition of the interaction has an insignificant  effect on the transverse flapping amplitudes, due to a relatively weaker coupling between the counter-rotating vortices emanating from the leading edge and trailing edge . However, the inhibition of the trailing edge vortex reduces the streamwise flapping amplitude, the flapping frequency and the net strain energy of foil. To further generalize our understanding of the LAF, we next perform low-Reynolds number ($Re\in[0.1,50]$) simulations for the identical foil properties to realize the impact of vortex shedding on the large amplitude flapping. Due to the absence of vortex shedding process in the low-$Re$ regime, the inverted foil no longer exhibits the periodic flapping. However, the flexible foil still loses its stability through divergence instability to undergo a large static deformation. 
From our study on the effect of foil aspect ratio without side effects, we find  that the foil aspect ratio has a minor impact on the LAF response at high Reynolds number.
Finally, we  introduce an analogous analytical model for the LAF  based on the dynamics of an elastically mounted flat plate undergoing flow-induced pitching 
oscillations in a uniform stream.
This study has implications on the development of novel control mechanisms for energy harvesting and propulsive devices.
\end{abstract}


\section{Introduction}\label{sec:intro}
Over the past few years, the self-induced flapping 
of an inverted foil immersed in an external flow field with its trailing edge (TE) clamped and leading edge (LE) free to vibrate has been a subject of active research experimentally \citep{kim2013,sader_2016}, analytically \citep{sader_2016,sader_inverted_rod} and numerically \revi{\citep{gurugubelli_JFM,ryu_2015,gilmanov_2015,mittal_2016}}. 
This particular configuration of flexible foil has superior abilities to harvest electrical current by converting the fluid kinetic energy into structural strain energy efficiently in comparison to its traditional counterpart with LE clamped \citep{tang2009c,michelin_2013}.
\changes{Apart from a practical relevance, the flapping dynamics of inverted foil has a 
fundamental value due to the richness of coupled fluid-structure physics associated with the complex 
interaction of wake dynamics with the flexible structure undergoing large deformation.
Furthermore, the formation of a leading-edge vortex (LEV) and the passive self-sustained large-amplitude motion can be important to understand 
the flight dynamics and locomotion of  small birds, bats and insects \citep{sane_insect,passiveInsect,passiveInsect_1}. 
The formation of LEV is a very common unsteady aerodynamic feature found in biological locomotion.
As the wing moves with a certain angle of incidence, the airflow rolls up to produce a stable vortex at LE and the streamlines tend to curve around the wing body due to the presence of LEV dynamics.
During the flight these living organisms stabilize the LEV over their wings to recapture the energy from vortex-flexible foil interaction to enhance the lift generation.}

While a pioneering theoretical analysis using inviscid potential flow was carried out by \cite{guo_jam} to predict the onset of linear flapping stability, experimental \citep{kim2013,sader_2016} and viscous numerical simulation studies \citep{gurugubelli_JFM,ryu_2015,tang_inverted_pof,mittal_2016} are performed to understand the self-induced flapping dynamics of an inverted foil. 
{The self-induced flapping dynamics of an inverted foil in a steady uniform flow can 
be characterized by three nondimensional parameters, namely Reynolds number $Re = {\rho^\mathrm{f} U_0 L}/{\mu^\mathrm{f}}$, mass-ratio $m^{*}= {\rho^\mathrm{s} h}/{\rho^\mathrm{f} L}$, and bending rigidity $K_B={B}/{\rho^\mathrm{f} U_0^2 L^3 }$. Here $U_0$, $\rho^\mathrm{f}$,  $\mu^\mathrm{f}$,
$L$, $W$, $h$,  $\rho^\mathrm{s}$, $E$, and $\nu$, 
denote the freestream velocity, the density of fluid, the dynamic viscosity 
of the fluid, the foil length, the foil width and the foil thickness, 
the solid density, Young's modulus, Poisson's ratio, respectively 
and $B$ represents the flexural rigidity defined as $B=Eh^3/12(1-\nu^2)$.} 
The experimental and numerical studies have shown that for a given $Re$ there exists a critical nondimensional bending rigidity $(K_B)_{cr}$ below which an inverted foil placed in an external flow loses its stability to undergo static deformation. \revi{\cite{kim2013}, \cite{gurugubelli_JFM} and \cite{sader_2016} have attributed this phenomenon to the divergence instability. \cite{sader_2016} presented an experimental and theoretical treatment to investigate the effects of foil aspect ratio and have shown that the foils with lower aspect ratios are more stable towards the divergence instability compared to the foils with higher aspect ratios. }

An inverted foil placed in a uniform flow exhibits three distinct dynamic flapping regimes as a function of reducing $K_B$ namely large amplitude flapping (LAF), deformed flapping and flipped flapping.  The transition from the static deformed state to the dynamic flapping modes is attributed to the flow separation at LE and the evolution of a large leading edge vortex (LEV) \citep{kim2013,gurugubelli_JFM}. \revi{The experiments of \cite{kim2013} first showed the existence of LAF and deformed flapping mode for aspect-ratios in the range [1,1.3] at $Re=30,000$ for two representatives $m^*$ corresponding to water and air. This experimental work has been followed up by a number of numerical studies \citep{ryu_2015,gurugubelli_JFM,tang_inverted_pof,gilmanov_2015,mittal_2016} which have confirmed the dynamic flapping modes reported by \cite{kim2013}. In addition to the dynamic flapping modes observed by \cite{kim2013}, the two-dimensional (2D) \citep{gurugubelli_JFM,mittal_2016} and three-dimensional (3D) \citep{tang_inverted_pof} numerical studies have demonstrated a new flipped mode for very low $K_B$. \cite{kim2013,gurugubelli_JFM,mittal_2016} have shown that $m^*$ has little an influence on the inverted foil flapping modes and response.
The 2D simulations of \cite{ryu_2015} and \cite{mittal_2016} have further shown that the unsteady flapping motion ceases for $Re<50$ and the foil undergoes a static deformation instead.} 

The LAF regime is of significant interest for their energy harvesting abilities compared to their conventional foil counterpart. \cite{sader_2016} studied the physical mechanism behind the LAF motion and has drawn an analogy to the vortex-induced vibration (VIV) in a flexibly mounted circular cylinder. The authors have attributed the LAF phenomenon to the synchronized vortex shedding from LE and TE. \revi{\cite{sader_2016} have constructed their analogy based on a number of similarities such as LAF requires vortex shedding, Strouhal number range ($St\approx0.1-0.2$) and the synchronization of the flapping motion with the periodic vortex shedding from the leading and trailing edges. However, the two-dimensional simulations of \cite{gurugubelli_JFM} and \cite{ryu_2015} have shown that LAF can exhibit complex vortex patterns involving as many as four pairs of counter-rotating vortices per flapping cycle. However, the role of complex vortex shedding patterns on the vortex-induced forces and their impact on the synchronization with flapping motion has not been explored.} In another related study, \cite{clapping_book} studied the clapping behavior exhibited by a stack of papers placed on the floor of a wind tunnel with their TE clamped. The experimental analysis showed that the papers lose their stability to exhibit only the deformed flapping regime and the stack of papers remain in the deformed state until the stack size is large enough for the wind to support. \revi{Interestingly, in this experimental study there is no trailing edge vortex (TEV) behind the stack of papers. However, one cannot strongly attribute the absence of LAF to the absence of interaction between the LEV and TEV vortices due to differences in the material properties of the foil and configurations.}
Despite a significant progress recently, the rise of LAF and the underlying fluid-structure dynamics remains unclear.

The objective of this work is to investigate the LAF mechanism and to examine the role of vortex shedding phenomenon on the origin of LAF motion. 
The present study extends the previous two-dimensional work of \cite{gurugubelli_JFM} to three dimensions.
Of particular interest is to understand the role of leading-edge vortex to sustain the large amplitude flapping response of 
inverted in a uniform flow. 
To investigate the complex fluid-structure interaction of LAF, 
we employ the recently developed variational body-conforming fluid-structure formulation 
based on the 3D Navier-Stokes and the nonlinear elasticity for large deformation. 
To handle strong inertial effects, the quasi-monolithic formulation is employed for a robust and stable treatment of 
fluid-structure interaction of a very light and thin flag-like structure.
To resolve the separated turbulent wake regime behind the inverted foil, we consider the
variational multiscale model for the simulation of two different configurations namely the standard inverted foil and  
the inverted foil with a long fixed splitter plate. 
Through the 3D coupled FSI simulations at high Reynolds number, we successfully predict the LE transverse amplitude and the maximum strain energy measured from the experiment of \cite{kim2013}. The turbulence spectra of kinetic energy in the wake of inverted foil exhibit the power law of -5/3 from the coupled FSI simulations.
We next characterize the coupled 
flapping dynamics by the wake and streamline topologies, the  force and response amplitudes and the frequency characteristics.
We also investigate the inverted foil for the laminar low-$Re$ regime to understand the impact of vortex shedding phenomenon on the LAF. In particular, we determine how an inverted foil continue to undergo static-divergence 
with large static deformation even for low-$Re$ flow.
We also examine the role of aspect ratio on the LAF response of inverted foil at high Reynolds number.
Finally, we present a list of similarities and differences between the LAF phenomenon of inverted foil and the VIV of a circular cylinder and provide an improved understanding of the mechanism of LAF.
We also provide the linkage between the LAF dynamics of a flexible inverted foil with the flow-induced oscillation of a flat plate mounted elastically on a torsional spring at the trailing edge  
and immersed in a flow stream. 
%
 
The structure of the paper is as follows: In Section 2, we present the variational fluid-structure model used in this study. While Section 3 describes the problem set-up, Section 4 is dedicated to the validation of the flow past an inverted foil at $Re = 30,000$. 
In Section 5, we present the flow fields to elucidate the generation of LEV and the interaction of LEV and TEV 
for both inverted foil configurations with and without splitter plate. We examine the vorticity dynamics,  the 3D flow structures, the frequency and response characteristics. 
We also expand our results by considering low $Re$ vortex shedding simulations, the effect of aspect ratio and the comparison 
of LAF with the well-known VIV phenomenon of circular cylinders. 
Finally, we conclude the present study in Section 6.

\section{Numerical Methodology}\label{sec:method}
To simulate the nonlinear fluid-structure interactions of thin flexible flag-like structures at high Reynolds number with large deformation, the stability of the coupled formulation
for low mass-ratio and its ability to capture the boundary layer effect are two crucial factors.
In the present coupled formulation, we adopt a quasi-monolithic scheme for the stable fluid-structure coupling
whereas the fluid system is solved on a deforming mesh that adapts to the
Lagrangian flexible body in a body-fitted manner via 
arbitrary Lagrangian-Eulerian (ALE) description. The body-conforming treatment of fluid-structure interface 
provides an accurate modeling
of the boundary layer and the vorticity generation over deformable surfaces.
Before proceeding to the  quasi-monolithic variational formulation,
we first present the governing fluid-structure equations employed for the numerical methodology.

\subsection{Fluid-Structure Equations}
\changes{The incompressible flow is governed by the Navier-Stokes equations and the equations in an arbitrary Lagrangian-Eulerian (ALE) reference frame are expressed as
\begin{align}
\rho^\mathrm{f}{\dd \bu^\mathrm{f} \over \dd t} +		\rho^\mathrm{f} \left(\bu^\mathrm{f} 
-\vec{w}\right)
\cdot \div \bu^\mathrm{f} = \div
\cdot \stress^\mathrm{f}+\force^\mathrm{f} \quad &\mbox{on}\quad \Otf, 
\label{eq:NS_mom}
\end{align}
\begin{align}
\div \cdot \bu^\mathrm{f} = 0 \quad &\mbox{on}\quad \Otf, 
\label{eq:NS_cont}
\end{align}
where $\bu^\mathrm{f}=\bu^\mathrm{f}(\xx,t)=(u,v,w)$ and $\vec{w}=\vec{w}(\xx,t)$ represent the fluid and mesh velocities respectively at a spatial point $\xx \in  \Otf$,
$\boldsymbol{f}^\mathrm{f}$ denotes the
body force applied on the fluid and $\boldsymbol{\sigma}^\mathrm{f}$ is the 
Cauchy stress tensor for the Newtonian fluid, written as
\begin{equation}
\stress^\mathrm{f} = -p \bI +\bT, \quad \bT = 2 \mu^\mathrm{f} \strain^\mathrm{f} 
(\bu^\mathrm{f} ), 
\quad \strain^\mathrm{f} (\bu^\mathrm{f})=\shalf \left[\div \bu^\mathrm{f}+
\left(\div \bu^\mathrm{f}\right)^T\right],
\label{eq:cauchyStress}
\end{equation}
where $p$ is fluid pressure, $\vec{\mathrm{I}}$ denotes second order identity tensor and $\vec{T}$ represents the fluid viscous stress tensor.
The dynamics of the inverted foil, $\Os$, is governed by the equation
\begin{equation}
\rho^\mathrm{s} \frac{\partial
	\vec{u^\mathrm{s}}}{\partial t} =\bm{\nabla} \cdot
\bm{\sigma}^\mathrm{s} + \vec{f}^\mathrm{s} \qquad
\mathrm{in}\quad\Omega^\mathrm{s},
\label{eq:navier}
\end{equation}
$\bu^\mathrm{s}=\bu^\mathrm{s}(\bz,t)$ is the structural velocity at a material point $\bz\in \Os_\mathrm{i}$, $\boldsymbol{f}^\mathrm{s}$ denotes the external forces acting on the solid and $\boldsymbol{\sigma}^s$ is the first Piola-Kirchhoff stress tensor.
In this study, the structural stresses are modeled using the Saint Venant-Kirchhoff model. Eventually, the fluid and structural solutions are  
coupled through the velocity and traction continuity along the fluid-solid interface, given by
\begin{align}
\int_{\vphi(\gamma,t)}\stress^\mathrm{f}(\bx,t)\cdot\vec{n}^\mathrm{f} da(\bx)+\int_{\gamma}\stress^\mathrm{s}(\bz,t)\cdot \vec{n}^\mathrm{s} da(\bz)=0
\quad \forall \gamma \subset \G,
\label{eq:bcsTraction}
\end{align}
\begin{equation}\label{Eq:VelocityCont}
\uf(\vphi(\bz,t),t)=\us(\bz,t) \quad \forall \bz\in \G,
\end{equation}
where $\boldsymbol{\mathrm{n}}^\mathrm{f}$ and $\boldsymbol{\mathrm{n}}^\mathrm{s}$ are, respectively, the outward normals to the deformed fluid and the undeformed solid interface boundaries, $\Gamma$ represents the interface between the fluid and the inverted foil at $t=0$, $\gamma$ is an edge on $\Gamma$ and 
$\boldsymbol{\varphi}^\mathrm{s}$ is the function that maps each Lagrangian point $\bz \in \Os$  
to its deformed position at time $t$.}
\subsection{Variational Quasi-Monolithic Formulation}\label{sec:quasiMonolithic}
We  next briefly present the variational fluid-structure formulation based on the Navier-Stokes and the nonlinear elasticity equations. \changes{We extend the 2D variational body-conforming quasi-monolithic technique presented by \cite{jie_CFEI} for flow-structure interactions to large-scale 3D turbulent flow-structure problems. The employed quasi-monolithic coupling technique is numerically stable even for very low structure-to-fluid mass ratio and has been used for simulating the flapping dynamics of thin flexible foil in \cite{bourlet2015,gurugubelli_JFM}. An energy-based mathematical proof for the numerical stability of the coupling formation for any mass-ratio is presented in \cite{jie_CFEI}. 
One of the main features of this quasi-monolithic formulation is that the kinematic and dynamic continuity conditions
which are to be satisfied along the common interface are absorbed into the formulation and are
satisfied implicitly.
Another attractive feature of the quasi-monolithic
formulation is that the mesh motion is decoupled from rest of the FSI solver variables by updating
the structural positions at the start of each iteration. Decoupling of the mesh motion allows
us to linearize the nonlinear Navier-Stokes without losing the stability and accuracy. Hence,
we solve the coupled system of equations only once per time step making it a computationally
efficient option. }

\changes{To derive the weak form of the Navier-Stokes Eqs.~(\ref{eq:NS_mom}) and (\ref{eq:NS_cont}), we consider a trial function space $\mathcal{S}^\mathrm{f}$ that satisfies the Dirichlet conditions and a test function space $\mathcal{V}^\mathrm{f}$ that is null along the Dirichlet boundaries. The variational form of the Navier-Stokes Eqs.~(\ref{eq:NS_mom}) and (\ref{eq:NS_cont}) in the ALE reference frame can be stated as\\
	Find $\{\vec{u}^\mathrm{f},p\} \in \mathcal{S}^\mathrm{f}$ such that $\forall \{\testf,\testp\}\in\mathcal{V}^\mathrm{f}$:
\begin{align}
\int_{\Otf} \rho^\mathrm{f}\left(\dd_t\bu^\mathrm{f}
+\left(\bu^\mathrm{f}-\vec{w}\right)\cdot\div\bu^\mathrm{f}\right)\cdot \testf \dO +
\int_{\Otf}\stress^\mathrm{f} : \div\testf\dO =\notag\\ \int_{\Otf}
\force^\mathrm{f} \cdot \testf \dO+\int_{\G^\mathrm{f}_\mathrm{h}(t)} \stress^{\mathrm{f}}_{\mathrm{b}} \cdot \testf \dG+\int_{\G\left(t\right)} \left(\stress^\mathrm{f}(\bx,t) \cdot \vec{\mathrm{n}}^\mathrm{f}\right) \cdot \testf (\bx) \mathrm{\dG}, \label{eq:WeakFormNavierStokes_mom}\\
\int_{\Omega^\mathrm{f}(t)}\div \cdot \bu^\mathrm{f} \testp \dO = 0,
\label{eq:WeakFormNavierStokes_cont}
\end{align}
where $\partial_t$ denotes the partial time derivative operator $\partial (\cdot)/\partial t$, $\G^\mathrm{f}_\mathrm{h}$ denotes the Neumann boundary along which $\vec{\sigma}^\mathrm{f}\cdot\vec{\mathrm{n}}^\mathrm{f}=\vec{\sigma}^\mathrm{f}_\mathrm{b}$. } 

\changes{Similar to the flow equations, to construct the 
weak variational form of the structural dynamics equation (\ref{eq:navier}) we consider the trial and test functional spaces $\mathcal{S}^\mathrm{s}$ that satisfy the Dirichlet conditions and $\mathcal{V}^\mathrm{s}$ that is null along the Dirichlet boundaries respectively. The weak form for the nonlinear elasticity of deformable foil is defined as follows.
Find $\vec{u}^\mathrm{s} \in \mathcal{S}^\mathrm{s}$ such that $\forall \tests\in\mathcal{V}^\mathrm{s}$:
\begin{align}
\int_{\Omega^\mathrm{s}} \rho^\mathrm{s}\dd_t\bu^\mathrm{s} \cdot \tests\dO+\int_{\Omega^\mathrm{s}}\stress^\mathrm{s} : \div\tests\dO =\notag\\ \int_{\Omega^\mathrm{s}}
\force^\mathrm{s} \cdot \tests \dO+\int_{\G^\mathrm{s}_{\mathrm{h}}} \stress^{\mathrm{s}}_{\mathrm{b}} \cdot \tests \dG + \int_{\G} \left(\stress^\mathrm{s}(\bz,t) \cdot \vec{\mathrm{n}}^\mathrm{s}\right) \cdot \tests (\bz) \mathrm{\dG},\label{Eq:WeakFormStructuralDynamics}
\end{align}
where $\G^\mathrm{s}_\mathrm{h}$ denotes the Neumann boundary along which $\vec{\sigma}^\mathrm{s}\cdot\vec{\mathrm{n}}^\mathrm{s}=\vec{\sigma}^\mathrm{s}_\mathrm{b}$. \changes{One of the key features of the quasi-monolithic formulation is that the kinematic and dynamic interface continuity conditions in Eqs.~(\ref{Eq:VelocityCont}) and (\ref{eq:bcsTraction}) are absorbed at the variational level by enforcing the condition $ \testf=\tests$ along the interface $\G$ and are satisfied implicitly. Such a condition can be realized by considering a conforming interface mesh. The weak form 
of the traction continuity condition along the fluid-structure interface Eq.~(\ref{eq:bcsTraction}) 
is given as}
\begin{align}
\int_{\G(t)} \left(\stress^\mathrm{f}(\bx,t) \cdot \vec{\mathrm{n}}^\mathrm{f}\right) \cdot \testf (\bx) \mathrm{\dG}
+\int_{\G} \left(\stress^\mathrm{s} (\bz,t) \cdot \vec{\mathrm{n}}^\mathrm{s}\right) \cdot \tests (\bz) \mathrm{\dG} = 0.
\label{eq:WeakFormTraction}
\end{align}}

\changes{Let the fluid domain $\Omega^\mathrm{f}$ be discretized into $\mathrm{n}_\mathrm{el}^\mathrm{f}$ number of three-dimensional Lagrange finite elements such that $\Omega^\mathrm{f} = \cup_{e=1}^\mathrm{n_{el}} \Omega^e$ and $\emptyset = \cap_{e=1}^\mathrm{n_{el}} \Omega^e$. A variational multi-scale (VMS)  based turbulence modeling for large eddy simulation (LES) has been considered in this work, wherein the trial function space $\mathcal{S}^\mathrm{f}$ is decomposed into coarse-scale space $\{\bar{\vec{u}}^\mathrm{f},\bar{p}\}\in\bar{\mathcal{S}}^\mathrm{f}$ and fine-scale space $\{(\vec{u}')^\mathrm{f},p'\}\in\mathcal{(S')}^\mathrm{f}$ \citep{hughes_multiscale,yuri_vms_2007}. Likewise, we can also decompose the test function space $\mathcal{V}^\mathrm{f}$ into the coarse and fine scale spaces $\{\testfbar,\testpbar\}\in\bar{\mathcal{V}}^\mathrm{f}$ and $\{\testfdash,\testpdash\}\in(\mathcal{V'})^\mathrm{f}$ respectively. Using  the weak forms of  Eqs.~(\ref{eq:WeakFormNavierStokes_mom}) and (\ref{eq:WeakFormNavierStokes_cont}) and applying the VMS decomposition,  then combining them with Eqs.~(\ref{Eq:WeakFormStructuralDynamics})-(\ref{eq:WeakFormTraction}),  we construct the weak form of coupled fluid-structure system  as follows:
\begin{align}\label{weakForm_VMS}
\left.
\begin{aligned}
\int_{\Of(t)} \rho^\mathrm{f}\left(\partial_t\bar{\vec{u}}^{\mathrm{f}}
+\left(\bar{\vec{u}}^{\mathrm{f}}-\vec{w}\right)\cdot\grad\bar{\vec{u}}^{\mathrm{f}}\right)\cdot \testfbar \dO 
+\int_{\Of(t)}{\bar{\stress}}^{\mathrm{f}} : \grad\testfbar  \dO
-\int_{\Of(t)}\grad \cdot \bar{\vec{u}}^\mathrm{f} q \dO 
\end{aligned}
\right\} A\notag\\
\left.
\begin{aligned}
+ {\sum_{e=1}^{n_\mathrm{el}} 
	\int_{\Omega^\mathrm{f}_\mathrm{e}(t)} \tau_m
	\left[\rho^\mathrm{f} \left(\bar{\vec{{u}}}^{\mathrm{f}}-\vec{w}\right)\cdot \nabla \testfbar + \nabla q \right] 
	\cdot} {\boldsymbol{\mathcal{R}}_\mathrm{m}({\bar{\vec{u}}}^\mathrm{f},\bar{p}) \dO}
\end{aligned} \right\} B \notag \\
\left.
+{\sum_{e=1}^{n_\mathrm{el}} \int_{\Omega^\mathrm{f}_\mathrm{e}(t)}  \nabla \cdot \testfbar \tau_c \nabla \cdot \bar{\vec{u}}^{\mathrm{f}} \dO } \right\} C\notag \\ 
\left.
+\int_{\Os} \rho^\mathrm{s}\dd_t\bu^\mathrm{s}\cdot \tests \dO +\int_{\Os}\stress^\mathrm{s}:\grad\tests \dO \right\} D \notag\\
\left.
-\displaystyle\sum_\mathrm{e=1}^\mathrm{n_{el}}\int_{\Omega^\mathrm{f}_\mathrm{e}(t)} \tau_\mathrm{m} \testfbar\cdot (\boldsymbol{\mathcal{R}}_\mathrm{m}({{\bar{\vec{u}}}}^\mathrm{f},{\bar{p}}) \cdot \nabla {\bar{\vec{u}}}^\mathrm{f}) \mathrm{\dO}\right\}E\notag \\
\left.
-\displaystyle\sum_\mathrm{e=1}^\mathrm{n_{el}}\int_{\Omega^\mathrm{f}_\mathrm{e}(t)} \nabla \testfbar:(\tau_\mathrm{m}\boldsymbol{\mathcal{R}}_\mathrm{m}({\bar{\vec{u}}}^\mathrm{f},\bar{p}) \otimes \tau_\mathrm{m}\boldsymbol{\mathcal{R}}_\mathrm{m}({\bar{\vec{u}}}^\mathrm{f},\bar{p})) \mathrm{\dO}=\right\}F\notag \\
\left.
\int_{\Of(t)}\force^\mathrm{f}\cdot\testfbar \dO
+\int_{\G^\mathrm{f}_\mathrm{h}}\sigma^\mathrm{f}_{\mathrm{h}} \cdot\testfbar d\G + \int_{\Os}\force^\mathrm{s}\cdot\tests \dO
+\int_{\G^\mathrm{s}_\mathrm{h}}\sigma^\mathrm{f}_{\mathrm{h}} \cdot\tests d\G,
\right\} G
\end{align}
In the above equation, the terms $A$ and $G$ represent the Galerkin weak-form of the coarse scale component for the Navier-Stokes Eqs.~(\ref{eq:WeakFormNavierStokes_mom}) and (\ref{eq:WeakFormNavierStokes_cont}). The term $B$ represents the element level 
Galerkin least-squared (GLS) stabilization consisting of both the convection and pressure stabilizations terms to damp the spurious oscillations and to circumvent the inf-sup condition respectively. The term $C$ is the least square stabilization on the incompressibility constraint, which provides additional stability for large $Re$ problems. The term $D$ represents the Galerkin weak-from of the nonlinear structural dynamics. The terms $E$ and $F$ correspond to the cross-stress and Reynolds stress terms. $\boldsymbol{\mathcal{R}}_\mathrm{m}(\bar{\boldsymbol{u}}^\mathrm{f},\bar{p})$ is the residual of the momentum equation at the element level, and the stabilizing parameters $\tau_m$ and $\tau_c$ in the terms $(B-E)$ are the least squares metrics \citep{hughes_gls,franca_1992}.}

\changes{As mentioned earlier, the quasi-monolithic formulation decouples the fluid mesh motion and explicitly determines the mesh velocity $\vec{w}$, which is carried out by determining the 
interface between the fluid and the structure explicitly for $\mathrm{n}^\mathrm{th}$ time step using 
the second order Adam-Bashforth method
\begin{equation}
\vphin= \vphinn + \frac{3\Delta t}{2}\usnn - \frac{\Delta t}{2} \usnnn,
\label{InterfaceAdvancing}
\end{equation}
where $\vphin$ is the interface position between the fluid and the flexible body for any time $t^\mathrm{n}$. 
The fluid mesh nodes on the domain $\Otf$ can be updated for the interface locations using a pseudo-elastic 
material model given by
\begin{align}
\label{MeshEq}
\div \cdot \stress^\mathrm{m} = \vec{0},
\end{align} 
where $\stress^\mathrm{m}$ is the stress experienced by the fluid mesh due to the strain induced by the 
interface deformation. Assuming that the fluid mesh behaves as a linearly elastic material, its experienced stress can be
written as
\begin{align}
\label{meshEq}
\qquad \stress^\mathrm{m} = (1+\tau_\mathrm{m})\left[\left(\div \vec{\eta}^\mathrm{f}+\left(\div
\vec{\eta}^\mathrm{f}\right)^T\right)+\left(\div \cdot \vec{\eta}^\mathrm{f}\right)\vec{\mathrm{I}} \right],
\end{align}
where $\tau_\mathrm{m}$ is a mesh stiffness variable chosen as a function of the element size to 
limit the distortion of the small elements located in the immediate vicinity of the fluid-structure 
interface. The mesh stiffness variable $\tau_\mathrm{m}$ has been defined as $\tau_\mathrm{m}=\frac{\max_\mathrm{i} |T_\mathrm{i}|-\min_\mathrm{i}|T_\mathrm{i}|}{|T_\mathrm{j}|}$, where $T_\mathrm{j}$ represents $j^\mathrm{th}$ element on the mesh $T$.}

\section{Problem Set-up}\label{subsec:problemStatement}
\begin{figure}
	\centering
	\begin{subfigure}{0.365\textwidth}
		\includegraphics[width=0.99\columnwidth]{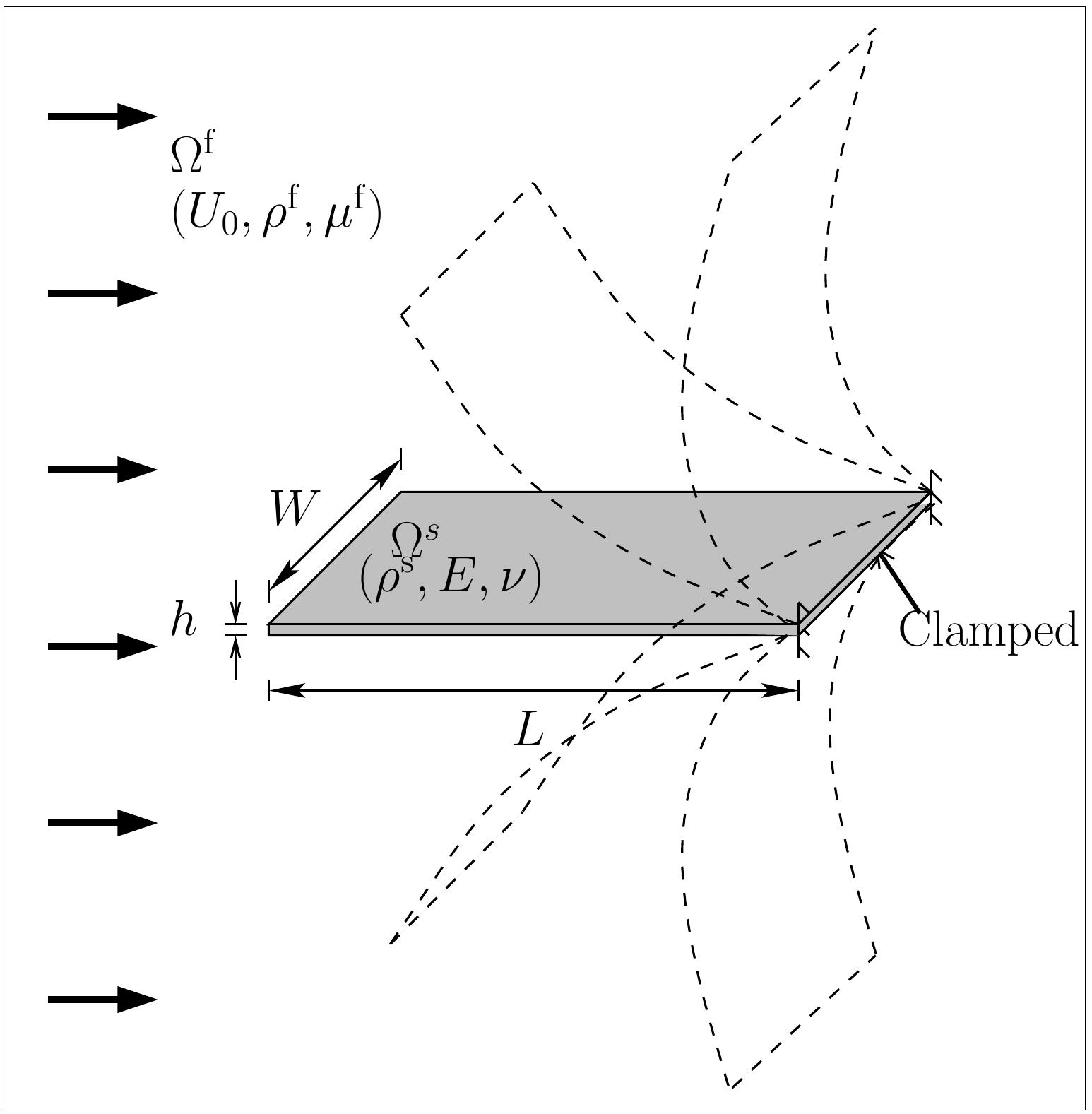}
		\caption{}
	\end{subfigure}
	\begin{subfigure}{0.42\textwidth}
		\includegraphics[width=0.99\columnwidth]{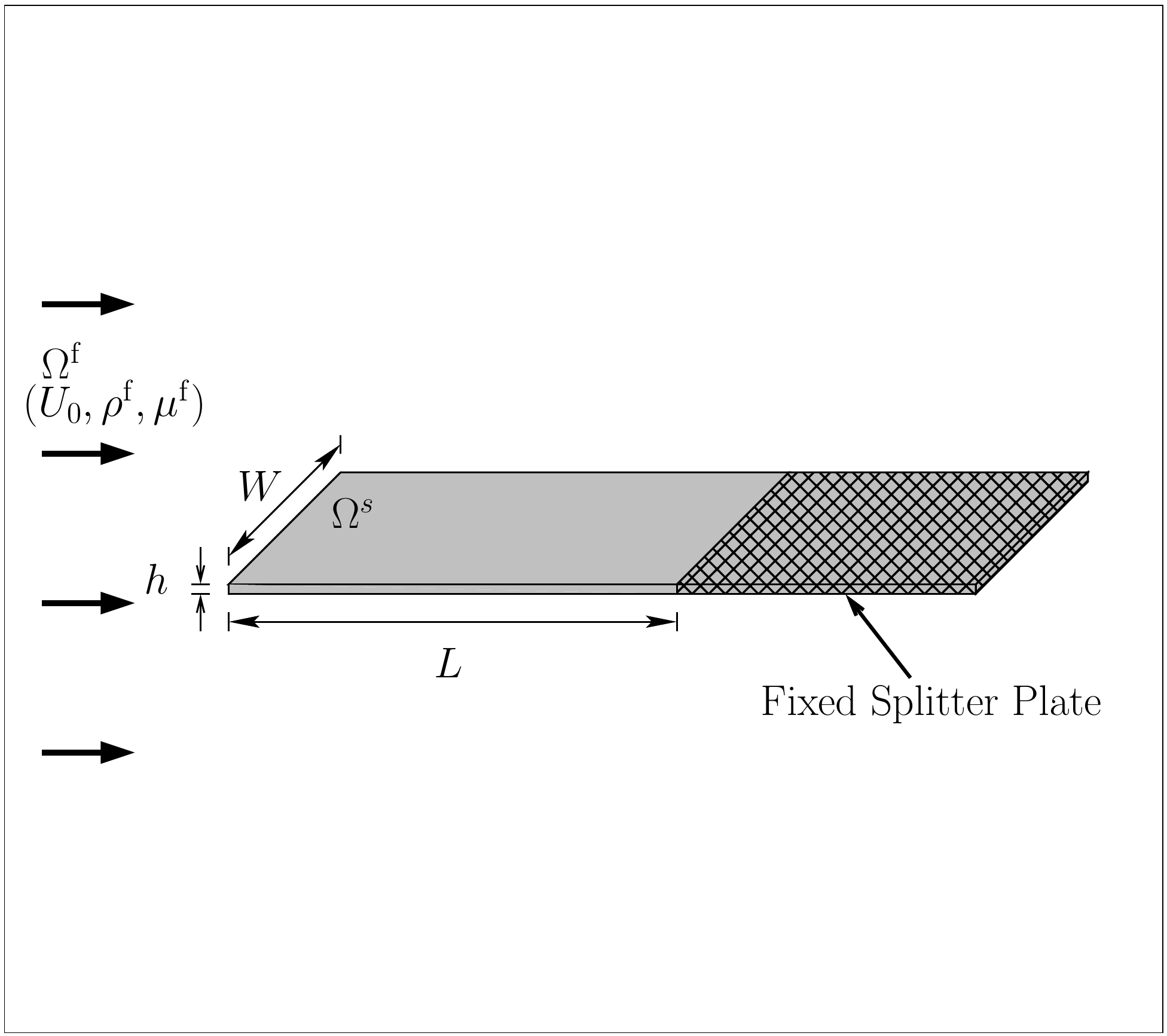}
		\caption{}
	\end{subfigure}
	\caption{Schematics of a flexible foil clamped at its TE and interacting with a uniform axial flow:
		(a) without splitter plate and (b) with a splitter plate. Dash-line ($---$) in (a) represents a typical LAF motion of inverted foil. }
	\label{3DSchematics_inv}
\end{figure}
\begin{figure}
	\centering
	\begin{subfigure}{0.77\textwidth}
		\includegraphics[trim=0mm 0mm 1mm 1mm,clip,width=0.99\columnwidth]{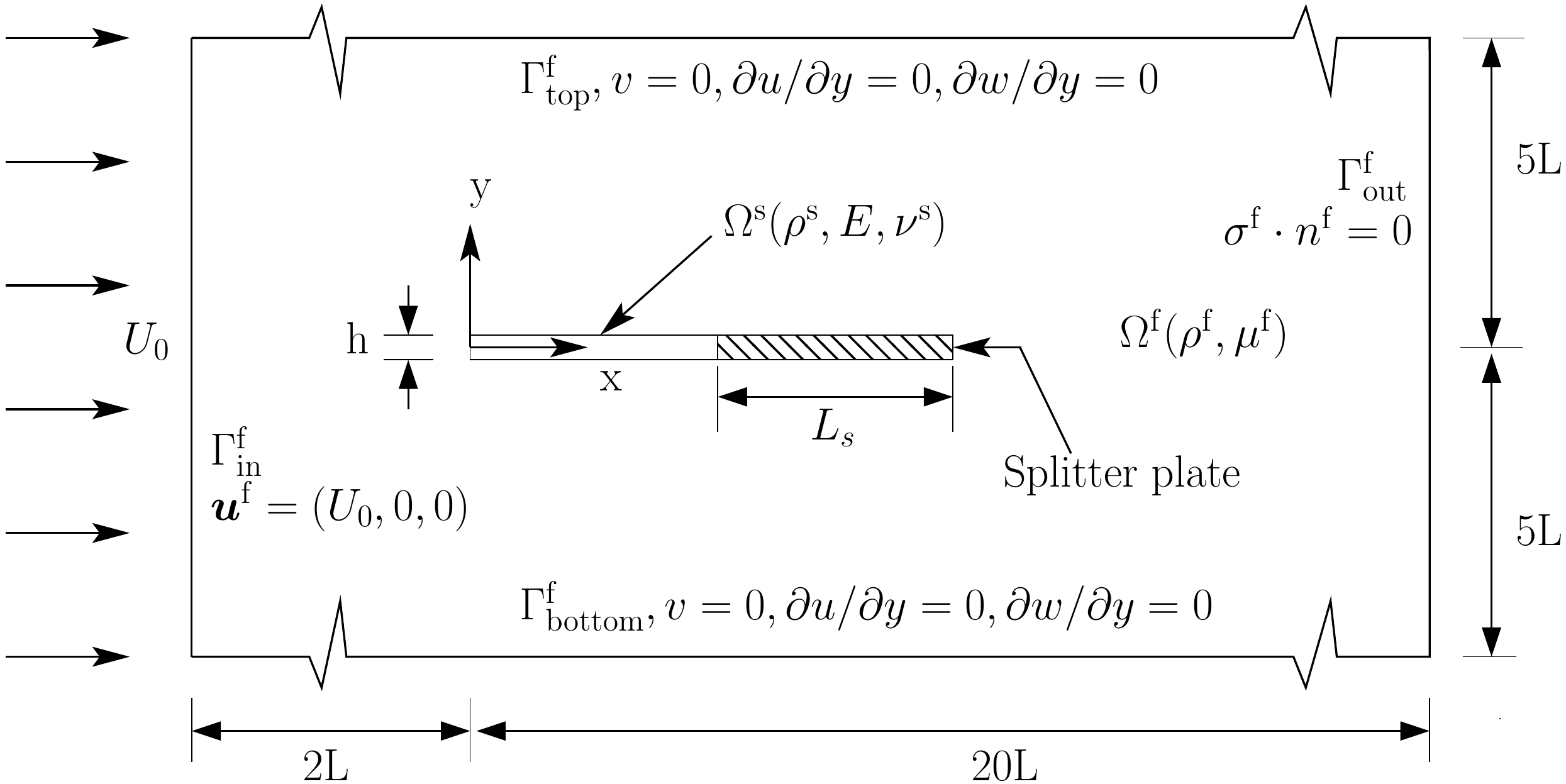}
		\caption{}
	\end{subfigure}
	\begin{subfigure}{0.21\textwidth}
		\includegraphics[trim=2mm 0mm 0mm 0mm,clip, width=1\columnwidth]{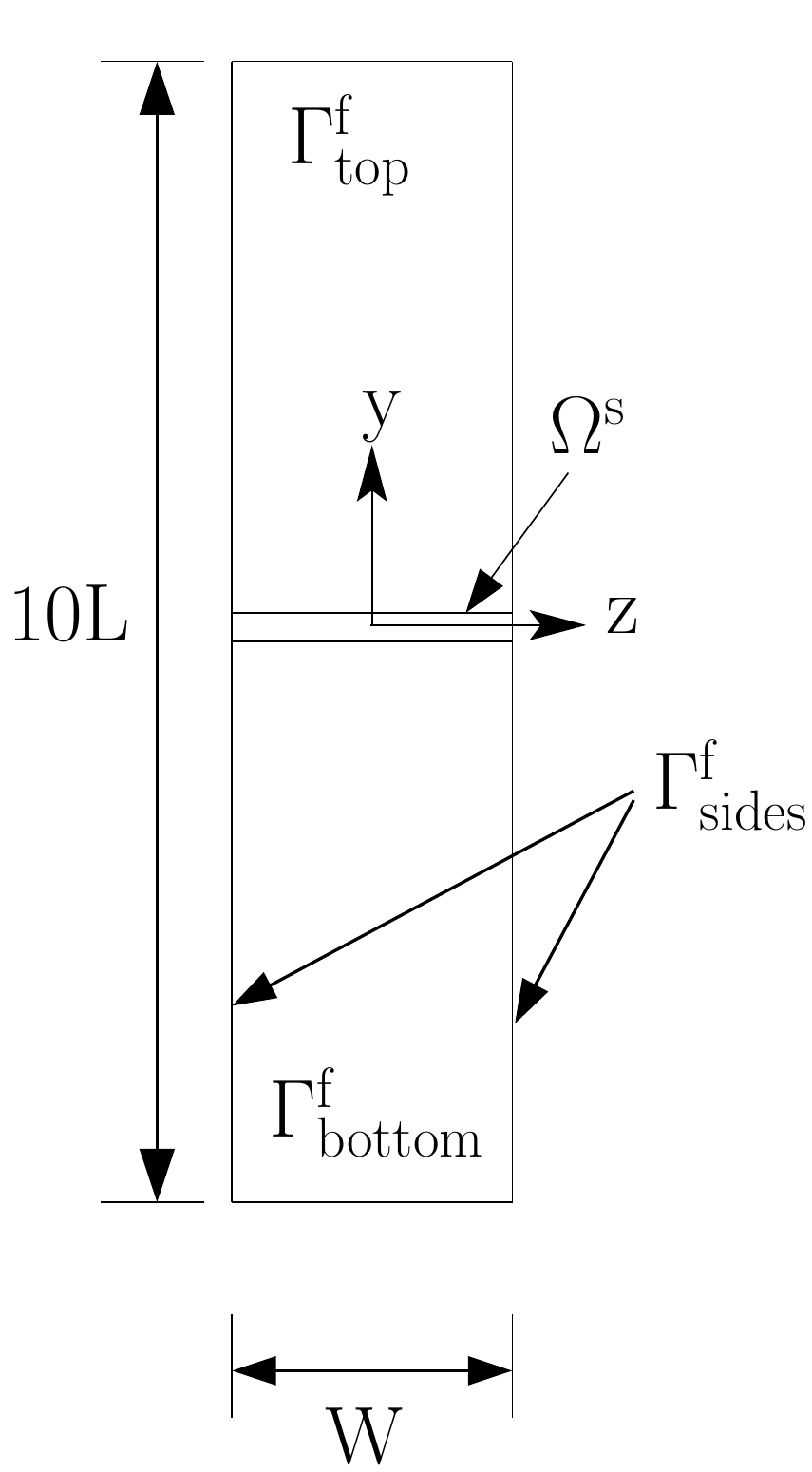}
		\caption{}
	\end{subfigure}\\
	\begin{subfigure}{0.4\textwidth}
		\centering
		\includegraphics[width=\columnwidth,trim=00mm 0mm 0mm 0mm,clip]{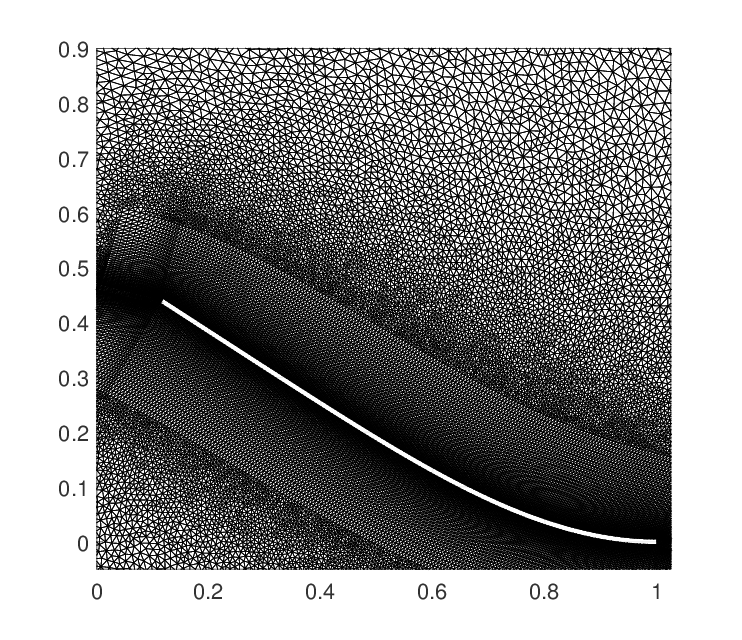}
	\end{subfigure}
	\begin{subfigure}{0.19555\textwidth}
		\centering
		\includegraphics[width=\columnwidth,trim=0mm 0mm 0mm 0mm,clip]{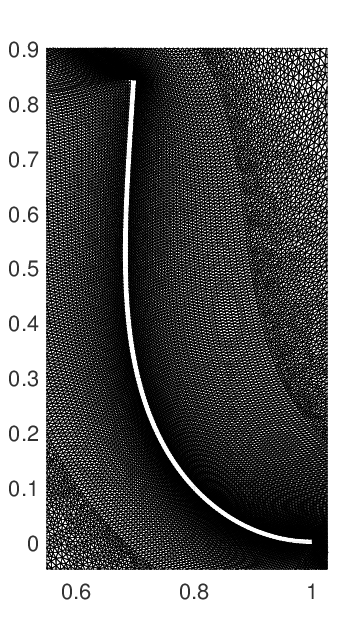}
	\end{subfigure}
	\begin{subfigure}{0.19555\textwidth}
		\centering
		\includegraphics[width=\columnwidth,trim=0mm 0mm 0mm 0mm,clip]{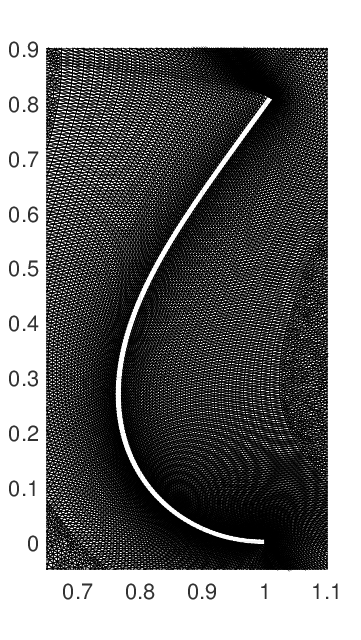}
	\end{subfigure}\\
	\begin{subfigure}{\textwidth}
		\centering
		\caption{}
	\end{subfigure}
	\caption{3D computational domain along with the boundary conditions: (a) front view, (b) side view, and 
		(c) a close-up view of the body-conforming boundary layer mesh around the inverted foil without splitter. 
	} \label{3DComputationDomain}
\end{figure}
We study the 3D nonlinear flapping dynamics of an inverted foil in a uniform axial flow, \changes{wherein the flow field and the foil can have spanwise variations,} for two different configurations,  as shown in figure~\ref{3DSchematics_inv}.  
In the first configuration, 
the  inverted foil interacts with an incompressible viscous flow with  the TE clamped. 
Whereas in the second configuration, we introduce a long fixed splitter plate 
at TE (figure~\ref{3DSchematics_inv}b) 
to inhibit the vortex-vortex interaction between the two counter-rotating 
vortices generated from TE and LE. \changes{In this study, we ignore the spanwise foil edge effects for simplicity 
and also from the viewpoint of computational efficiency. } 
Figures~\ref{3DComputationDomain}a and \ref{3DComputationDomain}b 
present the 3D \changes{spanwise periodic} computational setup of an inverted foil with length $L$ and thickness $H=0.01L$ clamped to a fixed splitter plate of length $L_s$ in Cartesian coordinate system. 
The size of the computational domain is \revi{$[22L\times 10L 
\times \mathrm{W}]$, where $\mathrm{W}$ is the width of the computational domain.} The foil is placed along the 
$x$-plane with its LE along the $y$-plane. 
At the inlet boundary $\G^\mathrm{f}_\mathrm{in}$, a stream of 
incompressible fluid enters into the domain  at a uniform velocity $U_0$. The slip-wall boundary condition \revi{$(v=0,\ \partial u/\partial y,\ \partial w/\partial y)$} is implemented along the top and bottom boundary surfaces $\G^\mathrm{f}_\mathrm{top}$ 
and $\G^\mathrm{f}_\mathrm{bottom}$,  respectively. The traction free 
condition is specified at the outflow plane $\G^\mathrm{f}_\mathrm{out}$ and
the computational domain is assumed to be periodic in the spanwise direction.
The body conforming quasi-monolithic formulation with exact interface tracking presented in Section~\ref{sec:quasiMonolithic} enables us to enforce the no-slip Dirichlet boundary condition along the deformable foil interface, which allows 
an accurate modeling of  the boundary layer on the foil.

%

\section{Validation}\label{sec:validation}

Before discussing the physical insight of LAF of an inverted foil, it is essential to establish the appropriate mesh resolution and validate 
the coupled fluid-structure solver.
\begin{figure}
	\centering
	\includegraphics[trim={0mm 0mm 5mm 12.5mm},clip,width=0.75\textwidth]{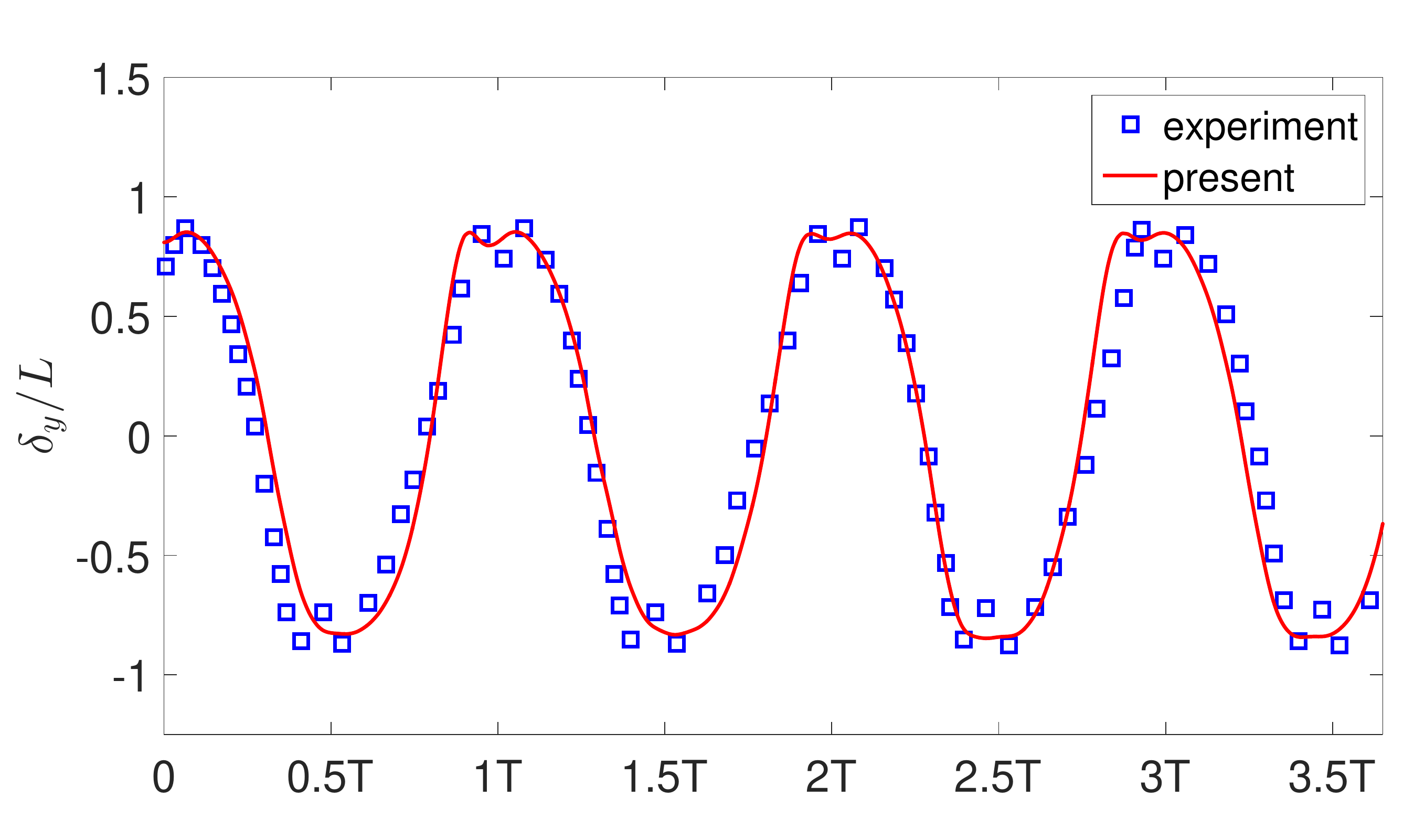}
	\includegraphics[trim={0mm 0mm 5mm 12.5mm},clip,width=0.75\textwidth]{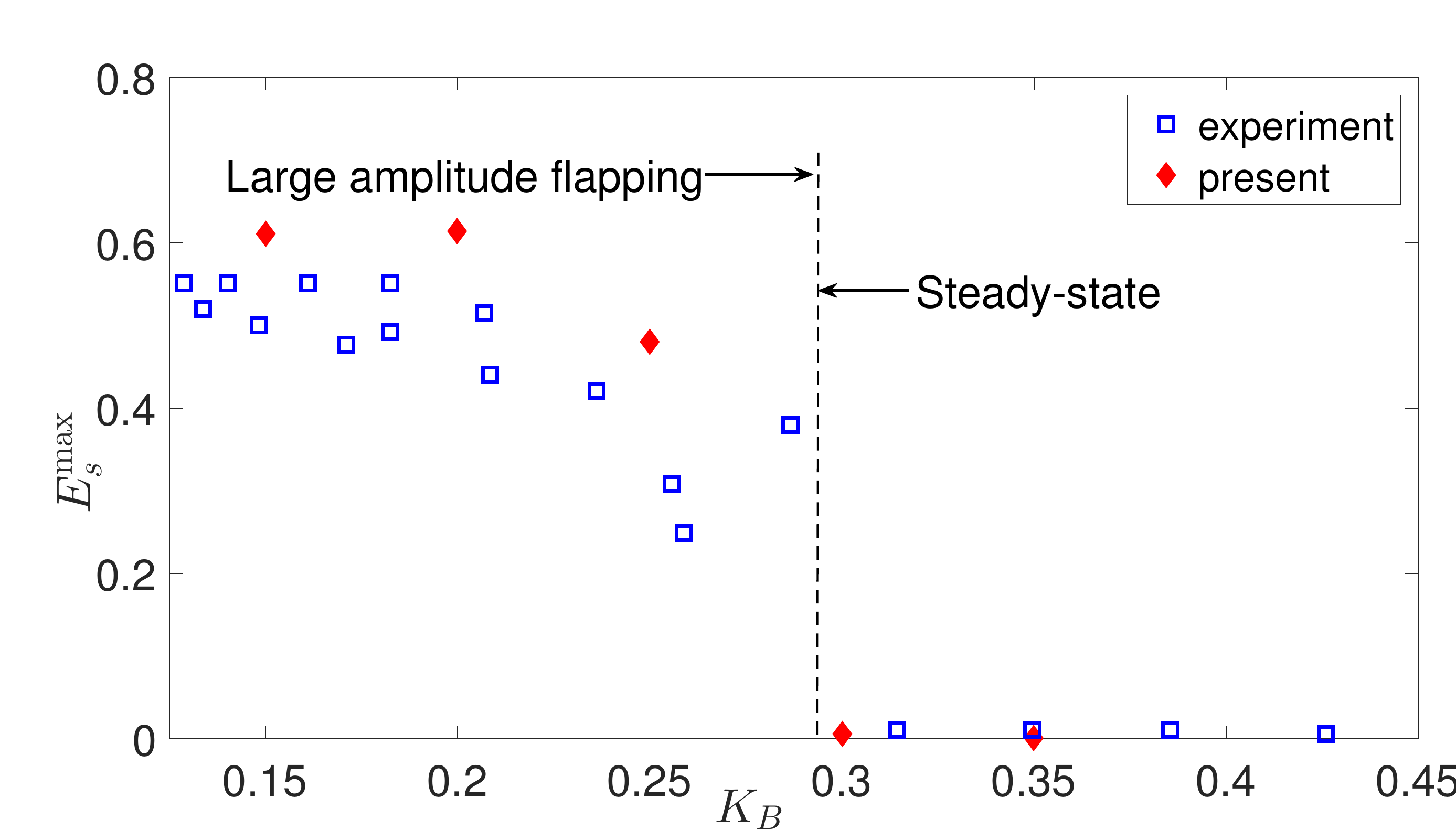}
	\caption{Comparison of LE transverse displacement of an inverted foil at $K_B=0.2$ (top) and maximum strain energy as function $K_B\in[0.15,0.35]$ (bottom) against experimental measurement \citep{kim2013} at $Re=30000$ and $m^*=1$. Here T represents the flapping time period. }\label{3dInvValidation}
\end{figure}
We validate the flapping dynamics of an inverted foil clamped at the TE without the splitter plate, i.e. $L_s=0$, for the nondimensional parameters 
$m^*=1, K_B \in [0.15,0.5]$ and $Re=30000$, which are consistent with the experimental 
conditions of \cite{kim2013}. \revi{The aspect ratio (AR$=\mathrm{W}/L$) of the foil is considered to be 0.5, a detailed analysis on the effect of AR on the flapping dynamics is presented in Section~\ref{sec:aspectRatio}.} The 3D simulations are performed on the 
finite element mesh with 1.9 million six-node wedge 
elements consisting of 21 layers of 2D finite element meshes with 46924 nodes each. 
Figure~\ref{3DComputationDomain}c presents a close-up view of the fluid boundary layer mesh around the foil at its maximum transverse displacement   for $K_B=0.2,\ Re=30000$ and $m^*=1$. To check the adequacy of the spatial
resolution, we have performed extensive grid refinement tests with different resolutions. Figure~\ref{3dInvValidation} (top)  presents a comparison between the experimental \citep{kim2013} LE cross-stream displacement and the current simulation at $K_B=0.2$. The LE displacement in the figure provides a good match with the experimental values. We also compare the maximum strain energy $E_s^\mathrm{max}$ developed due to the elastic deformation of the foil as a function of $K_B$ with experimental data \citep{kim2013} in figure~\ref{3dInvValidation} (bottom). The strain energy $E_s$ is evaluated as a function of curvature $\kappa$ using 
\begin{equation}
E_\mathrm{s}= {\frac{1}{2}\int_{0}^{l}EI \kappa^2 \mathrm{d}l}/{\rho^\mathrm{f}U_0^2 L^2} \quad \mbox{and} \quad \kappa = {\left|{\partial^2 f(x)}/{\partial x^2}\right|}/{\left[1+\left(\partial f(x)/\partial x\right)^2\right]^{3/2}},\label{strainEnergy}
\end{equation}
where $f(x)$ is a 
piecewise polynomial function of $6^\mathrm{th}$ order that has been constructed to define the deformed foil profile at each time instant. Figure~\ref{3dInvValidation} (bottom) clearly shows that our 3D simulations correctly predict the onset of flapping instability and the post-critical $E_s^\mathrm{max}$ follows a similar trend to that of the experiment. Some difference in the prediction of maximum strain energy $E_s^\mathrm{max}$ can be attributed to the spanwise end effects of the inverted foil in the experiment.

\begin{figure}
	\centering
	\begin{subfigure}{0.25\textwidth}
		\includegraphics[trim= 60mm 20mm 65mm 0mm,clip, width=0.99\columnwidth]{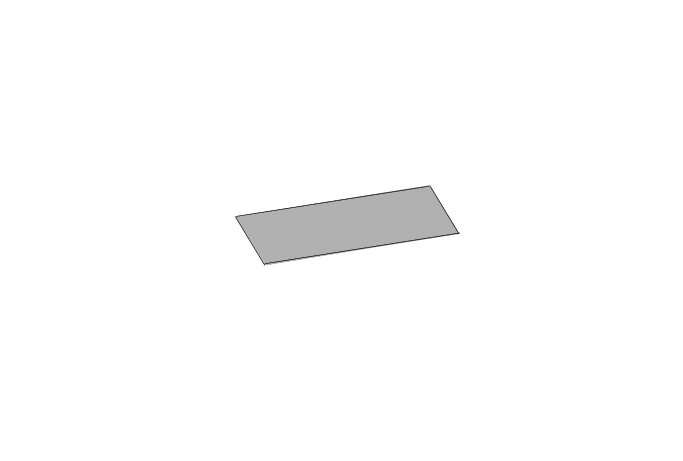}
		\caption{$K_B=0.35$}
	\end{subfigure}
	\begin{subfigure}{0.25\textwidth}
		\includegraphics[trim= 60mm 20mm 65mm 0mm,clip, width=0.99\columnwidth]{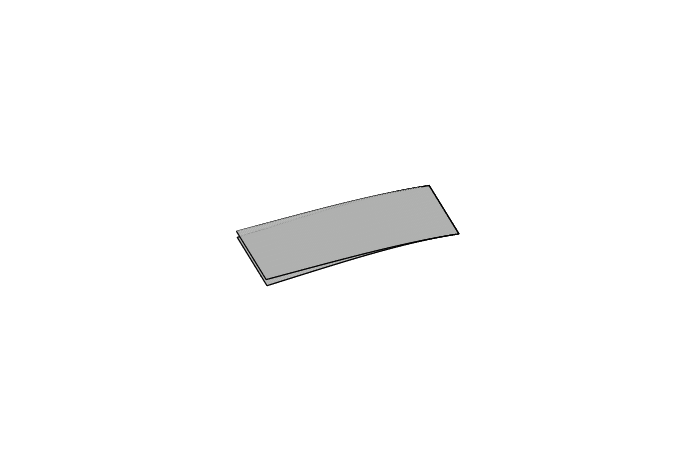}
		\caption{$K_B=0.3$}
	\end{subfigure}
	\begin{subfigure}{0.25\textwidth}
		\includegraphics[trim= 60mm 20mm 65mm 0mm,clip, width=0.99\columnwidth]{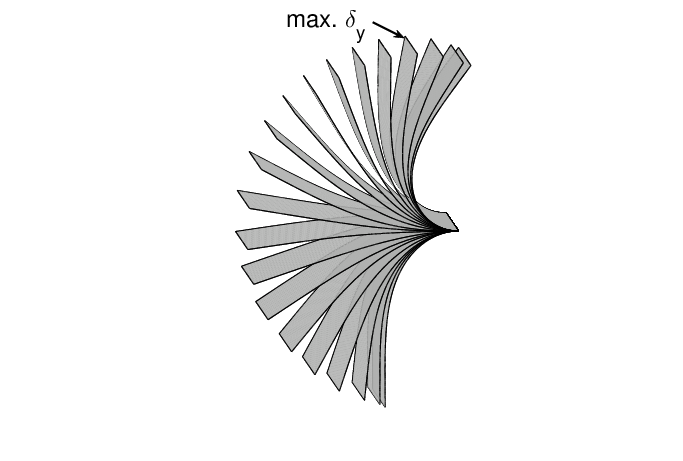}
		\caption{$K_B=0.2$}
	\end{subfigure}
	\caption{3D inverted flexible foil profiles over a flapping cycle at $m^*=1$ and $Re=30000$ for $K_{B}=$  (a) 0.35, (b) 0.3, and (c) 0.2.}\label{3DInvFoilProfile}
\end{figure}

Figure~\ref{3DInvFoilProfile} summarizes the coupled flapping modes exhibited by an inverted foil as a function of $K_B$. The foil loses it stability for $K_B \approx 0.3$ to perform the flapping motion that is primarily biased towards one side. This observation is consistent with the previous 2D analysis presented in \cite{gurugubelli_JFM}. The onset of flapping instability can be attributed to the combination of divergence and the flow separation at LE. By further decreasing $K_B$, 
the foil begins to exhibit the LAF motion (figure~\ref{3DInvFoilProfile}c) and the foil profiles for $K_B=0.2$ in figure~\ref{3DInvFoilProfile}c show that the foil does not recoil at its maximum transverse displacement. Instead, the foil deforms further downstream before it recoils and the foil  goes through two humps-like patterns in the vicinity of maximum transverse displacement, as shown in figure~\ref{3dInvValidation} (left).

\begin{figure}
	\begin{subfigure}{0.49\textwidth}
		\centering
		\includegraphics[width=0.98\columnwidth,trim=0cm 0cm 2cm 0cm,clip]{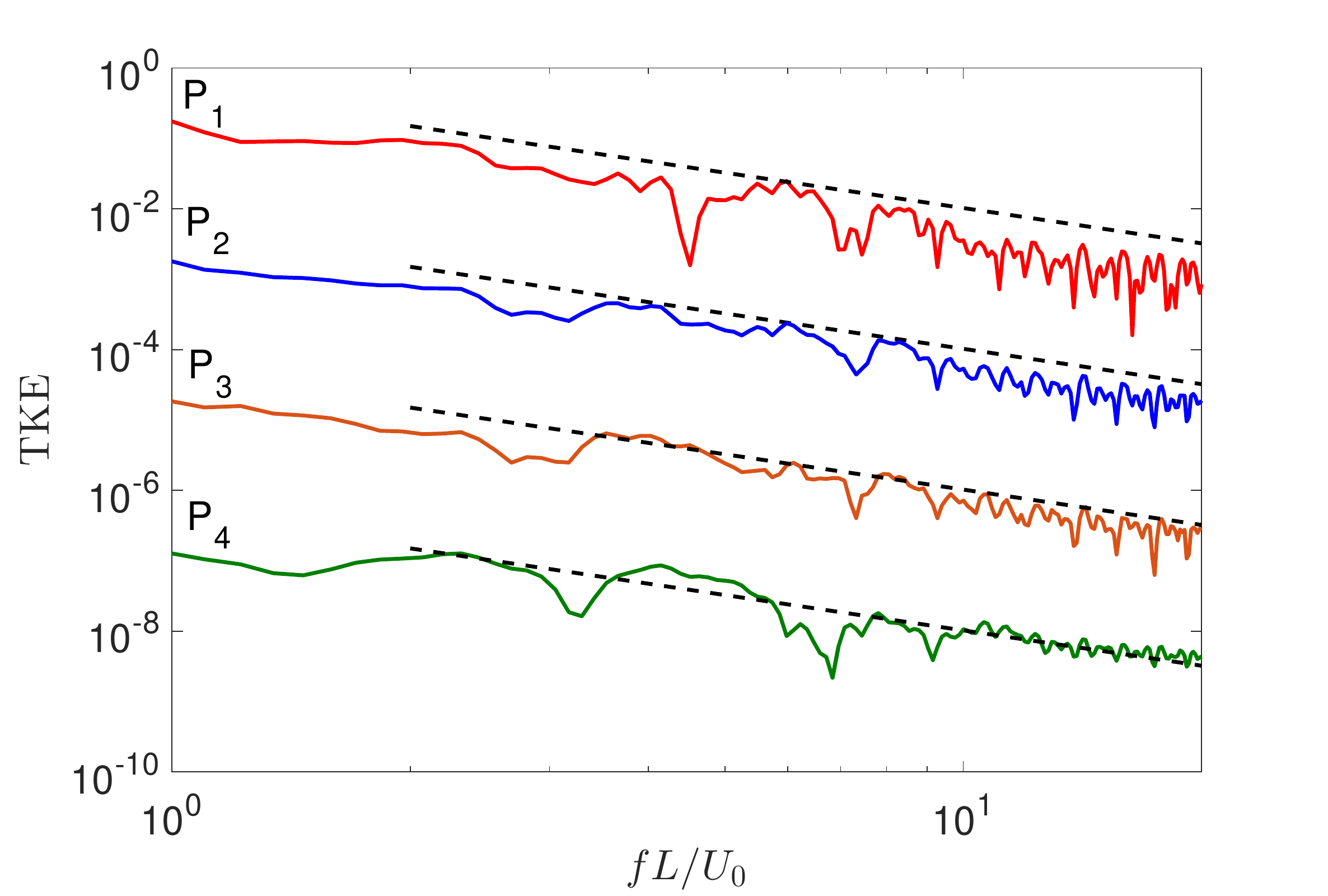}
	\end{subfigure}
	\begin{subfigure}{0.49\textwidth}
		\centering
		\includegraphics[width=0.98\columnwidth,trim=0cm 0cm 2cm 0cm,clip]{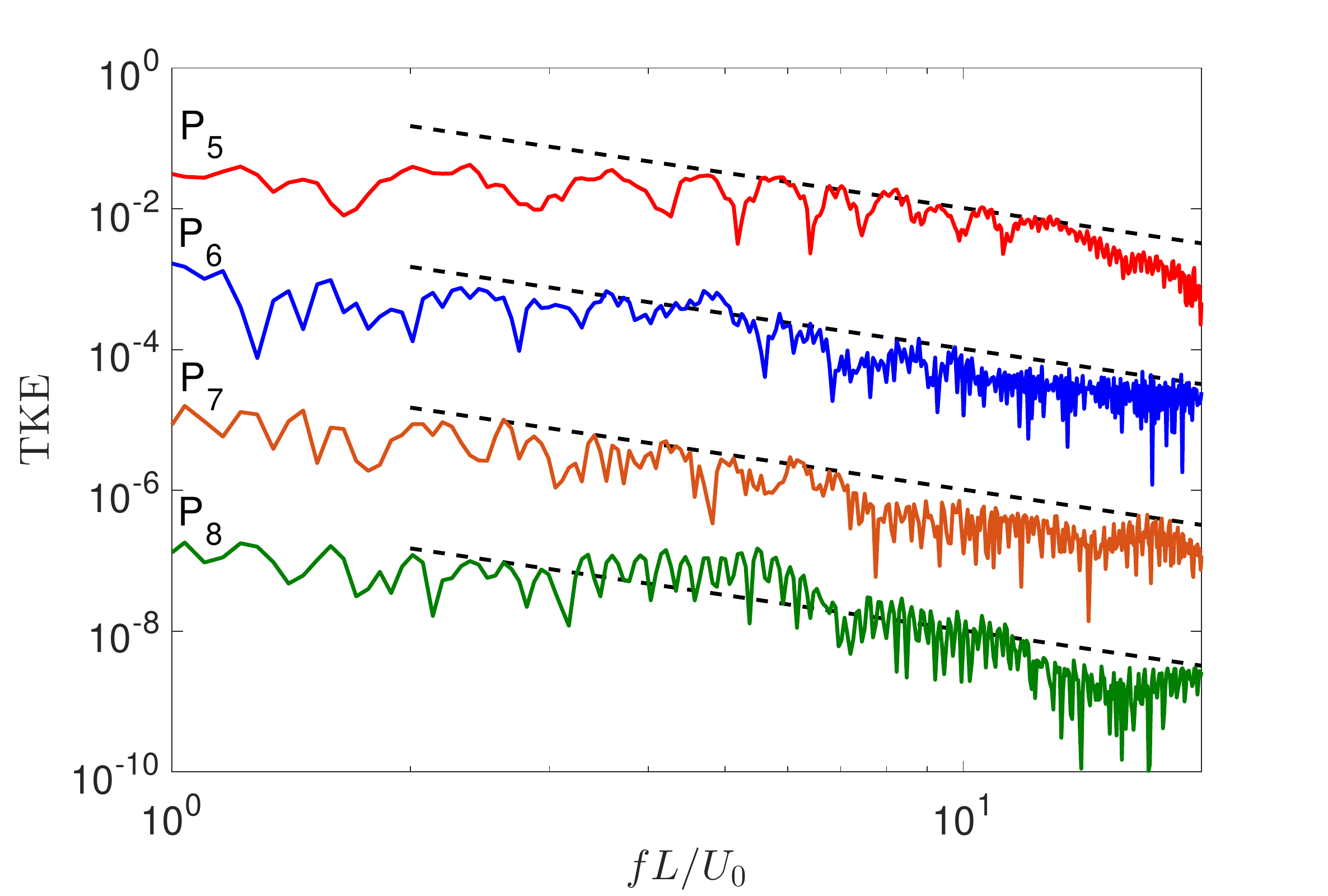}
	\end{subfigure}
	\caption{Spectral distribution of turbulent kinetic energy (TKE) at different positions in the wake for $K_B=0.2,\ m^*=1$ and $Re=30000$: $\mathrm{P}_1=(2L,0.75L),\ \mathrm{P}_2=(2L,0.5L)\ \mathrm{P}_3=(2.5L,0.75L),\ \mathrm{P}_4=(2.5L,0.5L),\ \mathrm{P}_5=(2L,-0.75L),\ \mathrm{P}_6=(2L,-0.5L),\ \mathrm{P}_7=(2.5L,-0.75L)$ and $\mathrm{P}_8=(2.5L,-0.5L)$. Except for the highest spectra, other spectral trends are shifted down for clarity. The dashed lines indicate the Kolmogorov's -5/3 power law.}
	\label{fig:Spectra}
\end{figure}

\revi{Although we have validated our numerical framework by comparing the time history of LE displacement with the experimental data, it is also important to verify if the mesh used here adequately captures the turbulent wake characteristics. For that purpose, we look into the turbulence power spectra in the wake region. 
Figure \ref{fig:Spectra} shows the spectral distribution of the turbulent kinetic energy (TKE) at eight points ($\mathrm{P}_1-\mathrm{P}_8$) in the wake behind the inverted foil. The figure shows that the kinetic energy decays with a slope of $-5/3$.
For locally homogeneous turbulence, 
Kolmogorov's -5/3 power states that the kinetic energy and the wave number, $k$, of integral scale eddies
follow the relation $TKE(k)\propto k^{-5/3}$ in the inertial range. 
This relationship is generally applied to the spatial 
distribution of turbulent kinetic energy ($TKE \equiv u_i u_i/2$) as a function 
of wavenumber spectrum. 
The $-5/3$ decay of the 
kinetic energy can be extended to frequency domain for the turbulence spectrum via the Taylor's frozen turbulence hypothesis 
\citep{TaylorSpectrum}. 
Therefore, the Kolmogorov's  $-5/3$ spectral decay of TKE versus the frequency in figure~\ref{fig:Spectra} confirms the adequacy of the fluid mesh for the VMS-based turbulence model presented in Section~\ref{sec:quasiMonolithic}.}
\section{Results}
\changes{The LCO flapping of inverted foil forms a large periodic obstacle to the impinging flow, which alters the fluid motion and generates a separated wake flow. Similar to a bluff body structure, the separated wake flow consists of vortices shed periodically from the deformed foil. The interaction between the vortices from the LE and TE,  and the flexible structure are strongly coupled to each other. The periodic vortex shedding in the separated wake flow can synchronize with the flapping motion. To realize the mechanism of LAF, we will investigate the onset of LAF, the response histories, the frequency characteristics,  the force dynamics, the vortex shedding patterns and the synchronization between the vortices shed during the flapping phenomenon.  
}
\subsection{Flow field}\label{sec:flowField}
\begin{figure}
	\centering
	\begin{subfigure}{0.49\textwidth}
		\includegraphics[width=0.9\columnwidth]{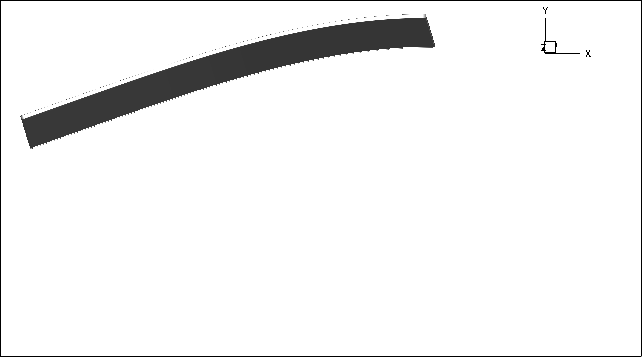}
	\end{subfigure}
	\begin{subfigure}{0.49\textwidth}
		\includegraphics[width=0.9\columnwidth]{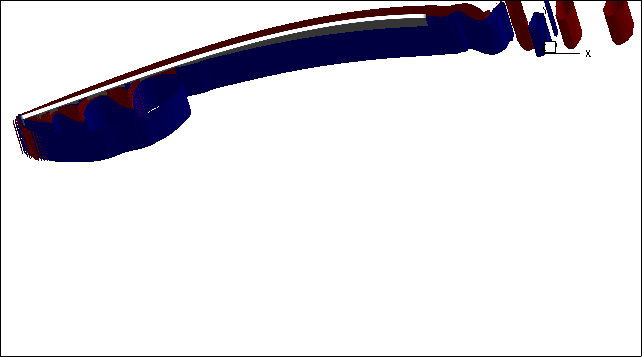}
	\end{subfigure}\\
	\begin{subfigure}{\textwidth}
		\centering
		\caption{$tU_0/L=7$}
	\end{subfigure}\\
	\begin{subfigure}{0.49\textwidth}
		\includegraphics[width=0.9\columnwidth]{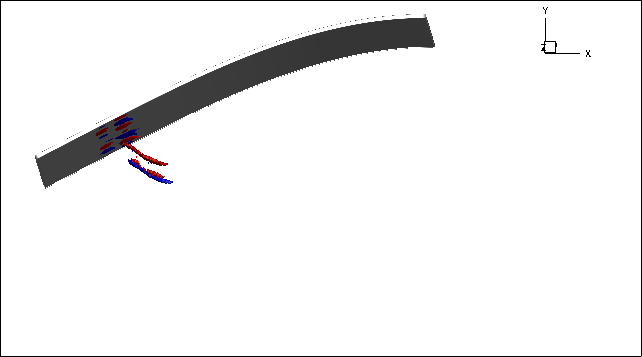}
	\end{subfigure}
	\begin{subfigure}{0.49\textwidth}
		\includegraphics[width=0.9\columnwidth]{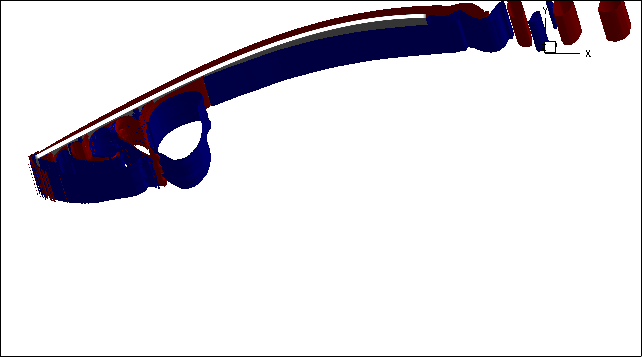}
	\end{subfigure}\\
	\begin{subfigure}{\textwidth}
		\centering
		\caption{$tU_0/L=7.5$}
	\end{subfigure}\\
	\begin{subfigure}{0.49\textwidth}
		\includegraphics[width=0.9\columnwidth]{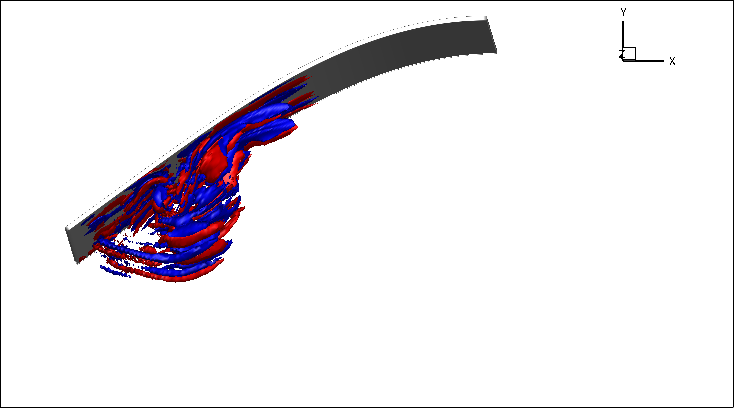}
	\end{subfigure}
	\begin{subfigure}{0.49\textwidth}
		\includegraphics[width=0.9\columnwidth]{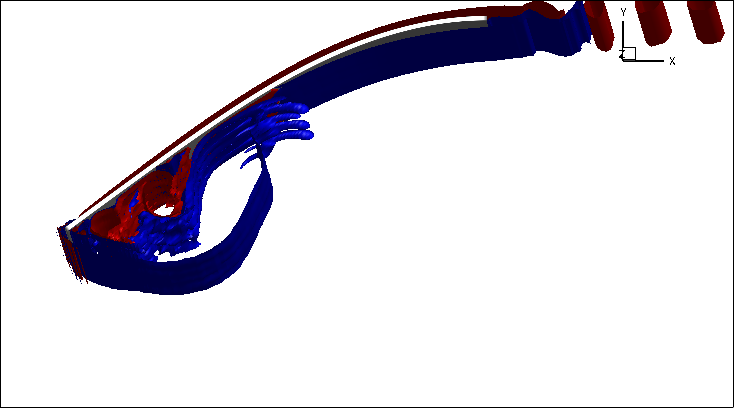}
	\end{subfigure}\\
	\begin{subfigure}{\textwidth}
		\centering
		\caption{$tU_0/L=8$}
	\end{subfigure}
	\caption{Temporal evolution of 3D-LEV over a deformable inverted foil in uniform flow: Iso-surfaces of streamwise $\omega_x$ (left) and spanwise $\omega_z$ (right) vorticity at $K_B=0.2,\ m^*=1$ and $Re=30000$. Here blue and red colors denote positive and  negative vorticity respectively.}\label{3D_LEV}
\end{figure}

\changes{Since the separated vortex structures play a key role in determining the periodic loading on a flexible body, we analyze the wake topology, more specifically we focus on the interaction of the LEV and the TEV develop behind the deformable flexible foil. The 2D simulations by \cite{gurugubelli_JFM} have shown that the onset of LAF is characterized by the LEV due to the flow separation at LE for sufficiently large deformation of the inverted foil. We first performed a series of 3D simulations at $Re=1000$ for $K_B=[0.4,0.25]$, $m^*=0.1$ and $\mathrm{AR}=0.5$ to find out the impact of flow 3D effects on the LAF response. Unlike the 2D simulations of \cite{gurugubelli_JFM} for the identical nondimensional parameters,  the 3D simulations at $K_B=0.4$ do not show any LAF phenomenon. The foil loses its stability only for $K_B\le0.3$ confirming that the foil 3D-effects play a significant role in the onset of LAF by stabilizing the onset of divergence instability and then flow separation at the leading edge. This observation conforms with the analysis presented by \cite{sader_2016}.}

\begin{figure}
	\centering
	\begin{subfigure}{0.75\textwidth}
		\centering
		\includegraphics[width=0.95\columnwidth]{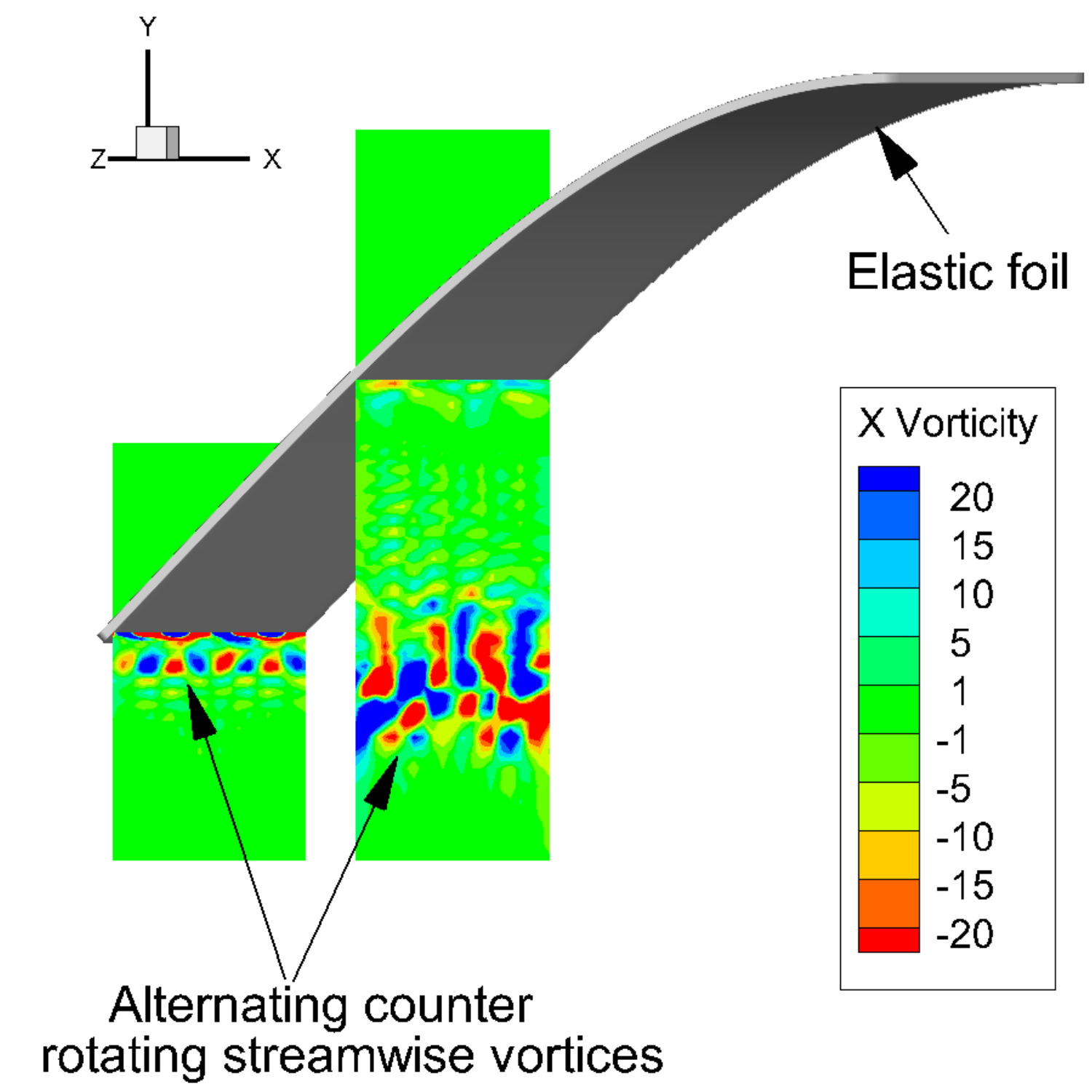}
	\end{subfigure}
	\caption{Instantaneous view of streamwise $\omega_x$ along the sectional planes at $x=0.2L$ and $0.45L$ for $K_B=0.2,\ Re=30000$ and $m^*=1$.}\label{streamwiseVorticityLEV}
\end{figure}

\changes{To visualize the role of 3D flow structures during the evolution of LAF,
in figure~\ref{3D_LEV}, we plot the evolution of the streamwise and spanwise vorticity of LEV at $K_B=0.2,\ Re=30000$, $m^*=1$ and $\mathrm{AR}=0.5$. As the foil deflects from its initial state due to divergence instability, at a certain angle of deformation a small 2D LEV develops behind the foil which can be seen in figure~\ref{3D_LEV}a for $tU_0/L=7$. This 2D LEV results in a low pressure region behind the foil which enhances the downward forces acting on the foil that in turn increases the foil downward deformation. The 2D LEV grows in size with the foil deformation and the first signs of 3D vortex structures are observed in figure~\ref{3D_LEV}b for $tU_0/L=7.5$. As the foil deflects downwards, it pushes the fluid below it normal to the surface with a velocity greater than the freestream velocity. This phenomenon typically represents jet in a cross-flow which tends to produce counter-rotating vortex pairs in the streamwise direction \citep{jetFlows,Jets_annRevie}. Figures~\ref{3D_LEV}b (left) and~\ref{3D_LEV}c (left)  also show the occurrence of similar counter-rotating streamwise vortex pairs. To further confirm this behavior, we plot the $x$-vorticity along the sectional planes at $x=0.2L$ and $0.45L$ for $tU_0/L=8$ in figure~\ref{streamwiseVorticityLEV}. The figure clearly shows the formation of 3D streamwise counter-rotating vortices along the foil span.} 

\begin{figure}
	\centering
	\begin{subfigure}{0.275\textwidth}
		\includegraphics[width=0.99\columnwidth]{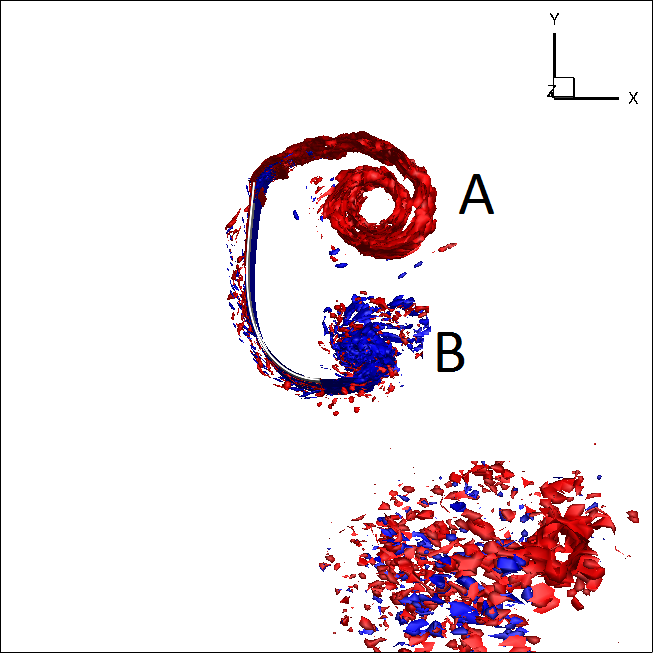}
		\caption{$tU_0/L=13.5$}
	\end{subfigure}
	\begin{subfigure}{0.275\textwidth}
		\includegraphics[width=0.99\columnwidth]{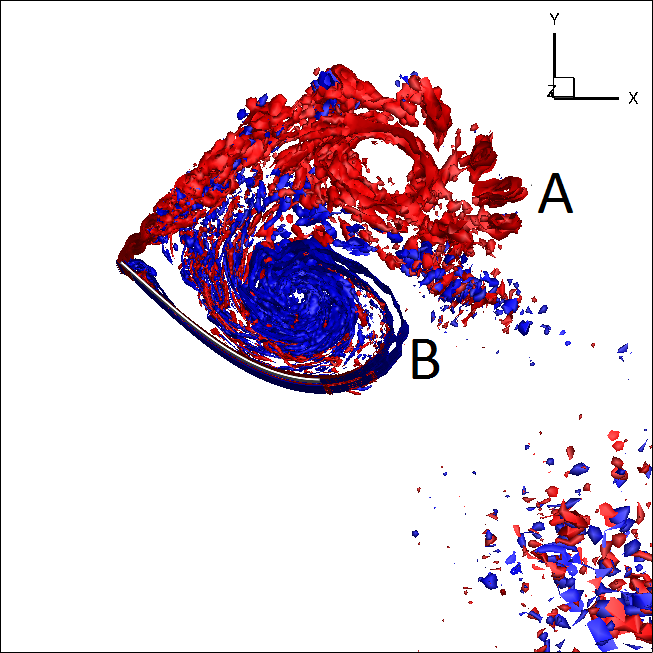}
		\caption{$14.5$}
	\end{subfigure}
	\begin{subfigure}{0.275\textwidth}
		\includegraphics[width=0.99\columnwidth]{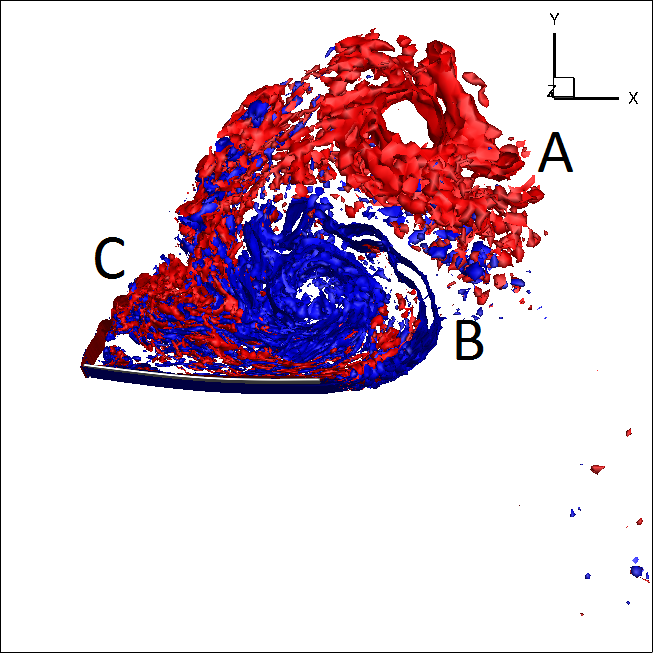}
		\caption{$15.75$}
	\end{subfigure}\\
	\begin{subfigure}{0.275\textwidth}
		\includegraphics[width=0.99\columnwidth]{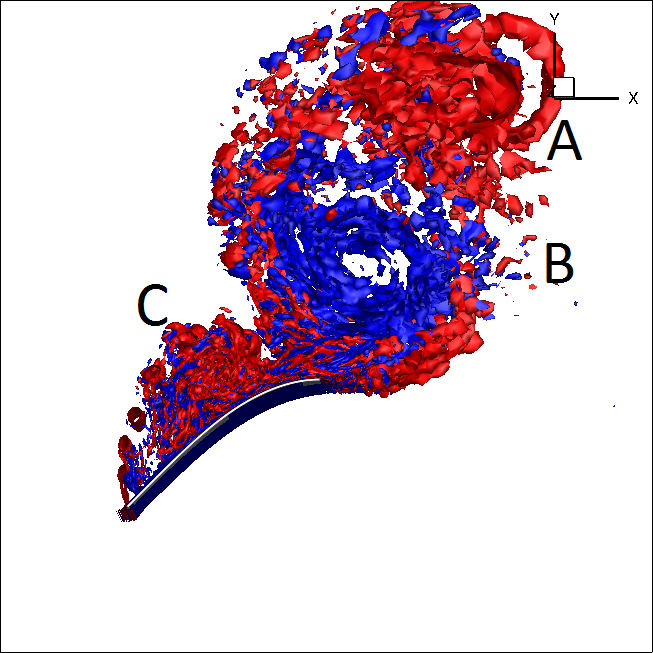}
		\caption{$16.25$}
	\end{subfigure}
	\begin{subfigure}{0.275\textwidth}
		\includegraphics[width=0.99\columnwidth]{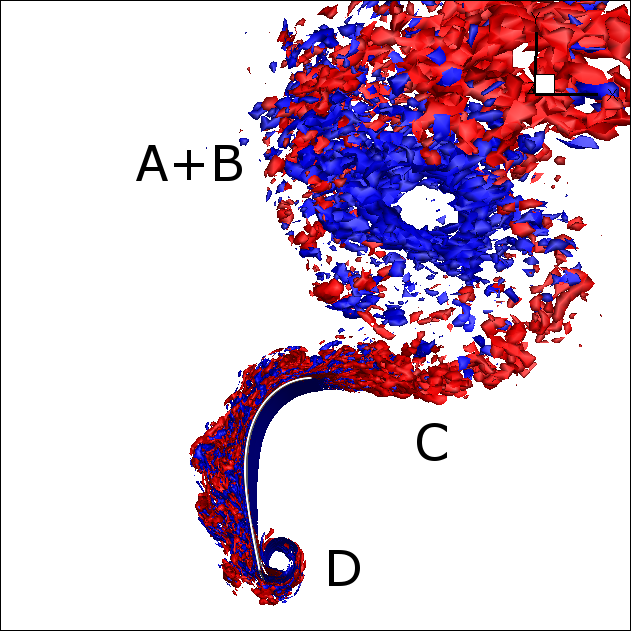}
		\caption{$tU_0/L=17$}
	\end{subfigure}
	\begin{subfigure}{0.275\textwidth}
		\includegraphics[width=0.99\columnwidth]{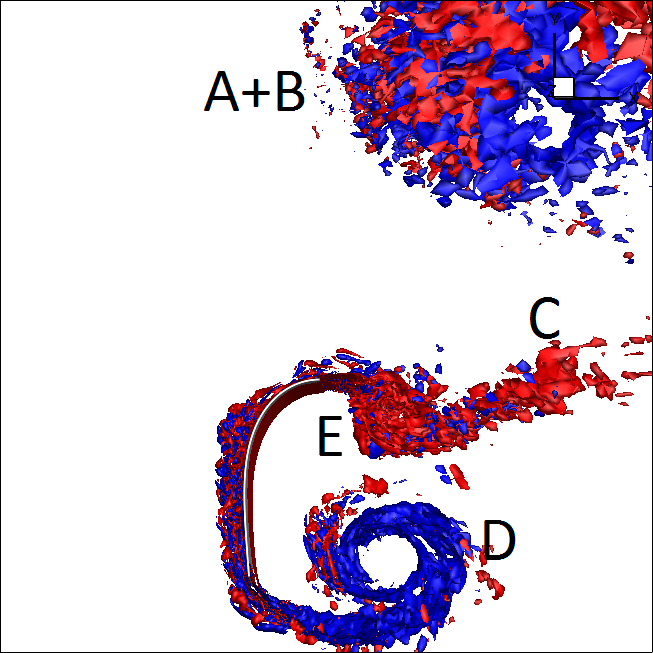}
		\caption{$tU_0/L=17.5$}
	\end{subfigure}
	\caption{Generation and interaction of LEV and TEV behind an inverted foil over a half flapping oscillation: Time evolution of nondimensional spanwise $\omega_z$ distribution (blue: $\omega_z=15$ and red: $\omega_z=-15$) at $K_B=0.2,\ m^*=1$ and $Re=30000$.}\label{3D_inv_vortex}
\end{figure}


We next present the 3D vortex organization for the LAF regime corresponding to
$(KB, Re, m^*,\mathrm{AR})= (0.2, 30000,1.0,0.5)$ over a half-cycle. 
Figure~\ref{3D_inv_vortex} presents the vortex mode exhibited by the LAF over one-half cycle of the periodic oscillation for the downstroke. 
Figures~\ref{3D_inv_vortex}a-c show the LEV `A' and an oppositely signed TEV `B', which are formed due to 
the roll of shear layer via Kelvin-Helmholtz instability, interact with each other to form a pair of counter-rotating vortices.
The counter-rotating vortex `B' grows in size and cuts-off the vortex `A' from the LE. Meanwhile, as the LEV `A' is shed, a new LEV `C' can be seen developing at the LE in figure~\ref{3D_inv_vortex}c and  this phenomenon is marked by a rise in the lift acting on the foil. However, this time foil inertia overcomes the lift acting on the foil and as the foil crosses the mean position, the LEV `C' convects along the foil surface and sheds into the wake. \changes{A detailed analysis describing the impact of the vortices on the force dynamics is described in the next paragraph.}
A new LEV `D' can be seen developing on the opposite surface of the foil once the foil crosses certain deformation which in turn pulls an oppositely signed TEV `E'. {Similar to the downstroke even the upstroke generate a pair of counter-rotating vortices. However, during the upstroke we do not observe the secondary LEV `C'.} \changes{The LAF mode for $K_B=0.2$ exhibits a combination of $P+S$ and $P$ vortex modes due to transition from the pure $P$ mode for $K_B=0.25$ to the pure $P+S$ mode for $K_B=0.15$. Similar $P$ and $P+S$ vortex modes per half-cycle have also been observed by \cite{ryu_2015,gurugubelli_JFM,mittal_2016} through 2D simulations. The vortex mode notation used here is consistent with the notation introduced by \cite{williamson1988} for an oscillating circular cylinder.}

\begin{figure}
	\centering
	\begin{subfigure}{0.49\textwidth}
		\includegraphics[width=0.99\columnwidth,trim=0mm 0mm 0mm 2mm,clip]{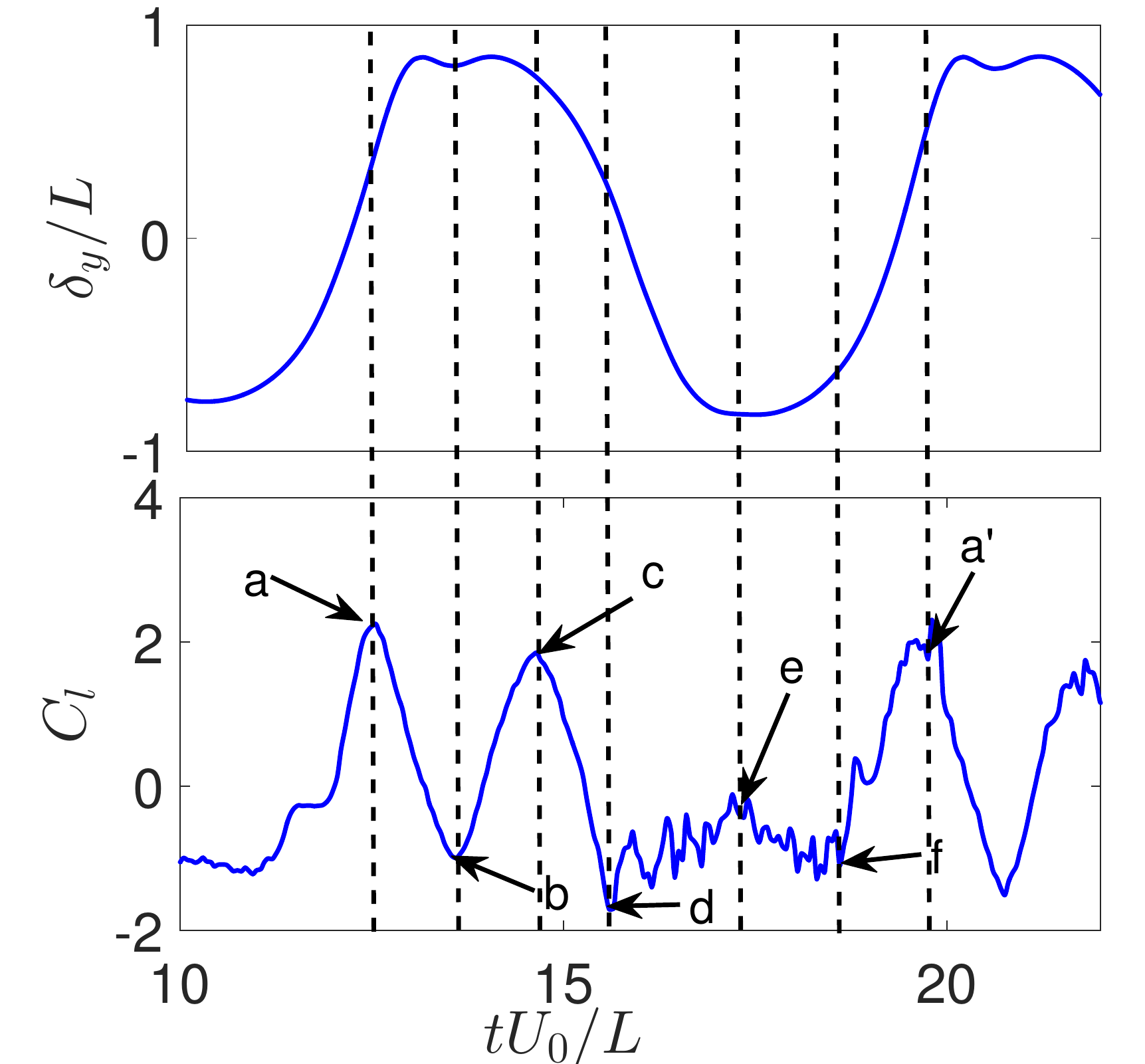}
	\end{subfigure}
	\begin{subfigure}{0.49\textwidth}
		\includegraphics[width=0.99\columnwidth,trim=0mm 0mm 0mm 2mm,clip]{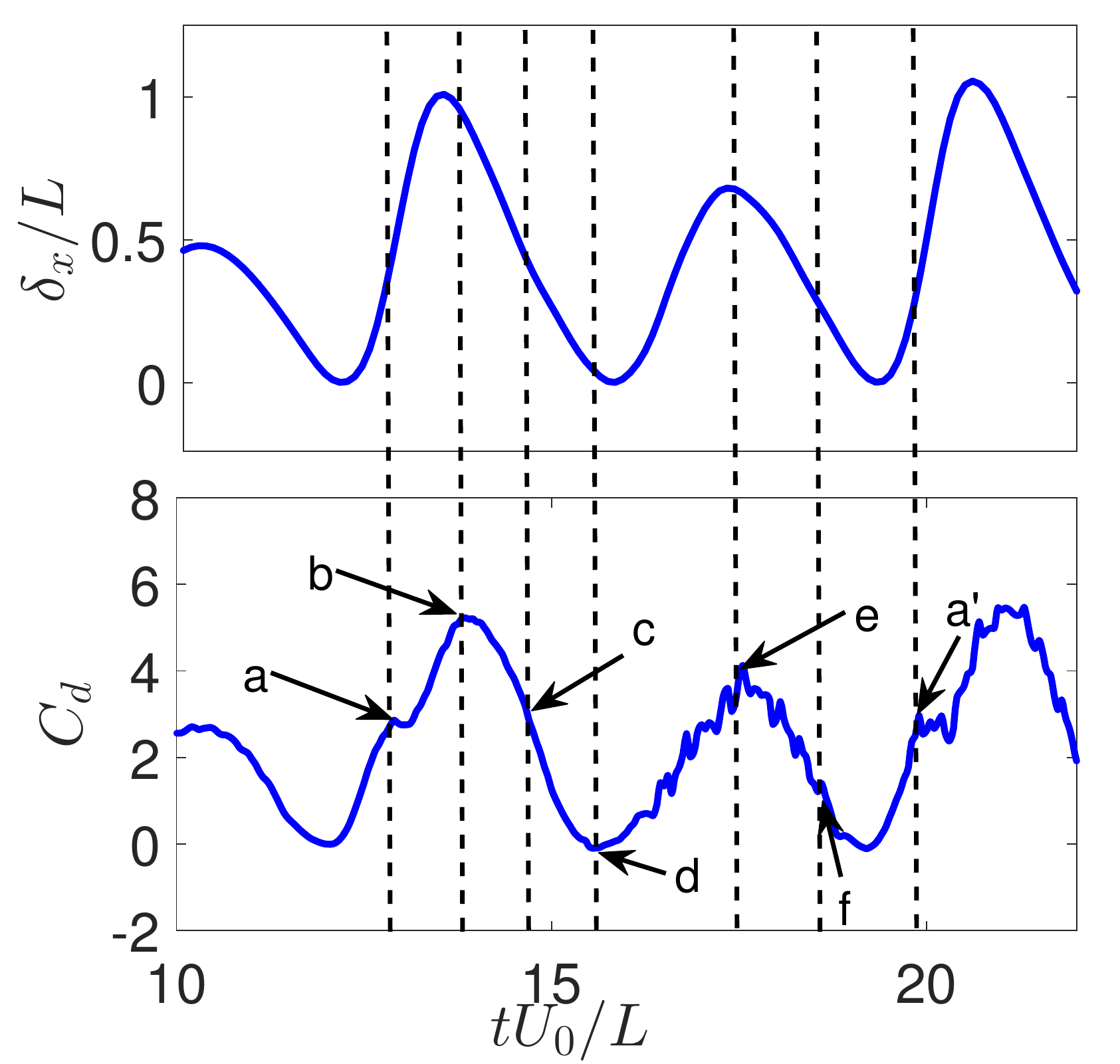}
	\end{subfigure}	
	\caption{Time histories of response amplitudes (top) and force coefficients (bottom) for 
		the inverted foil at  nondimensional parameters $K_B=0.4,\ Re=30000,\ m^*=1$ and $AR=0.5$:
		In this plot points (a,b,c,d,e,f,a$\mathrm{'}$) correspond to the time instances $tU_0/L$=(12.6, 13.5, 14.65, 15.65, 17.5, 18.6, 19.8).}\label{comparisonPlot}
\end{figure}

\changes{To investigate the influence of the vortex modes on the flapping motion and forces acting on the inverted foil, we plot the time traces 
of the LE displacements in both transverse and  streamwise directions  and contrast them against the lift and drag coefficients in figure~\ref{comparisonPlot}. 
The figure shows that the lift acting on the foil reaches the maximum at point `a' even before it reaches the maximum transverse displacement. 
Between points `a' and `b', while the transverse lift force acting on the foil experiences a sudden drop, the drag force and streamwise displacements continue to increase. 
In this time window, the transverse amplitude increases continuously and reaches to the maximum value.
We can attribute this observation to the large foil deformation like in figure~\ref{3D_inv_vortex}a, where the majority of the fluid force acting on the foil contributes 
to the horizontal drag rather than to the vertical lift force. In other words, the deformed foil develops a maximum projected area to the oncoming flow stream. 
 Above a critical foil deformation, the flapping dynamics is dominated by the drag acting on the foil. 
Point `b' in the figure represents the time instance at which the inverted foil recoils from the peak streamwise position of the LE . 
 During the initial stages of the downstroke, i.e. between points `b' and `c', the lift acting on the foil begins to increase because 
the lift force is recovered back due to the elastic recoil of the foil. 
It should be noted that at this stage the LEV `A' is not yet shed and is still attached to the LE. At point `c', the LEV `A' separates from the LE. 
The separation of the LEV `A' is characterized by a sudden drop in the lift and the drag. 
The drag acting on the foil reaches the minimum as it crosses the mean position. On the other hand,  around the same time the lift acting on the foil starts to increase slightly 
due to the formation of the secondary LEV `C' at `d'. 
%
While no sharp changes in the lift even when the LEV `D' is formed, one can observe oscillations in the lift plot between the points `d' and `e'. 
This is because the secondary LEV `C' in front of the foil minimizes the effect of the LEV `D' behind the foil and thus we do not observe any sharp changes in the lift plot. 
Thereby resulting in a lower streamwise flapping amplitude and drag for the downstroke compared to the upstroke. 
The inverted foil again recoils back at the point `e'. Similar to the recoil phenomenon during the downstroke, the lift force acting on the foil is recovered before the LEV `D' separates from the LE. The separation is again characterized by a sharp change in the lift curve at point `f'. Since no secondary LEV is observed during the upstroke, the lift acting on the foil increases until the point `$\mathrm{a}'$'. This phenomenon continues and repeats itself over each flapping cycle.}

The formation of counter-rotating vortex pair of `A+B' in figure~\ref{3D_inv_vortex}a represents a close resemblance to the flow over a circular cylinder wherein the interaction between the counter-rotating asymmetric vortices plays a significant role in the  vortex-induced vibration when the frequency of vortex formation is relatively close to the natural frequency of the structure. 
The VIV lock-in of an elastically mounted vibrating circular  is characterized  by the matching of the
periodic vortex shedding frequency and the oscillation frequency of the body \cite{sarpkaya2004}.
\changes{The VIV response characteristics of transversely vibrating circular bodies become significantly influenced if we disturb the synchronization between the shedding phenomenon or suppress either of the two alternate vortices.}  The prevention of vortex-vortex interaction and the frequency lock-in can reduce the large transverse amplitudes. However, the role of the \changes{vortex organization} and the interaction between two counter-rotating vortices on the LAF is not systematically explored earlier. To understand the impact of the vortices on the LAF, we introduce a splitter plate at the trailing edge to suppress the TEV.
\subsection{Effect of splitter plate behind inverted flapping}\label{sec:splitterPlate}
\begin{figure}
	\centering
	\begin{subfigure}{0.275\textwidth}
		\includegraphics[width=0.99\columnwidth]{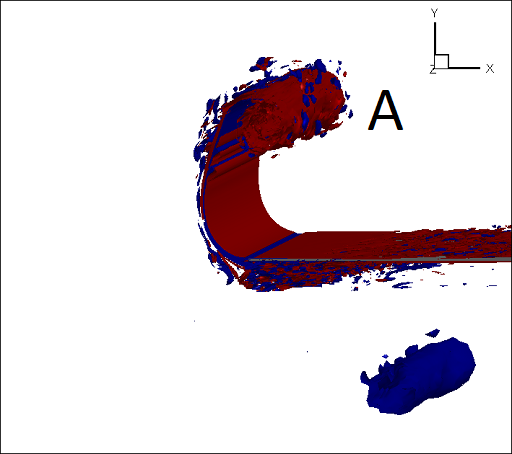}
		\caption{$tU_0/L=15$}
	\end{subfigure}
	\begin{subfigure}{0.275\textwidth}
		\includegraphics[width=0.99\columnwidth]{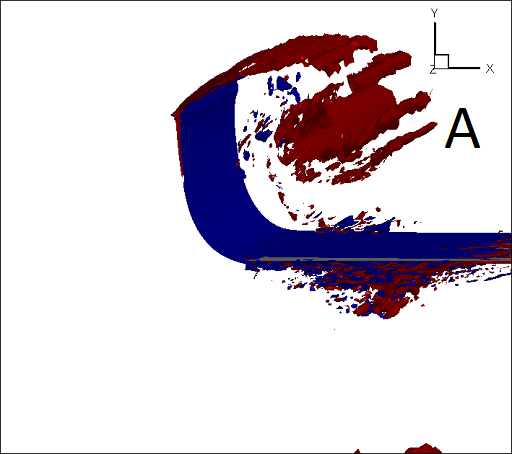}
		\caption{$tU_0/L=16$}
	\end{subfigure}	
	\begin{subfigure}{0.275\textwidth}
		\includegraphics[width=0.99\columnwidth]{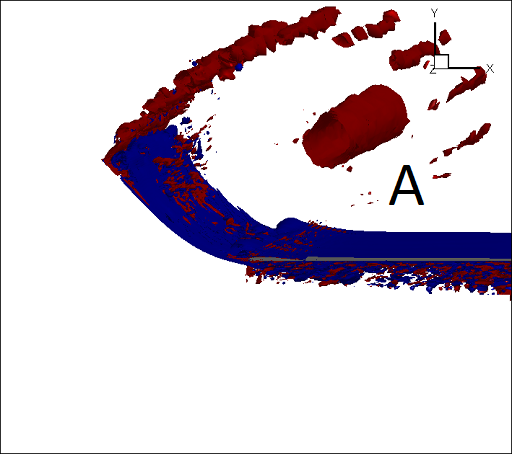}
		\caption{$tU_0/L=17$}
	\end{subfigure}
	\begin{subfigure}{0.275\textwidth}
		\includegraphics[width=0.99\columnwidth]{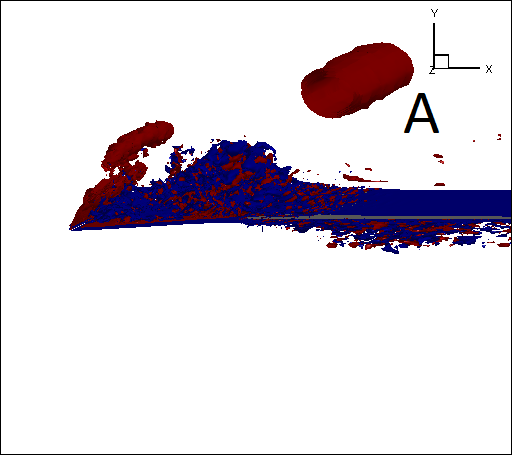}
		\caption{$tU_0/L=18$}
	\end{subfigure}
	\begin{subfigure}{0.275\textwidth}
		\includegraphics[width=0.99\columnwidth]{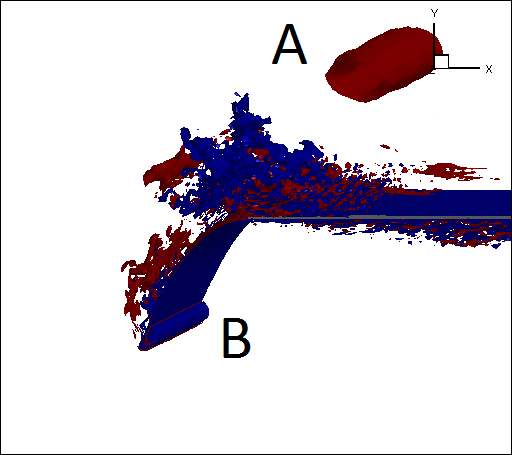}
		\caption{$tU_0/L=18.8$}
	\end{subfigure}
	\begin{subfigure}{0.275\textwidth}
		\includegraphics[width=0.99\columnwidth]{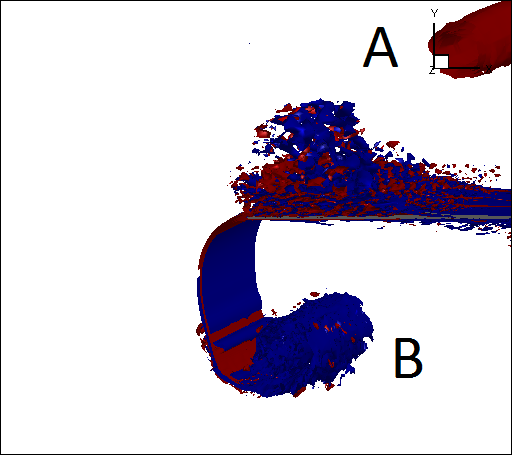}
		\caption{$tU_0/L=19.8$}
	\end{subfigure}
	\caption{Generation and interaction of LEV and TEV behind an inverted foil with splitter plate over a half flapping oscillation: Time evolution of $\omega_z$ at nondimensional parameters $K_B=0.2,\ m^*=1$ and $Re=30000$. Here blue and red denote positive and negative vorticity respectively and the flow is from left to right.}\label{vorticity_splitter}
\end{figure}

We replicate the 3D simulation at $Re=30000,\ K_B=0.2,\ m^*=1$ and $\mathrm{AR}=0.5$ by considering a typical splitter plate of length $L_s=4L$ at TE. 
Similar to the circular cylinder, the splitter plate will inhibit the vortex-vortex interaction between the counter-rotating vortices from the LE and the TE. Figure~\ref{vorticity_splitter} presents the isometric vortex mode observed for an inverted foil with a splitter plate at the trailing edge. The figure shows that unlike the inverted foil without splitter plate which exhibited $2P+2S$ vortex mode per flapping cycle, in this case only two counter-rotating vortices `A' and `B' are shed over the flapping cycle i.e. $2S$ vortex mode. Additionally, by comparing figures~\ref{3D_inv_vortex}~and~\ref{vorticity_splitter}, it can be observed that the time taken over one half cycle of the inverted foil with splitter is 20\% more than that of a simple inverted foil without a splitter and this increase in the time period is mainly due to the increase in time taken by the foil to reach the mean position. However, for both the inverted foil configurations, the time taken by the foil to deform from its mean position to the point of recoil remains nearly identical. Lower time-periods during the first half of the downstroke in the case of inverted foil without splitter plate can be attributed to the greater induced velocity due to the interaction between the counter-rotating vortices `A+B'. As a result of this, the drag force which is the main source of the inverted foil bending reduces more rapidly than over the foil with the splitter plate. Detailed analysis on the impact of the vortex-vortex interaction on the flapping amplitudes and the forces is presented in the following paragraphs.

\begin{figure}
	\centering
	\begin{subfigure}{0.49\textwidth}
		\includegraphics[width=0.99\columnwidth,trim=10mm 10mm 0mm 2.5mm,clip]{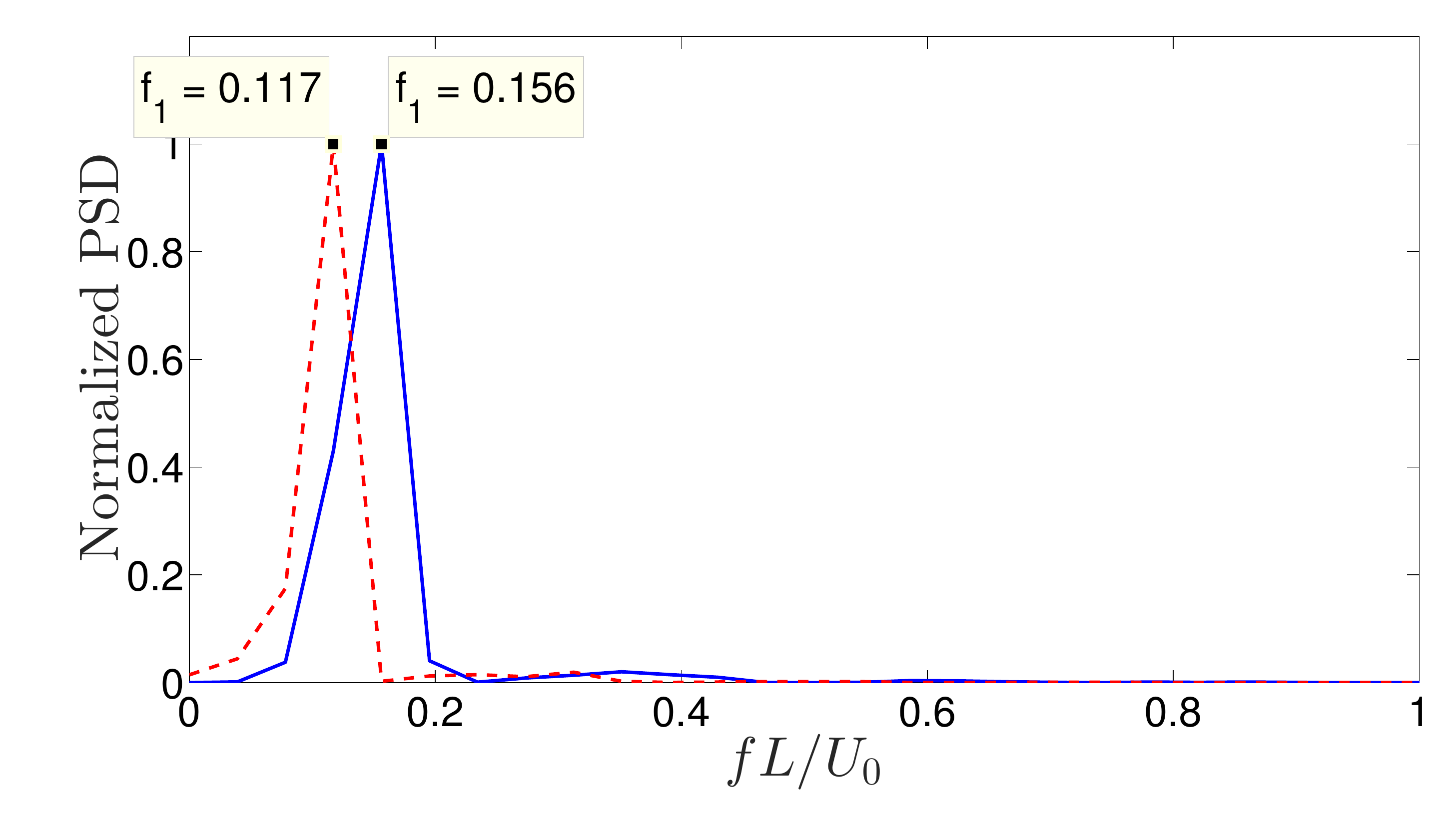}
		\caption{transverse frequency}
	\end{subfigure}
	\begin{subfigure}{0.49\textwidth}
		\includegraphics[width=0.99\columnwidth,trim=10mm 10mm 0mm 2.5mm,clip]{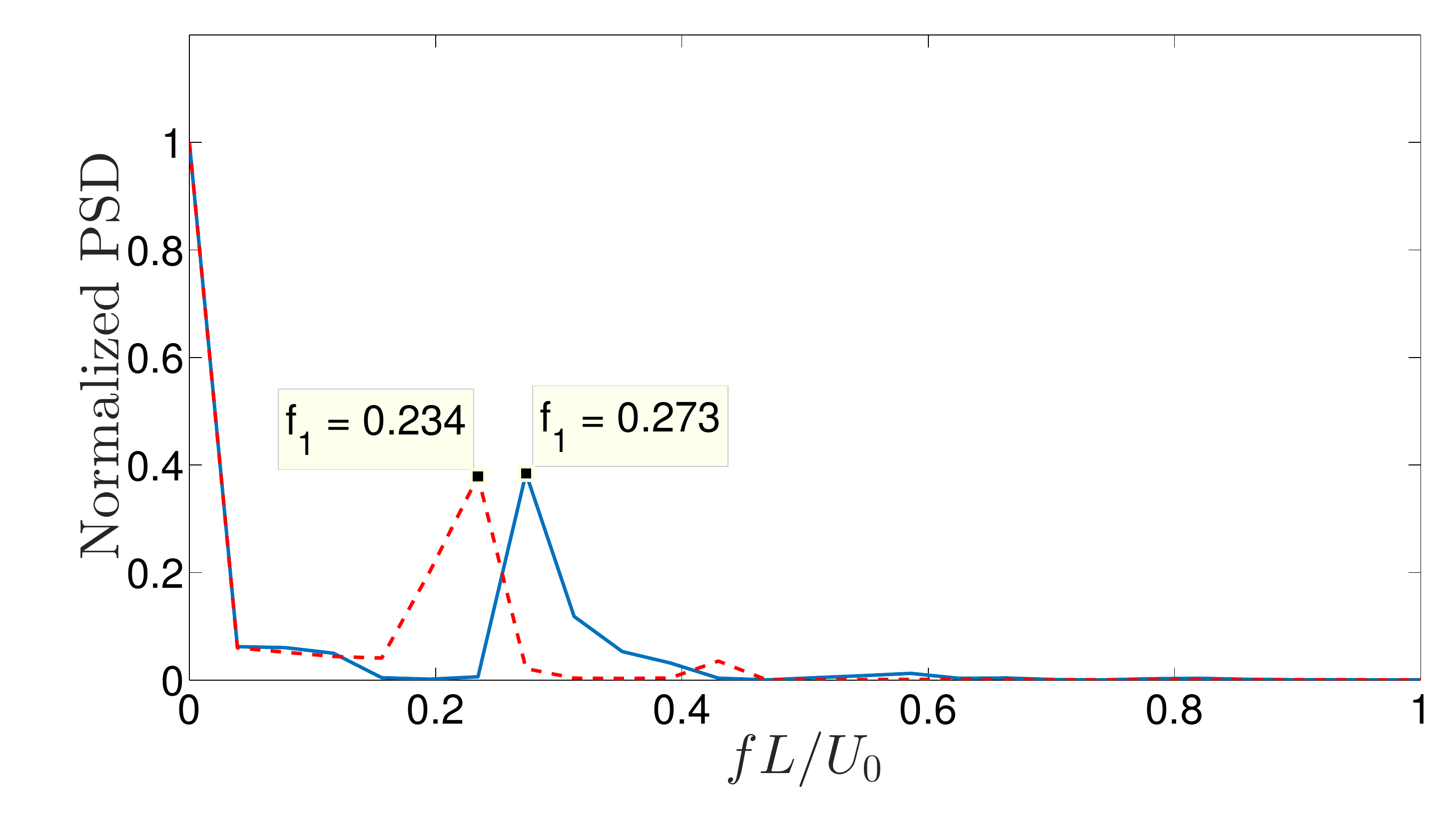}
		\caption{streamwise frequency}
	\end{subfigure}	\\
	\begin{subfigure}{0.49\textwidth}
		\includegraphics[width=0.99\columnwidth,trim=10mm 5mm 0mm 2.5mm,clip]{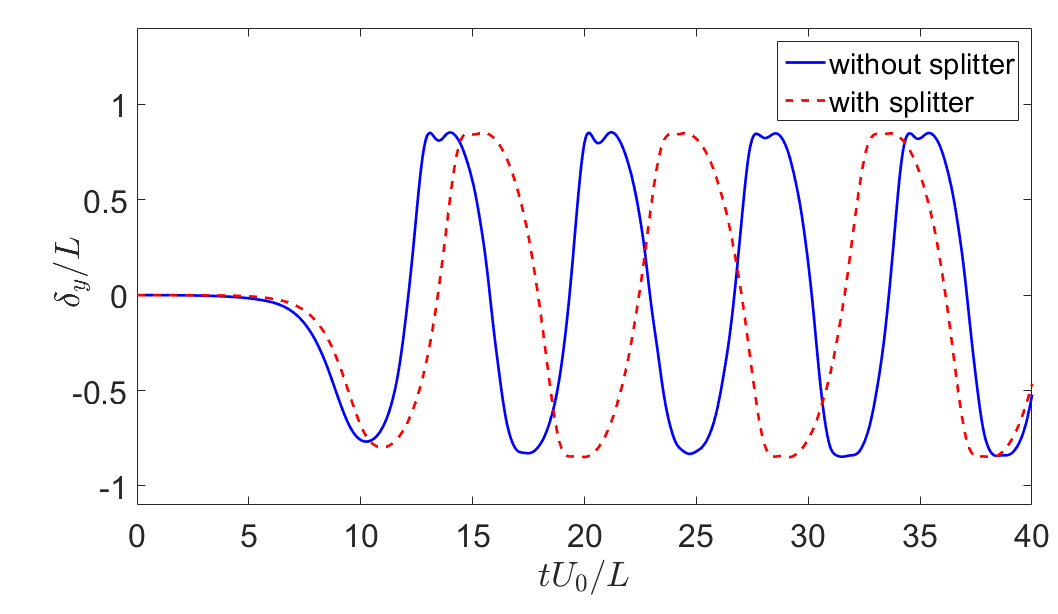}
		\caption{transverse displacement}
	\end{subfigure}
	\begin{subfigure}{0.49\textwidth}
		\includegraphics[width=0.99\columnwidth,trim=10mm 5mm 0mm 2.5mm,clip]{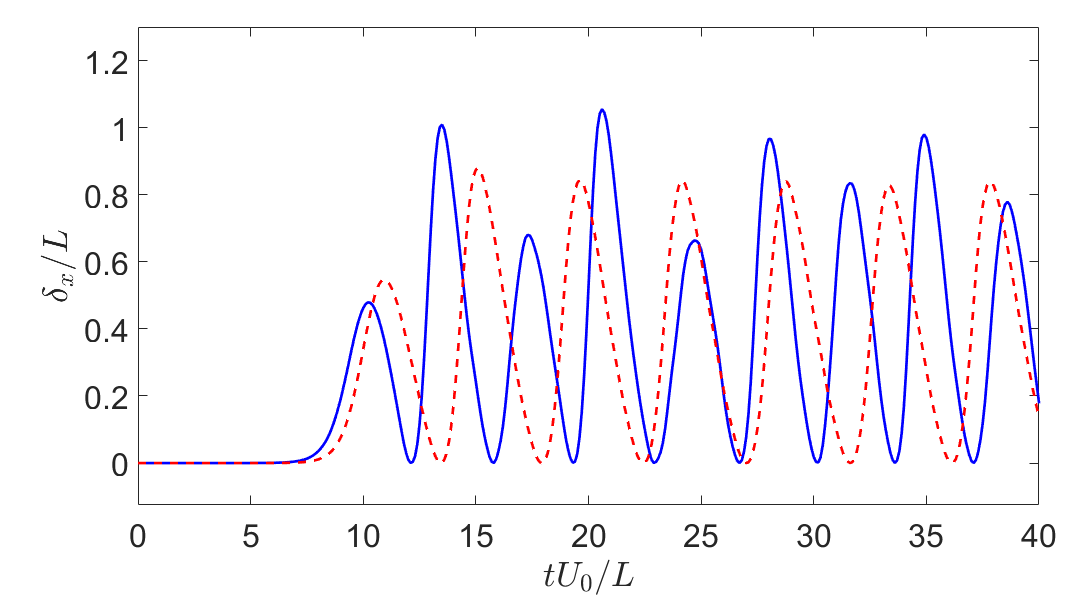}
		\caption{streamwise displacement}
	\end{subfigure}
	\caption{Frequency spectra and the time histories of LE response amplitudes for inverted foil: (a) transverse frequency, (b) streamwise frequency, (c) transverse displacement response and (d) streamwise displacement response for the inverted foil with (\textcolor{red}{-\hspace{1mm}-\hspace{1mm}-}) and without (\textcolor{blue}{---}) the fixed splitter plate for identical physical conditions $Re=30000,\ m^*=1$ and $K_B=0.2$.}\label{ComparisonBetweenInvFoilWithAndWithOutSplitter}
\end{figure}
\begin{figure}
	\centering
	\begin{subfigure}{0.22\textwidth}
		\centering
		\includegraphics[width=\columnwidth,trim=95mm 120 110mm 0,clip]{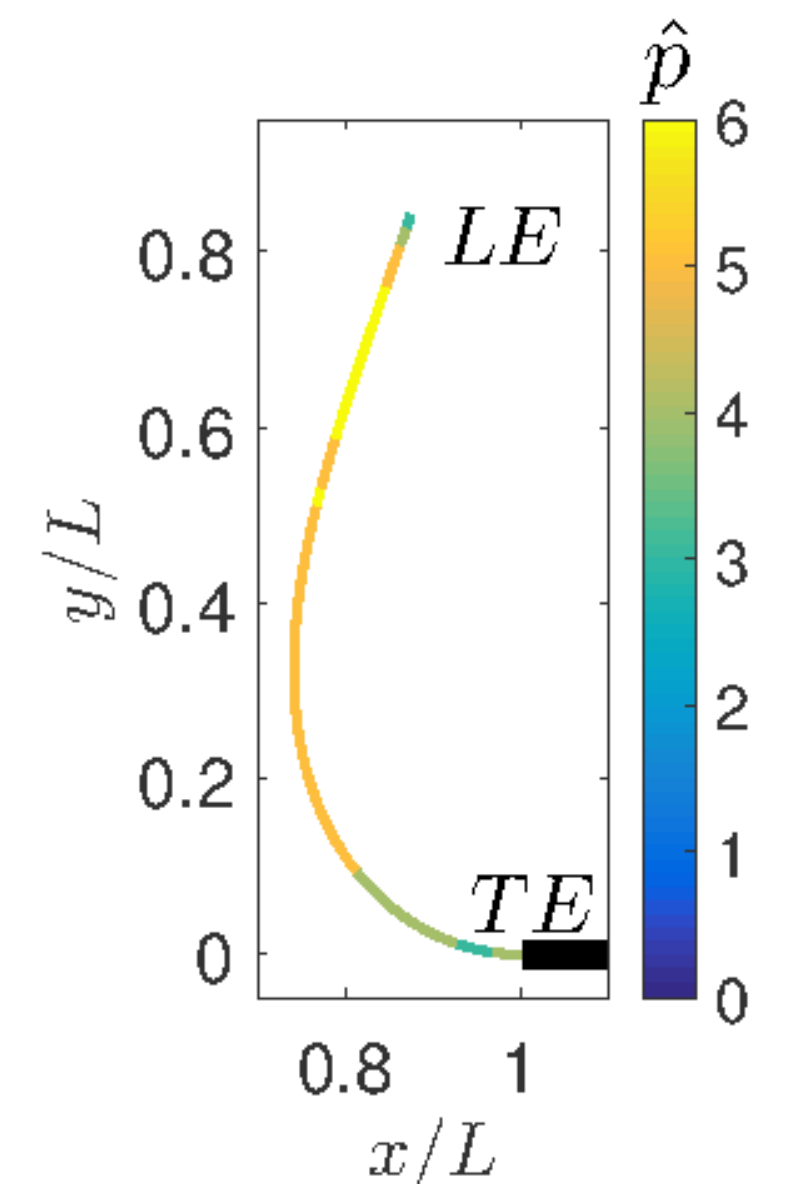}
		\caption{}
	\end{subfigure}
	\begin{subfigure}{0.22\textwidth}
		\centering
		\includegraphics[width=\columnwidth,trim=95mm 120 110mm 0,clip]{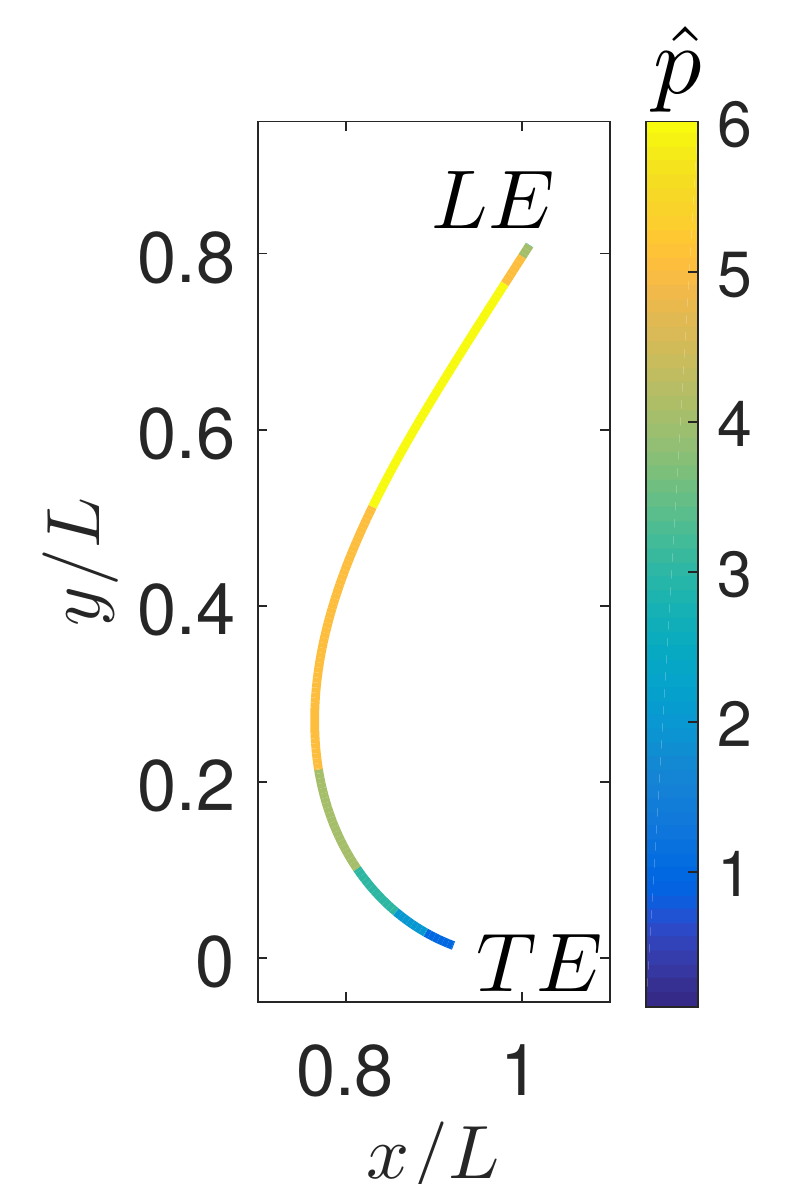}
		\caption{}
	\end{subfigure}
	\begin{subfigure}{0.54\textwidth}
		\includegraphics[width=\columnwidth,trim=10mm 40mm 80mm 7.5,clip]{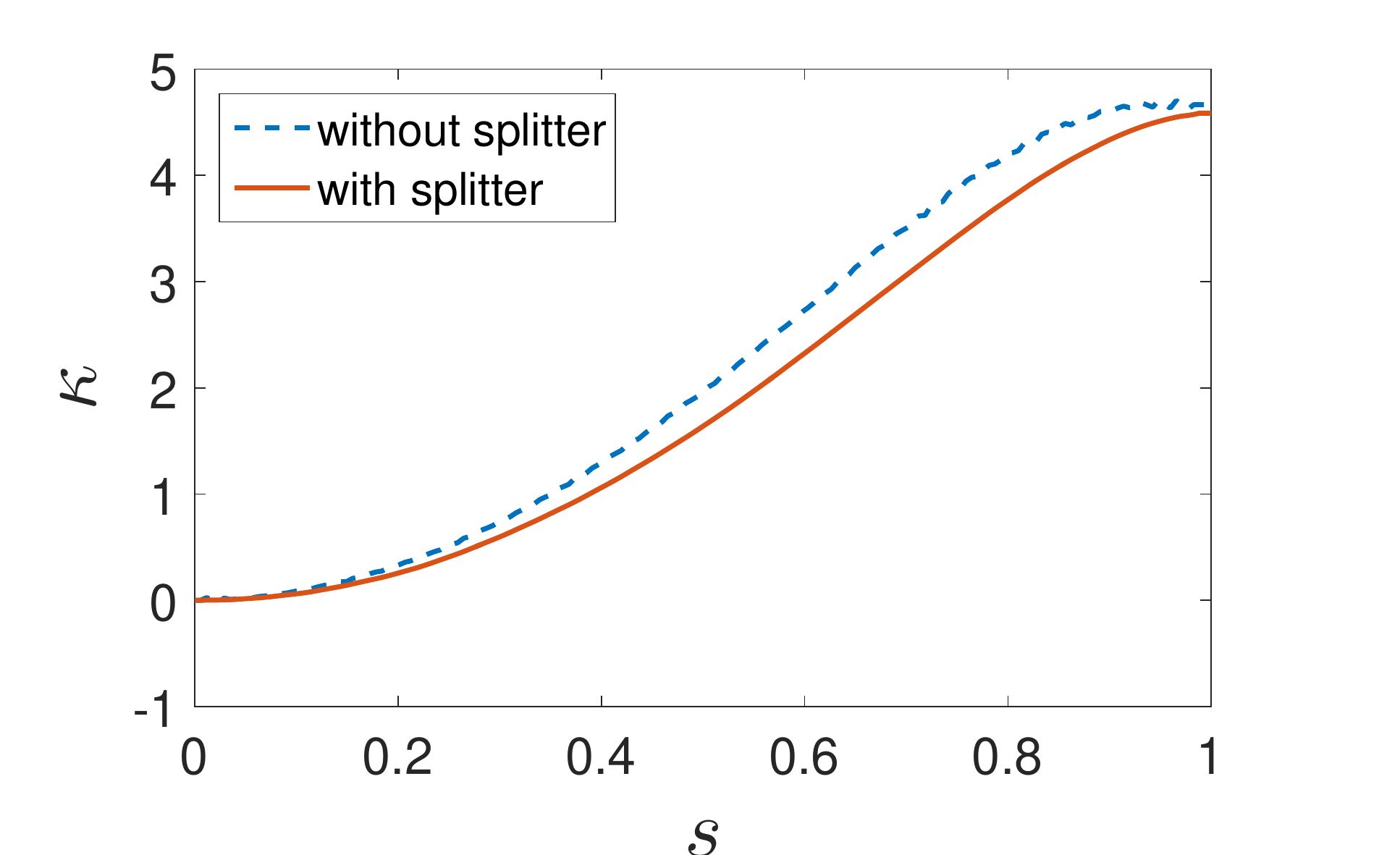}
		\caption{}
	\end{subfigure}
	\caption{Contours of nondimensional pressure distribution along two inverted foil configurations: (a) with, (b) without splitter plate at their maximum streamwise deformation, and (c) comparison of the variation of curvature ($\kappa$) along the foils at maximum deformation. $s$ represents the curvilinear coordinate with 0 at $LE$ and 1 at $TE$.}\label{pressureAndCurvature}
\end{figure}

To begin, figures~\ref{ComparisonBetweenInvFoilWithAndWithOutSplitter}a~and~\ref{ComparisonBetweenInvFoilWithAndWithOutSplitter}b  present the frequency characteristics of the LE displacements for the inverted foil configurations with and without the splitter plate. The figures confirm the reduction in the flapping frequency due to the inhibition of the vortex-vortex interaction by the splitter plate and the flapping frequency of the foil without splitter is 14.26\% greater than its counterpart with the splitter.
Figures~\ref{ComparisonBetweenInvFoilWithAndWithOutSplitter}c and~\ref{ComparisonBetweenInvFoilWithAndWithOutSplitter}d compare the LE transverse and streamwise displacements of the inverted foil with and without the splitter plate, respectively.
It is evident from figure~\ref{ComparisonBetweenInvFoilWithAndWithOutSplitter}c that the transverse flapping amplitudes remain similar for both the inverted foils with and without the splitter plate. 
 From figure~\ref{ComparisonBetweenInvFoilWithAndWithOutSplitter}d, the inhibition of the interaction between the vortices reduces the maximum streamwise flapping amplitudes and makes them more regular. To explain the reduction in the streamwise flapping amplitudes, we examine the nondimensional pressure distribution along the foil \revii{$(\hat{p}(x,y)=\Delta p(x,y)/0.5 \rho^\mathrm{f}U_0^2)$} for both the inverted foils presented in figures~\ref{pressureAndCurvature}a and \ref{pressureAndCurvature}b, \revii{where $\Delta p(x,y)$ denote the net pressure acting on any point $(x,y)$ on the foil surface}. Due to the formation of TEV, $\hat{p}$ at the TE is significantly lower for the inverted foil without a splitter plate. As a result of this, the foil without a splitter experiences 11.5\% lower and 2.4\% greater mean drag compared to the foil with a splitter for the regions $0.5\le s \le 1$ and $0\le s < 0.5$ respectively. Due to the greater drag close to the LE, the inverted foil without splitter experiences greater bending moments compared to its counterpart. The reduction in the streamwise flapping amplitudes due to the inhibition of the vortex-vortex interaction by the splitter affects the curvature of foil. Figure~\ref{pressureAndCurvature}c presents the variation of curvature along the foils for both the configurations. The figure clearly shows the area under the curvature curve for the inverted foil without splitter plate is greater than the area under the foil with a splitter. \revi{Since the strain energy is proportional to the square of the curvature (Eq.~\ref{strainEnergy})},  we can deduce that the energy harvesting ability of an inverted foil can be manipulated by controlling the vortex-vortex interaction without directly altering the foil properties. 

\begin{figure}
	\centering
	\begin{subfigure}{0.49\textwidth}
		\includegraphics[width=0.99\columnwidth,trim=10mm 0mm 17mm 2.5mm,clip]{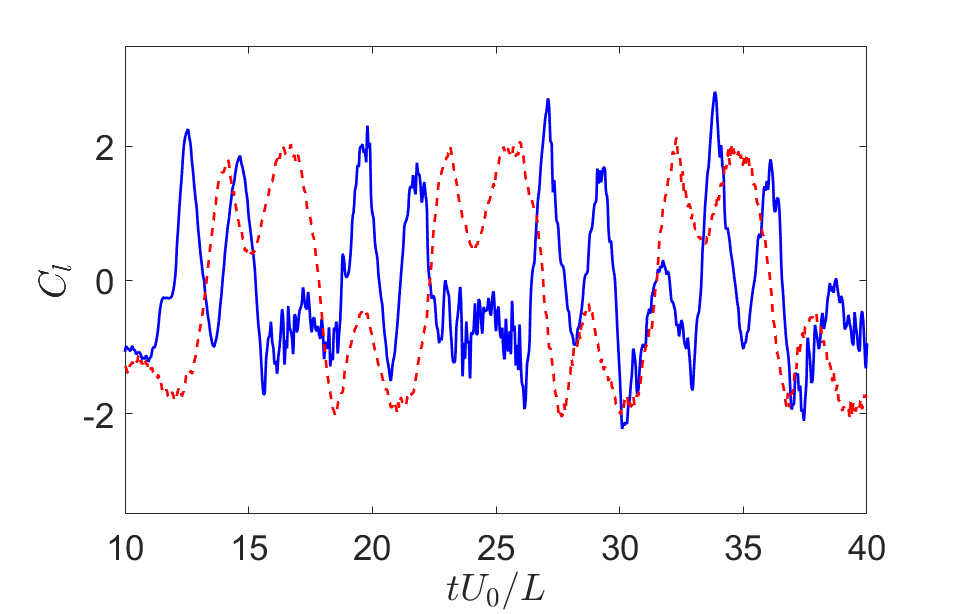}
		\caption{}
	\end{subfigure}
	\begin{subfigure}{0.49\textwidth}
		\includegraphics[width=0.99\columnwidth,trim=10mm 0mm 17mm 0mm,clip]{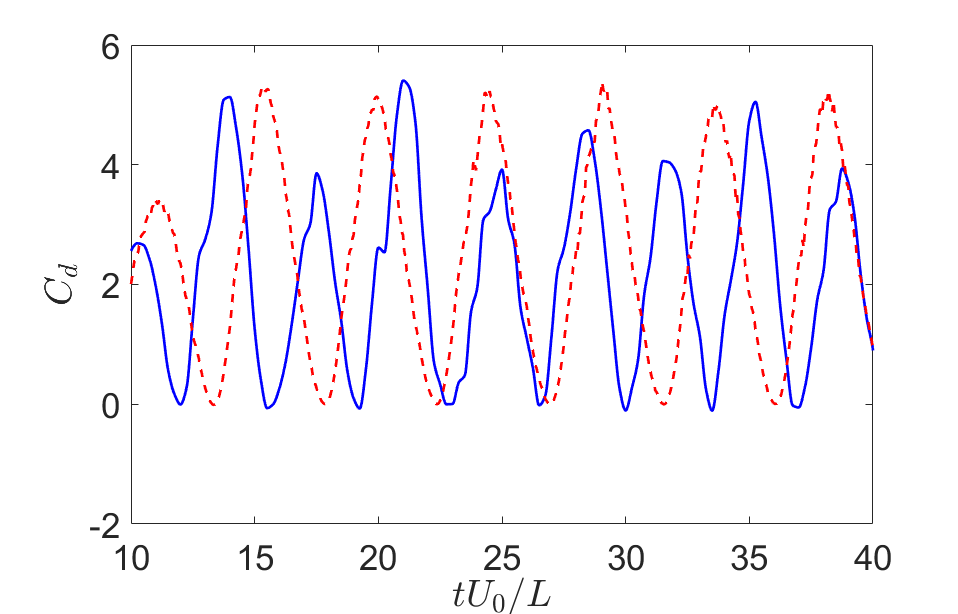}
	\caption{}
	\end{subfigure}\\
	\begin{subfigure}{0.49\textwidth}
		\includegraphics[width=0.99\columnwidth,trim=10mm 2mm 17mm 2.5mm,clip]{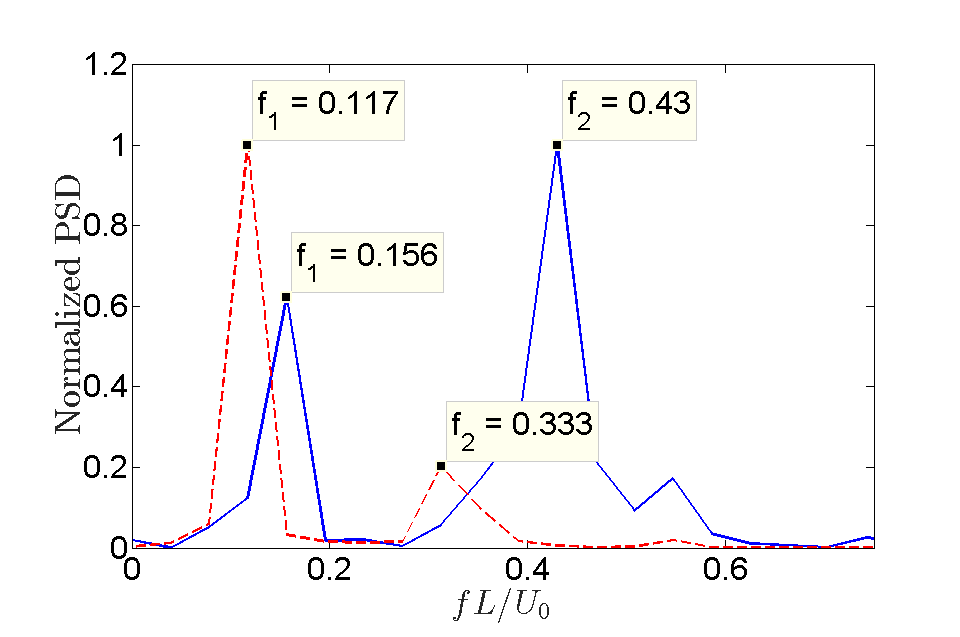}
		\caption{}
	\end{subfigure}
	\begin{subfigure}{0.49\textwidth}
		\includegraphics[width=0.99\columnwidth,trim=10mm 2mm 17mm 2.5mm,clip]{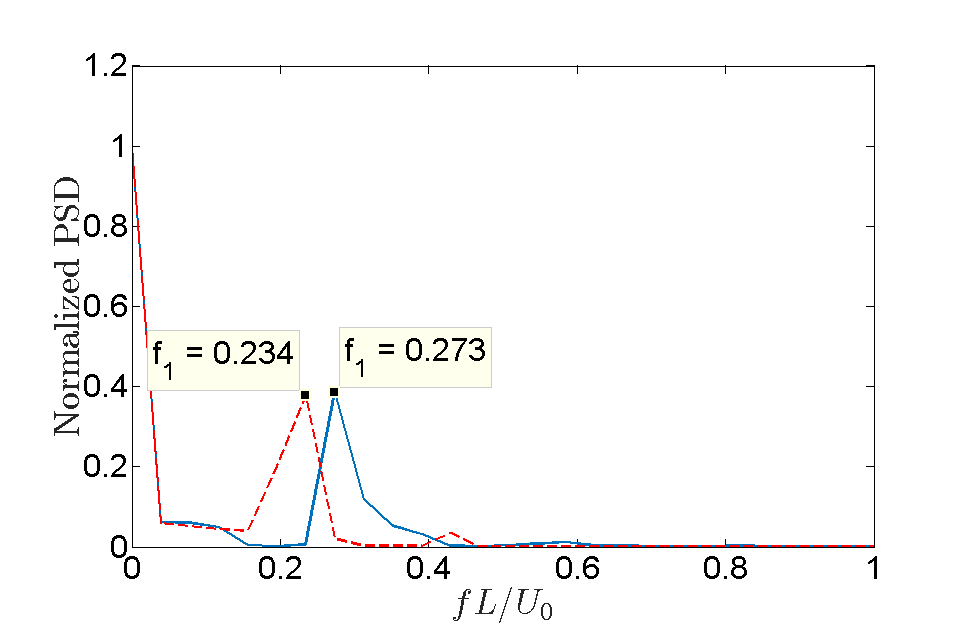}
		\caption{}
	\end{subfigure}
	\caption{Time histories of force coefficients and frequency spectra for inverted foil with and without splitter plate  for the identical physical conditions of $Re=30000,\ m^*=1$ and $K_B=0.2$: (a) lift, (b) drag, frequency response of lift (c) and (d) drag. Here inverted foil with (\textcolor{red}{-\hspace{1mm}-\hspace{1mm}-}) and without (\textcolor{blue}{---}) a splitter plate are shown.}\label{forceDynamics}
\end{figure}
\begin{figure}
	\centering
	\begin{subfigure}{0.49\textwidth}
		\includegraphics[width=0.99\columnwidth,trim=0mm 0mm 0mm 2mm,clip]{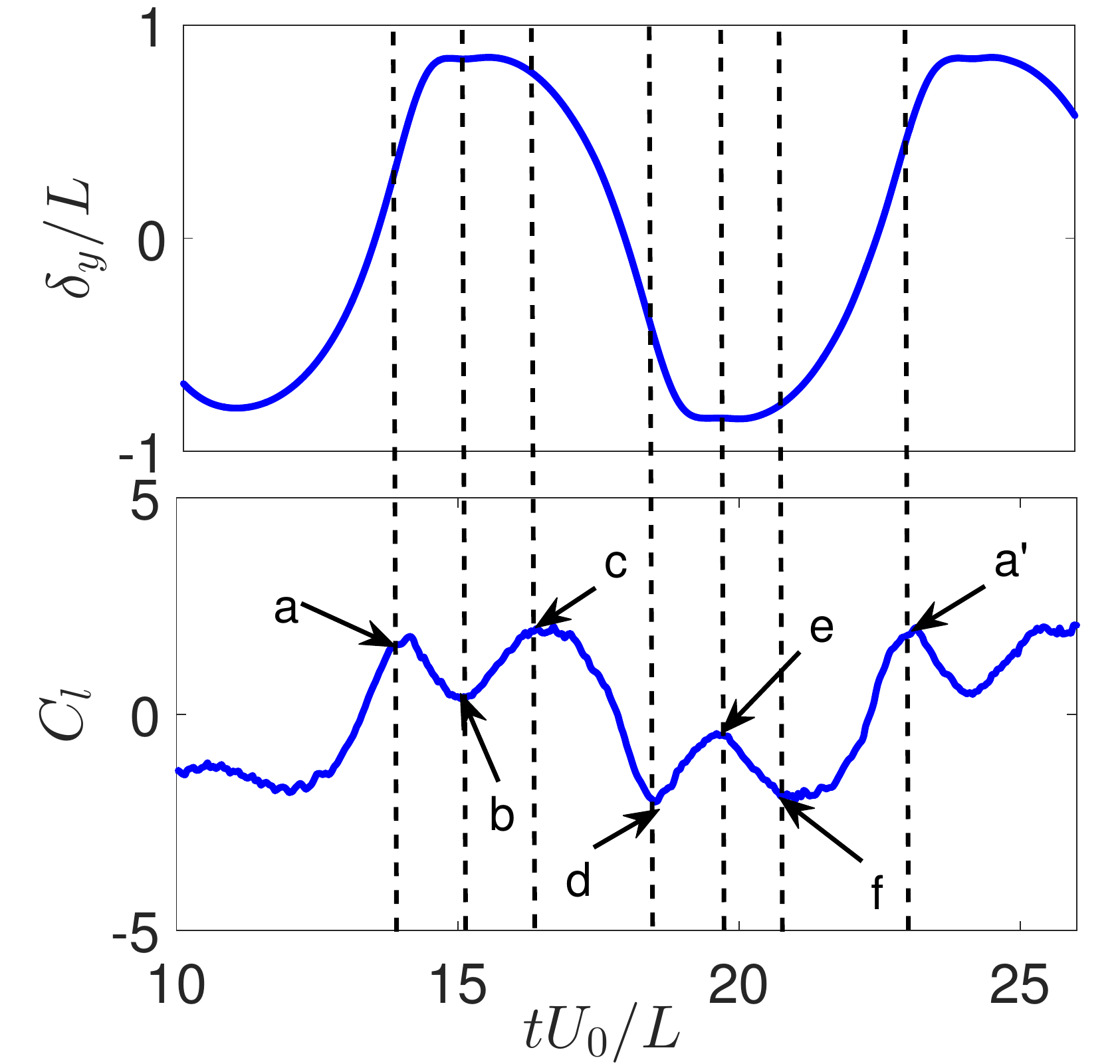}
	\end{subfigure}
	\begin{subfigure}{0.49\textwidth}
		\includegraphics[width=0.99\columnwidth,trim=0mm 0mm 0mm 2mm,clip]{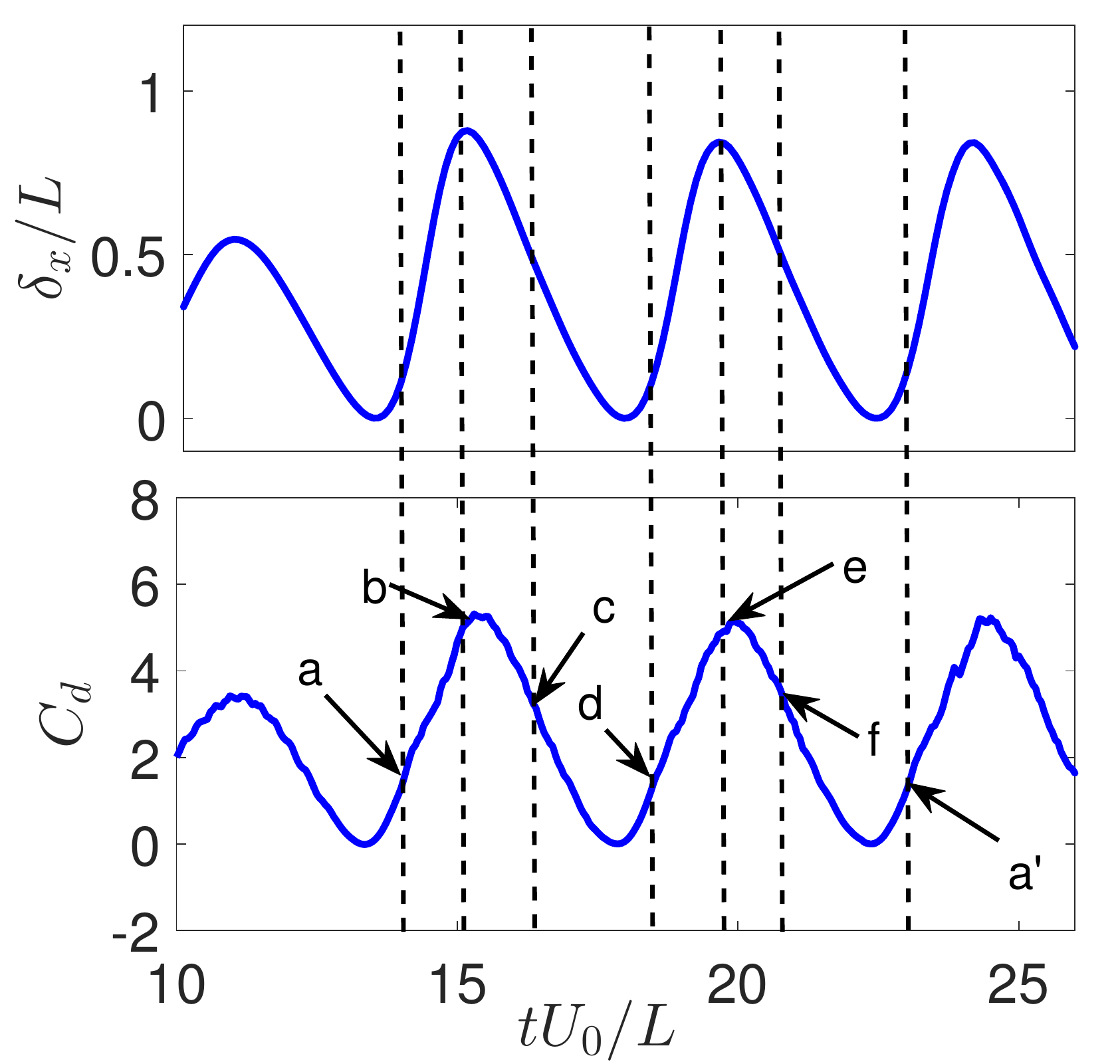}
	\end{subfigure}	
	\caption{Comparison of the evolutions of LE transverse and streamwise displacements against the lift ($C_l$) and drag ($C_d$) coefficients  for the inverted foil with splitter at nondimensional parameters $K_B=0.4,\ Re=30000,\ m^*=1$ and $AR=0.5$. In this plot, points (a,b,c,d,e,f,a$\mathrm{'}$) correspond to the time instances $tU_0/L$=(13.95, 15.0, 16.4, 18.5, 19.8, 20.8, 23).}\label{comparisonPlot_splitter}
\end{figure}
Inhibition of the vortex-vortex interaction has a profound impact on the drag and lift forces (figure~\ref{forceDynamics}). Due to the formation of the counter-rotating vortex at TE of the inverted foil without splitter, both the drag and lift forces acting on the foil drop more rapidly compared to its counterpart with the splitter plate. As a consequence of this, the foil without the splitter travels more quickly from its maximum deformation to the mean position thereby resulting in lower flapping period and higher flapping frequency compared to the foil with the splitter plate. 
The lift forces exhibit an additional secondary frequency $f_2$, which has $3^\mathrm{rd}$ harmonic component in the flapping frequency in addition to the fundamental frequency $f_1$ corresponding to the transverse response.
Unsurprisingly, figure~\ref{forceDynamics}c shows $f_2$ as the dominant frequency for the inverted foil without splitter because three vortices (`A+B', `C' in figure~\ref{3D_inv_vortex}) are shed over one-half cycle. However, there is no synchronization between the periodic vortex induced dominant $f_2$ and transverse flapping. On the other hand, the dominant vortex induced frequency $f_1$ synchronizes with the transverse flapping frequency for the inverted foil with the splitter plate because there is only one vortex (`A') shed over  one-half cycle of flapping oscillation and no secondary LEV are observed. For complex vortex modes like $2P+S$ observed in our study for inverted foil without splitter and $2P+2S$ or $4P$ modes observed in \cite{ryu_2015} and \cite{gurugubelli_JFM}, the dominant vortex induced forces may not synchronize with the transverse flapping frequency. \revii{To further understand the impact of the suppression of TEV on the force dynamics, we plot the time histories of LE transverse and streamwise displacement responses and contrast against the lift and drag acting on the foil over a flapping cycle in figure~\ref{comparisonPlot_splitter}. Qualitatively if we compare figures~\ref{comparisonPlot} and \ref{comparisonPlot_splitter}, both the figures share similar characteristics. The primary difference between the two figures is the point `d'. In figure~\ref{comparisonPlot}, the point `d' represented the time instant at which the lift acting on the foil increases due to the formation of the secondary LEV `C'. Since there is no secondary LEV for the inverted foil with the splitter, the lift acting on the foil drops until the foil attains a large enough deformation so that drag can sustain the deformed foil. The point `d' in figure~\ref{comparisonPlot_splitter},  represents the point of transition from the lift dominated to the drag dominant foil deformation. Therefore, the mechanism of the foil motion from the maximum transverse deformation to the mean position depends on the foil inertia due to the elastic restoring forces and shedding of the LEV `A'. The TEV only enhances this motion of the foil and hence the foil travels faster when the TEV is present as compared  to the foil where it is suppressed. }

\begin{figure}
	\begin{subfigure}{0.48\textwidth}
		\centering
		\includegraphics[width=0.98\columnwidth]{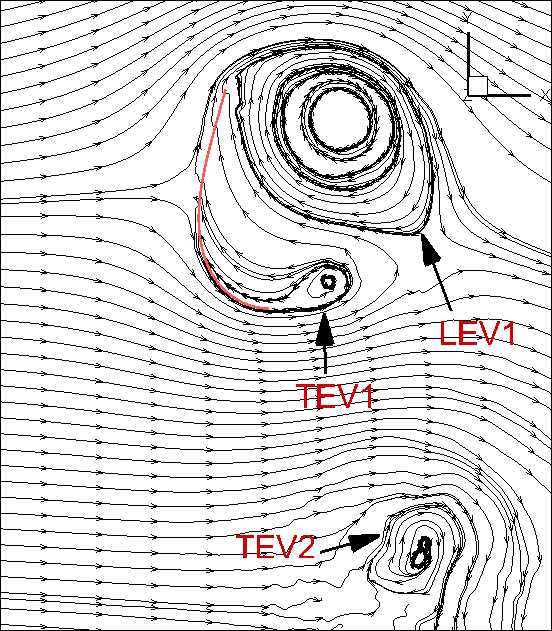}
	\end{subfigure}
	\begin{subfigure}{0.48\textwidth}
		\centering
		\includegraphics[width=0.98\columnwidth]{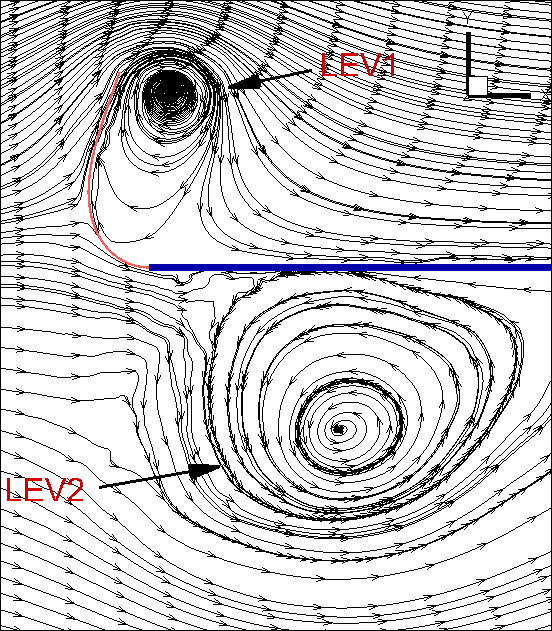}
	\end{subfigure}
\caption{Streamline topology for the inverted foil without (left) and with (right) splitter plate attached at TE for the identical nondimensional parameters $K_B=0.2,\ Re=30000$ and $m^*=1$. }\label{streamline_inverted}
\end{figure}

\changes{Figure~\ref{streamline_inverted} shows the streamline topology for the inverted foil with and without the splitter, where LEV1 (TEV1) and LEV2 (TEV2) represent the LEV (TEV) shed during current and previous half-cycles respectively. The figure conforms with the vortex modes presented in figures~\ref{3D_inv_vortex} and \ref{vorticity_splitter}. In addition to the suppression of TEV, the figure also presents one more distinct feature. In the case of the inverted foil without splitter, (LEV+TEV) vortex pair travel away from the foil at an angle inclined to the freestream velocity. Similar observations can be seen in 2D vortex modes presented in \cite{ryu_2015,gurugubelli_JFM}. On the other hand, the LEV in the inverted foil with the splitter travels along the splitter plate and grows in size. The wall surfaces of splitter plate act as the source of vorticity which feeds into the LEV. To further demonstrate the role of vortex shedding on the LAF, we next perform numerical simulations for low-$Re$ flows at identical $K_B$ and $m^*$.}

\subsection{The role of vortex shedding on inverted foil flapping at low $Re$}
\begin{figure}
	\centering
	\begin{subfigure}{0.49\textwidth}
		\includegraphics[width=0.99\columnwidth,trim=7.5mm 10mm 20mm 5mm,clip]{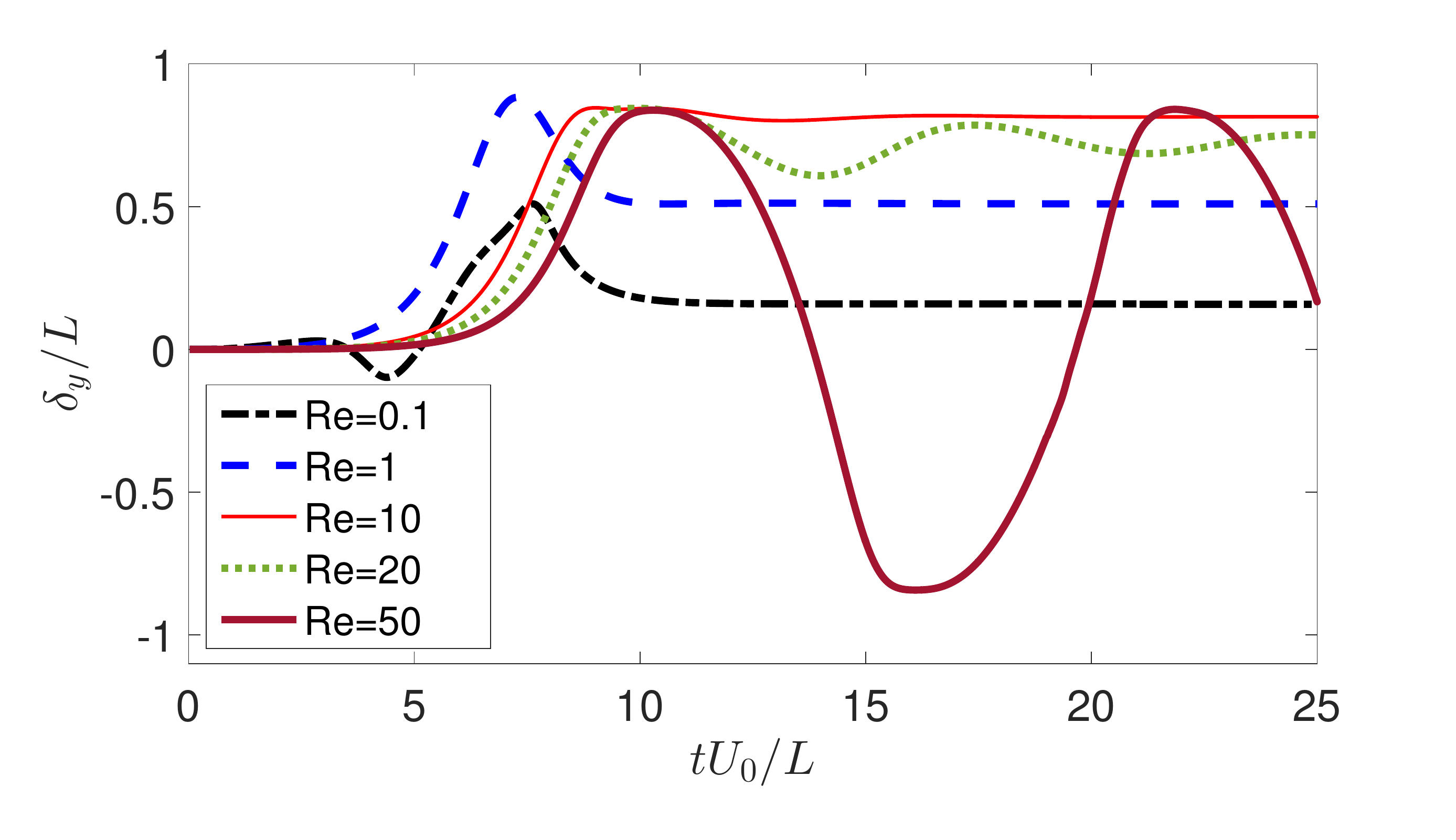}
		\caption{}
	\end{subfigure}
	\begin{subfigure}{0.49\textwidth}
		\includegraphics[width=0.99\columnwidth,trim=7.5mm 10mm 20mm 5mm,clip]{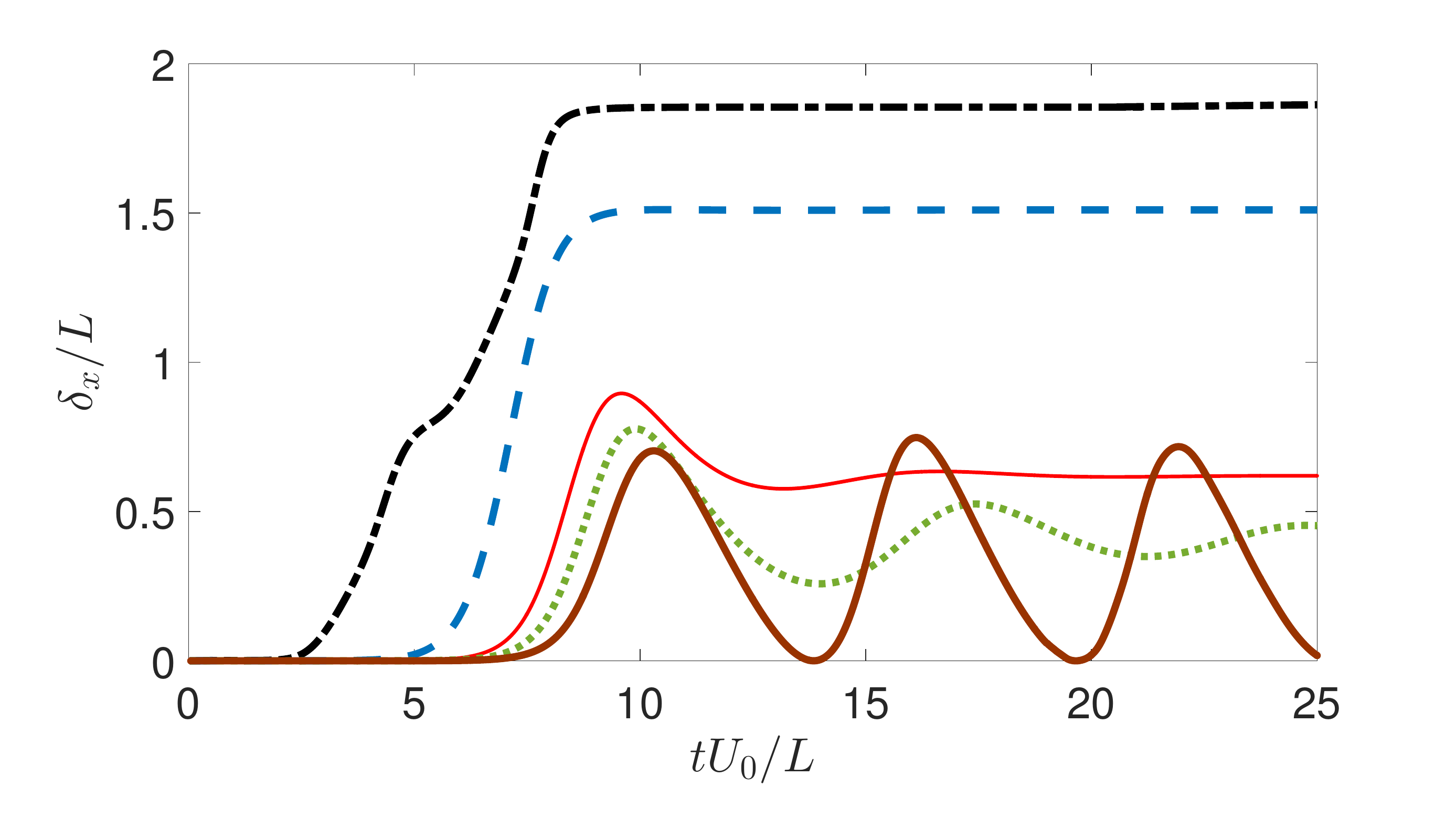}
		\caption{}
	\end{subfigure}\\
	\begin{subfigure}{0.65\textwidth}
		\includegraphics[width=0.99\columnwidth,trim=25mm 30mm 50mm 5mm,clip]{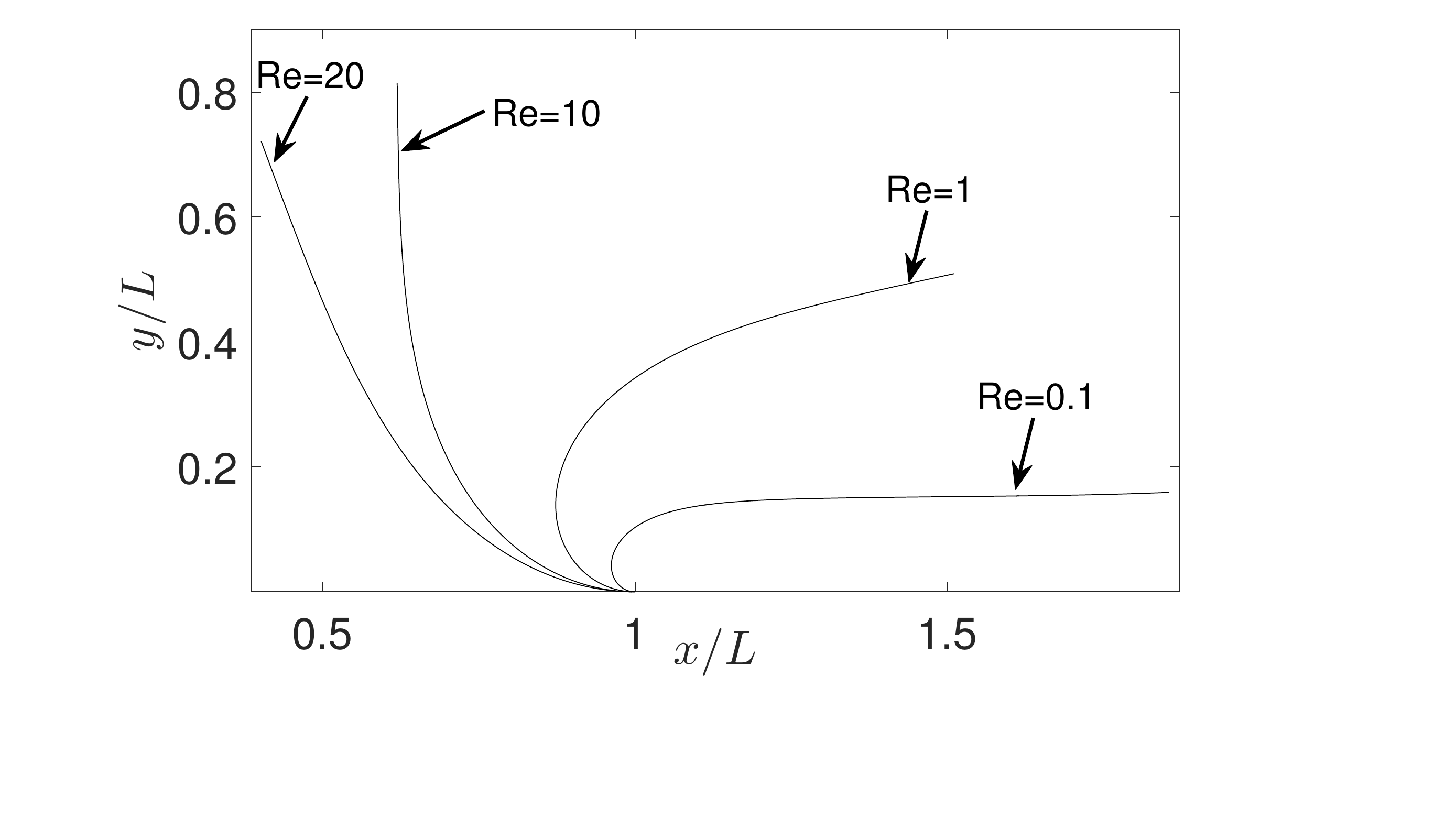}
		\caption{}	
	\end{subfigure}
	\caption{Time traces of the LE response amplitudes of  inverted foil at $K_B=0.2$ and $m^*=1$ for low-$Re$ laminar flow: (a) transverse,  (b) streamwise, 
and (c) steady-steady foil profiles.}
\label{lowRe_response}
\end{figure}
\begin{figure}
	\centering
	\begin{subfigure}{0.49\textwidth}
		\includegraphics[width=0.99\columnwidth,trim=0mm 0mm 12mm 0mm,clip]{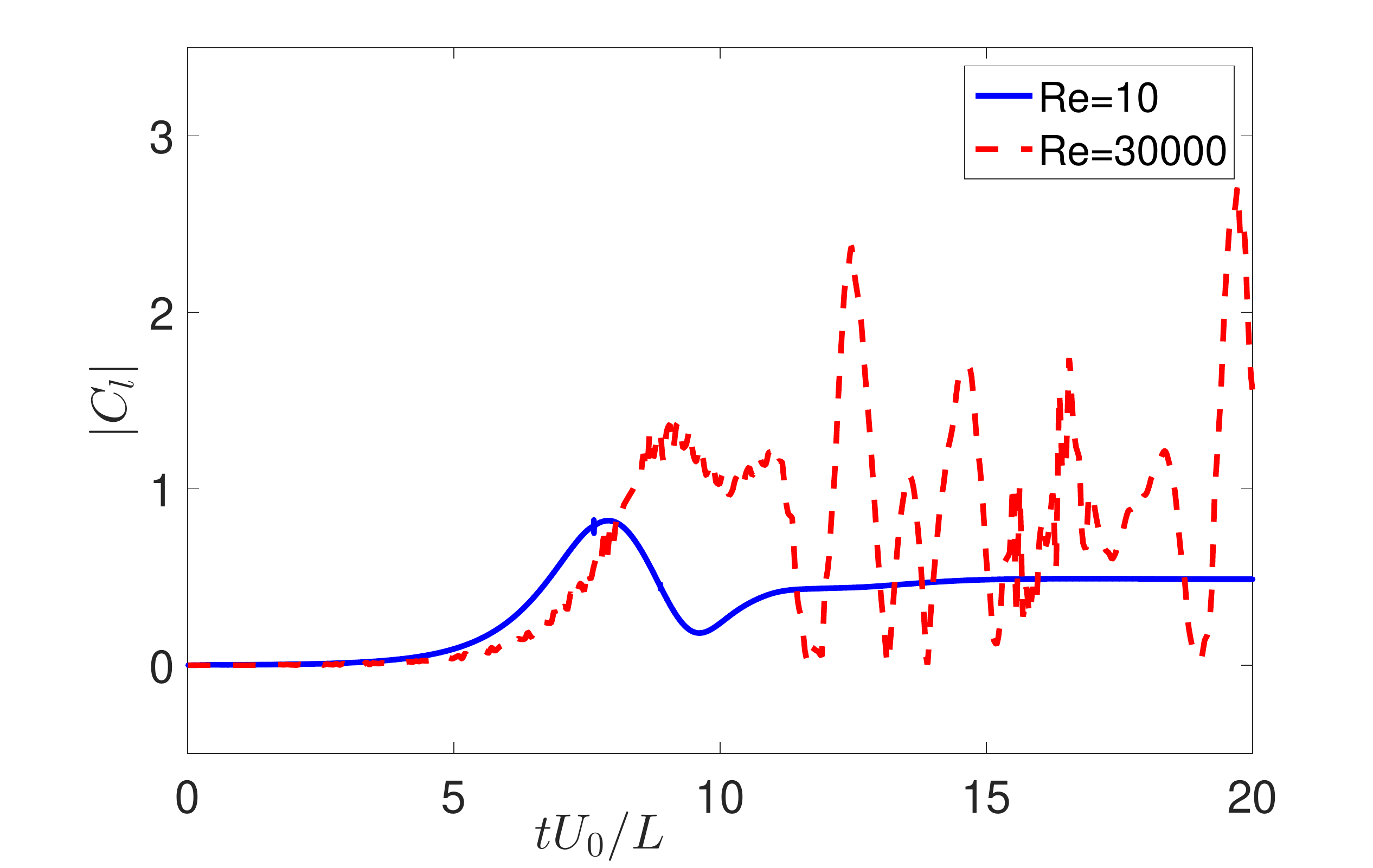}
	\caption{}
	\end{subfigure}
	\begin{subfigure}{0.49\textwidth}
		\includegraphics[width=0.99\columnwidth,trim=0mm 0mm 12mm 0mm,clip]{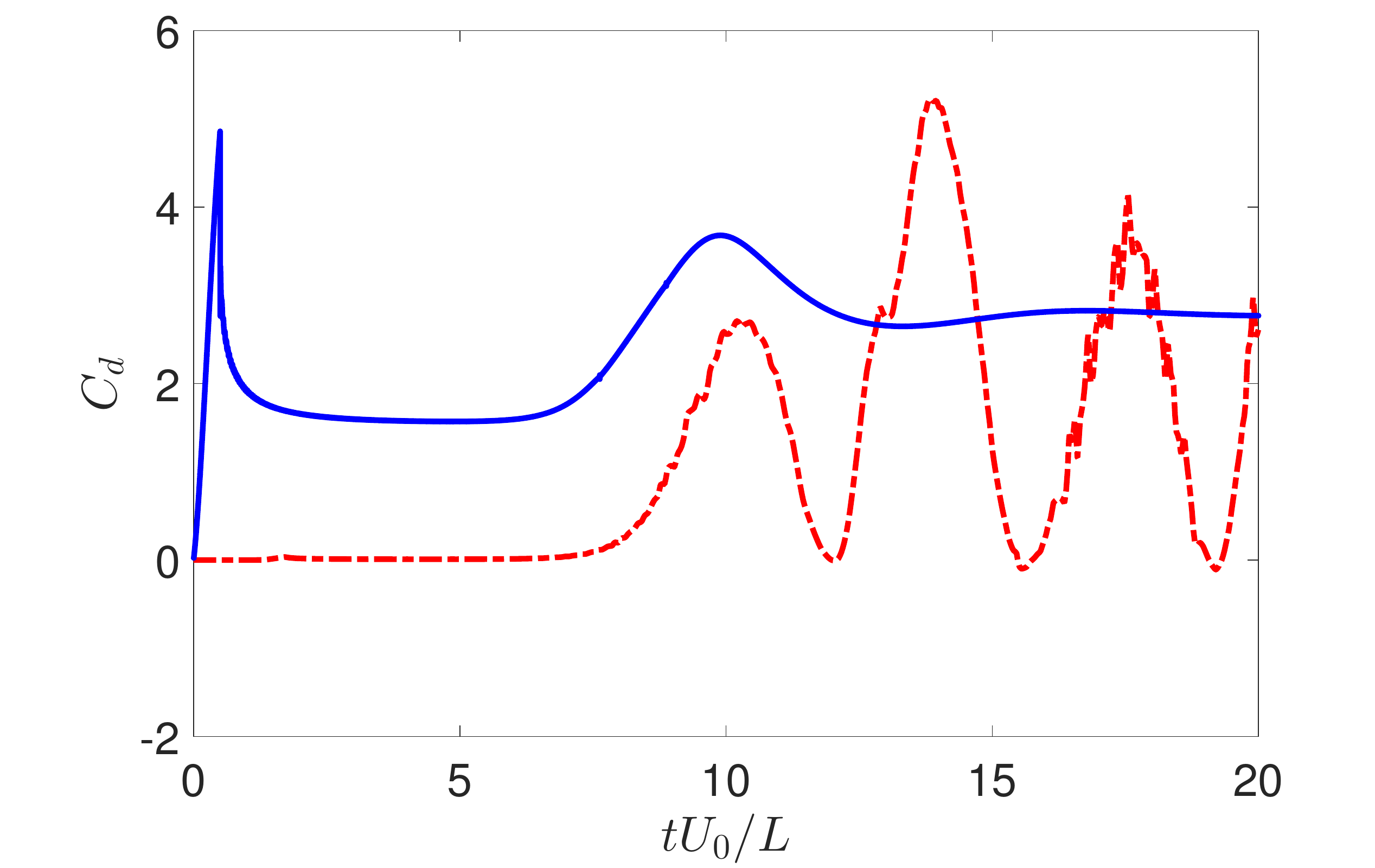}
	\caption{}
	\end{subfigure}
	\caption{Comparison of the time histories of lift (left) and drag (right) during the onset of instability at $K_B=0.2$ and $m^*=1$ for $Re=$ 30000 and 10. }\label{lowRe_highRe}
\end{figure}
\begin{figure}
	\centering
	\begin{subfigure}{0.49\textwidth}
		\includegraphics[width=0.99\columnwidth,trim=0mm 0mm 12mm 0mm,clip]{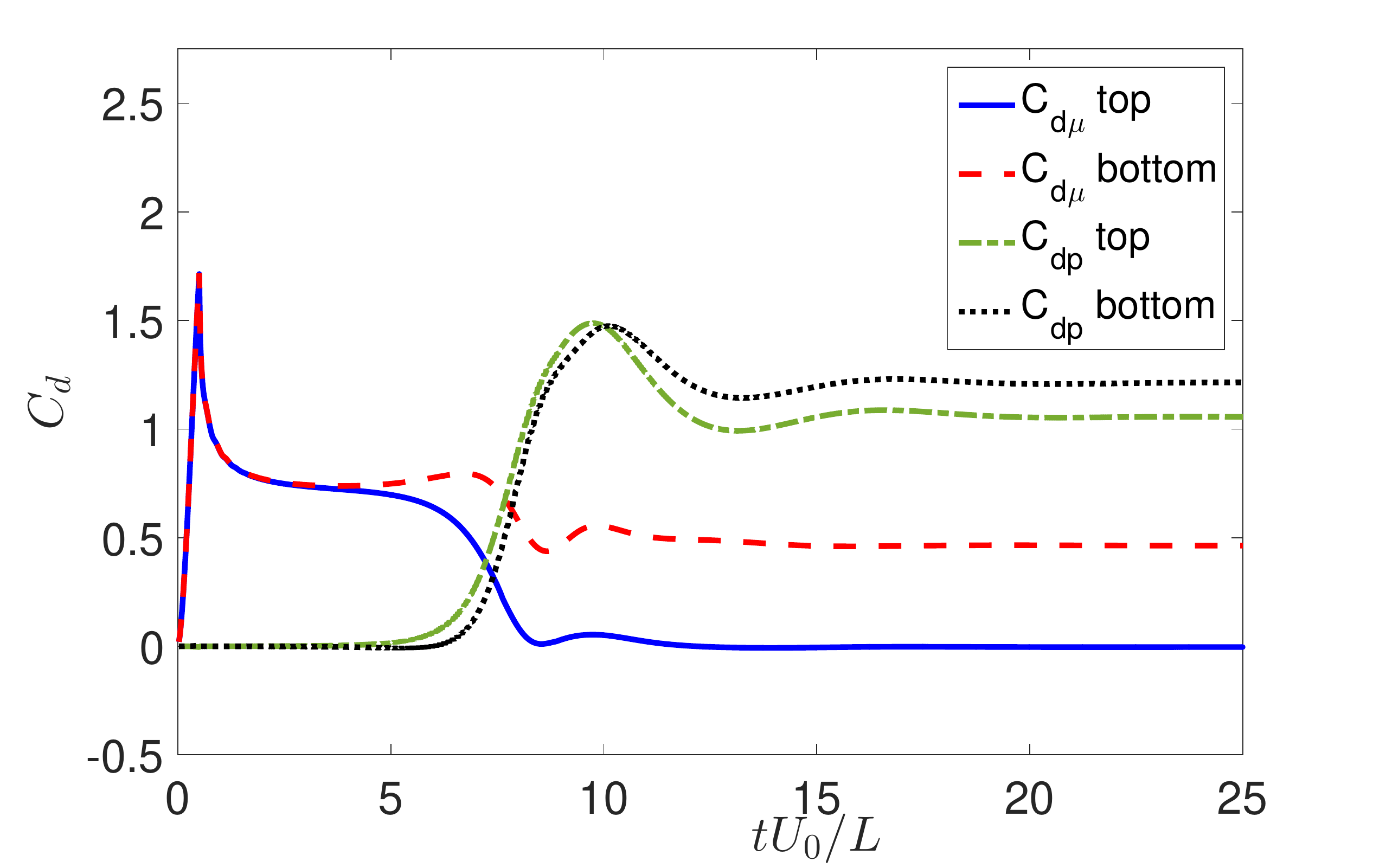}
		\caption{}
	\end{subfigure}
	\begin{subfigure}{0.49\textwidth}
		\includegraphics[width=0.99\columnwidth,trim=0mm 0mm 12mm 0mm,clip]{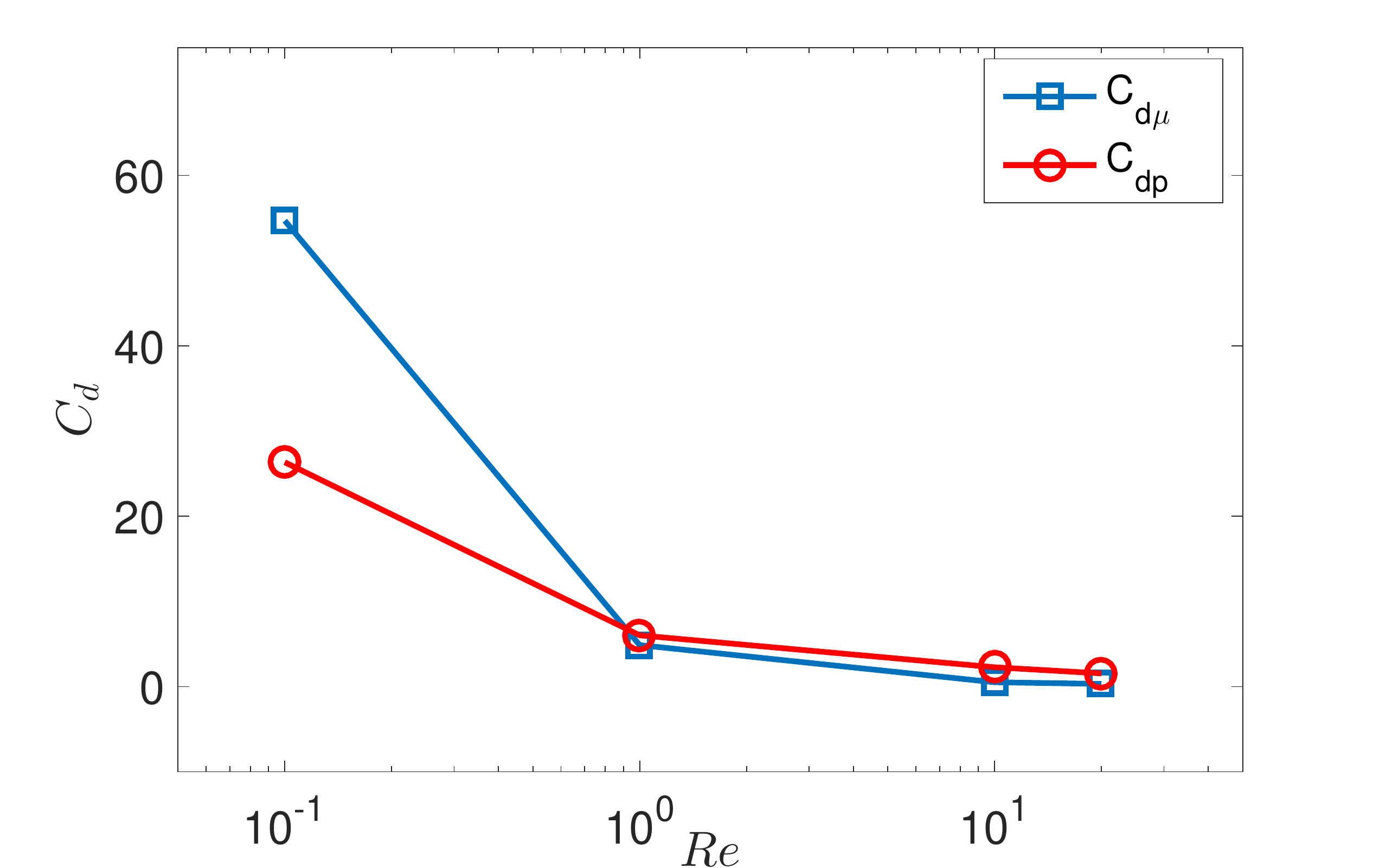}
		\caption{}
	\end{subfigure}
	\caption{Decomposition of drag force of the inverted foil for low $Re$ laminar flow at $K_B=0.2$ and $m^*=1$: (a) time evolution of drag decomposed into viscous and pressure components acting on the top and bottom inverted foil surfaces for $Re=10$, (b) summary of viscous and pressure components of drag as a function of $Re$.}\label{lowRe_pressure_viscous}
\end{figure}

We perform a series of numerical experiments on the simple inverted foil without the splitter for low-$Re\in[0.1,50]$ regime to realize the effects of vortex shedding phenomenon on the LAF and the onset of flapping instability. 
As a function of $Re$ based on the initial foil length,  figure~\ref{lowRe_response} summarizes the  time traces of LE displacements and the steady-state deformed foil profiles as a function of $Re$. From the LE displacements in figures~\ref{lowRe_response}a and \ref{lowRe_response}b, we can see that there exists a critical $Re$ below which the LAF motion ceases to exist. \revi{However, the inverted foil continues to undergo static-divergence even for very small $Re$ and undergoes a large static deformation. Similar large static deformations for $Re\in[20,50]$ have been reported by \cite{ryu_2015} and \cite{mittal_2016}.} This phenomenon is distinctly different from the flexible foil with its LE clamped where the foil remains stable for very low-$Re$ \citep{shelley_review}. For $Re<10$, the inverted foil flips about the clamped TE and exhibits a flipped state as shown in figure~\ref{lowRe_response}c. For the flipped state, the LE of the foil aligns itself with the flow direction.
%
\revii{To examine
how an inverted foil continue to undergo static-divergence with large static deformation even for low $Re$, we plot the time evolution of lift and drag during the onset of instability in figures~\ref{lowRe_highRe}a and ~\ref{lowRe_highRe}b respectively for $Re=30000$ and $10$ at $K_B=0.2$ and $m^*=1$.
Even though the maximum lift acting on the foil for $Re=10$ is significantly lower than the lift for $Re=30000$ (figure~\ref{lowRe_highRe}a),  the maximum transverse amplitude for $Re=10$ and $30000$ is approximately same  (figures~\ref{ComparisonBetweenInvFoilWithAndWithOutSplitter}c and \ref{lowRe_highRe}a). For low $Re$, the large static deformation can be attributed to the relatively large static drag ($\bar{C}_d=2.78$) acting on the foil which is greater than the mean ($\bar{C}_d=2.19$) and the root-mean square ($C_d^\mathrm{rms}=2.68$) drag at $Re=30000$. We have earlier shown in Section~\ref{sec:flowField} that the large deformation of the foil depends on both the lift and drag forces. For low $Re$, the lift acting on the inverted foil is just sufficient to break its symmetry and the drag bends it through large deformations. To understand the role of viscous stress and LEV on the large drag at low $Re$, we plot the viscous and pressure drag components on top and bottom foil surfaces in figure~\ref{lowRe_pressure_viscous}a. 
The viscous and pressure drag components acting on the foil, respectively, are computed using
\begin{equation*}
C_{d\mu} = \int_{\Gamma(t)}(\vec{T}^\mathrm{f}\cdot \vec{\mathrm{n}}^\mathrm{f})\cdot \vec{\mathrm{n}}_\mathrm{x}\ \mathrm{d} \vec{\Gamma} \qquad \mbox{and} \qquad C_{dp} = \int_{\Gamma(t)}(-p\vec{\mathrm{I}}\cdot \vec{\mathrm{n}}^\mathrm{f})\cdot \vec{\mathrm{n}}_\mathrm{x}\ \mathrm{d} \vec{\Gamma},
\end{equation*}
where $\vec{T}$ is the fluid viscous stress defined in Eq.~(\ref{eq:cauchyStress}) and $\vec{\mathrm{n}}_\mathrm{x}$ represents the unit vector in the streamwise direction.
Both top and bottom surfaces experience a large viscous drag initial untill the foil achieve sufficiently large transverse amplitude and the flow separates at LE to form the LEV. After the formation of LEV, the viscous drag acting on the foil drops and the pressure induced drag due to LEV increases. Therefore, both the viscous and pressure drag components act jointly to produce the large foil deformation at low $Re$. Figure~\ref{lowRe_pressure_viscous}b summarizes the viscous and pressure drag components as a function of $Re$. As $Re$ decreases both viscous and pressure drag components acting on the foil increases.} {Three key points to highlight from this low Re study are: (i) vortex shedding at LE of an inverted foil is necessary for sustaining unsteady periodic flapping; (ii) unlike the flexible foil with LE clamped which remains stable for the low-$Re$ regime, the inverted foil loses its stability to undergo large static deformation for $K_B<(K_B)_{cr}$; and (iii) 
as flow speed decreases further $(Re < 1)$ the foil undergoes $180^0$ static deformation.}

\begin{figure}
	\centering
	\begin{subfigure}{0.275\textwidth}
		\includegraphics[width=0.99\columnwidth]{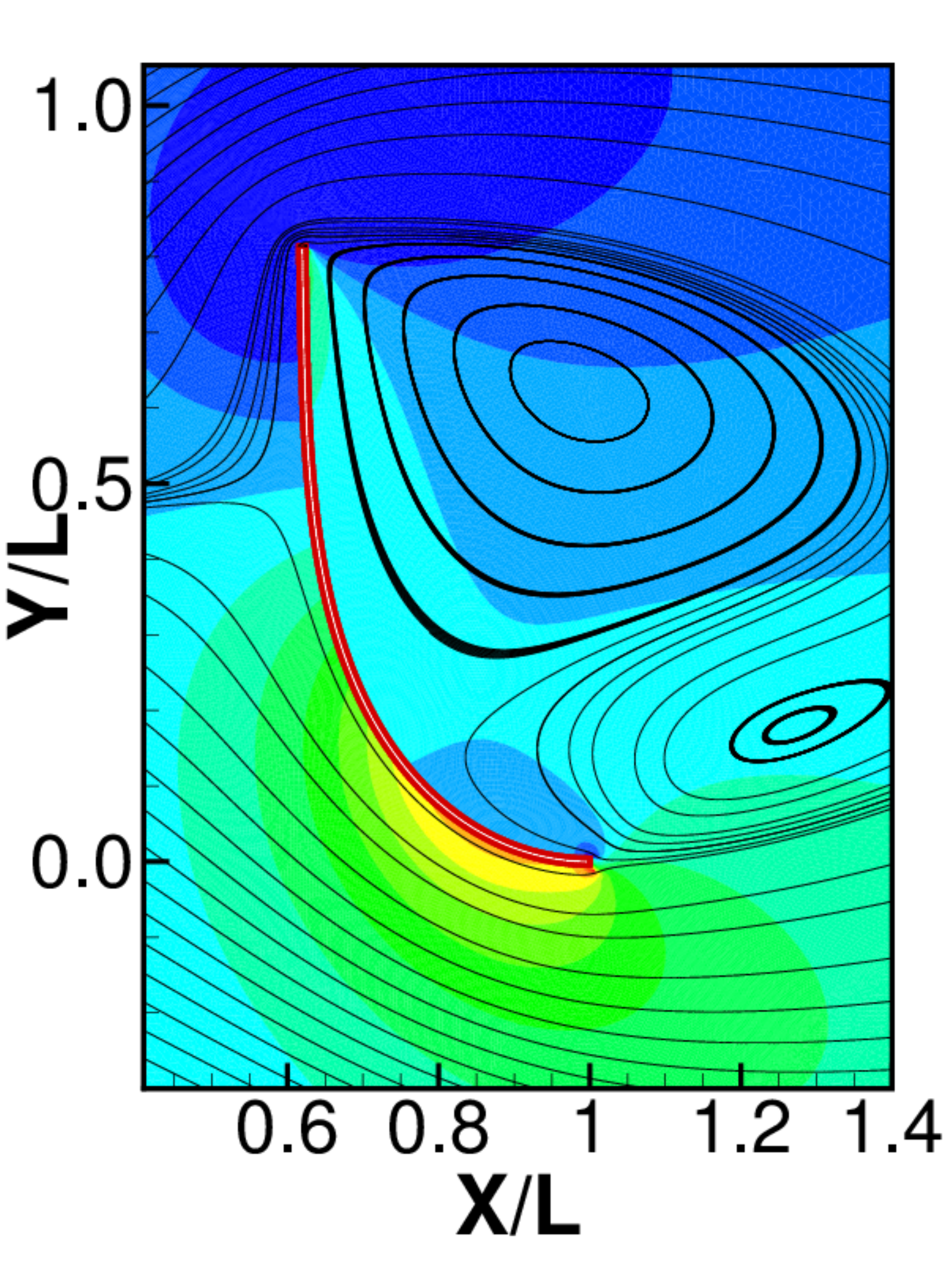}
		\caption{}
	\end{subfigure}
	\begin{subfigure}{0.275\textwidth}
		\includegraphics[width=0.99\columnwidth]{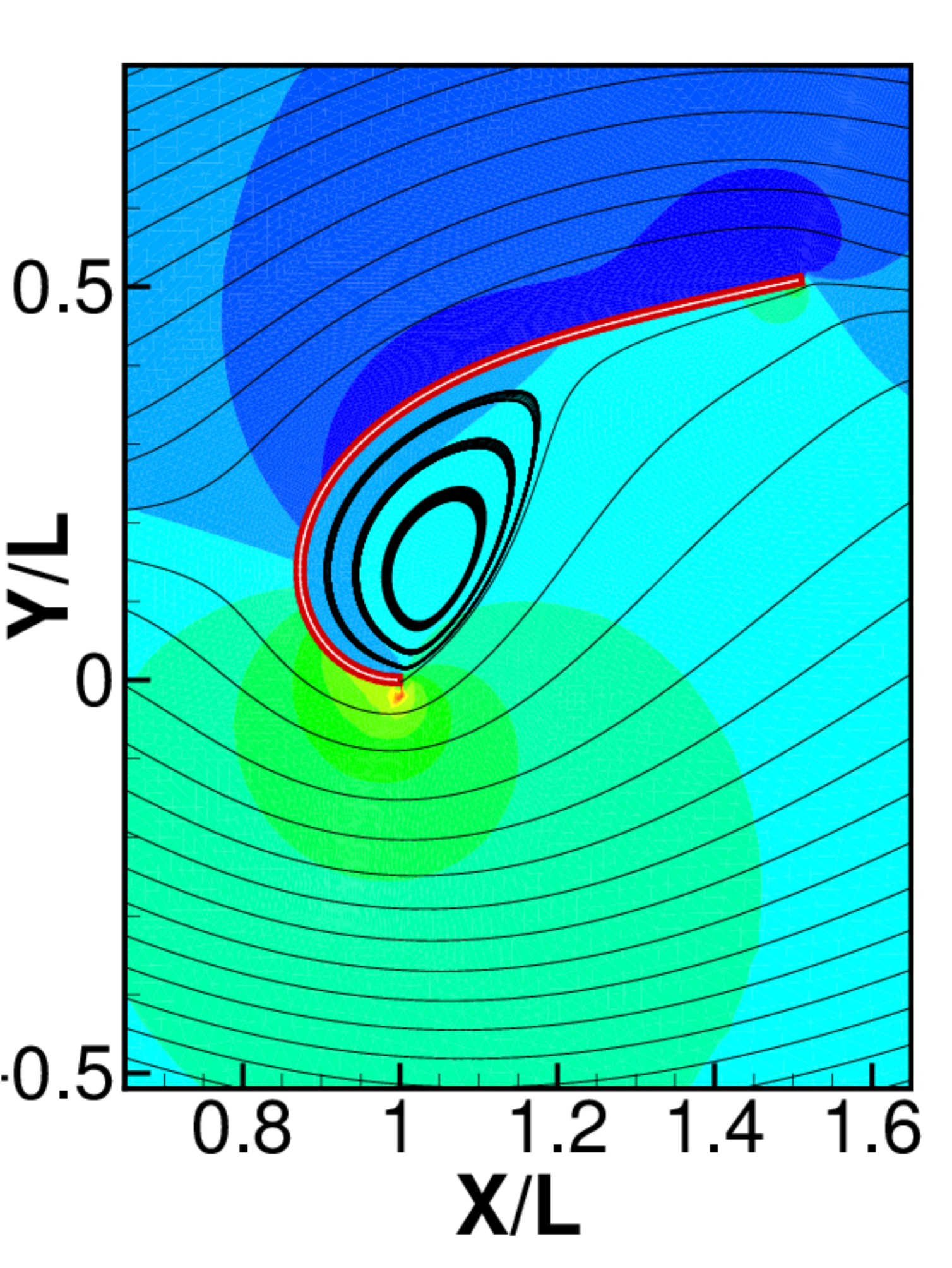}
		\caption{}
	\end{subfigure}
	\begin{subfigure}{0.275\textwidth}
		\includegraphics[width=0.99\columnwidth]{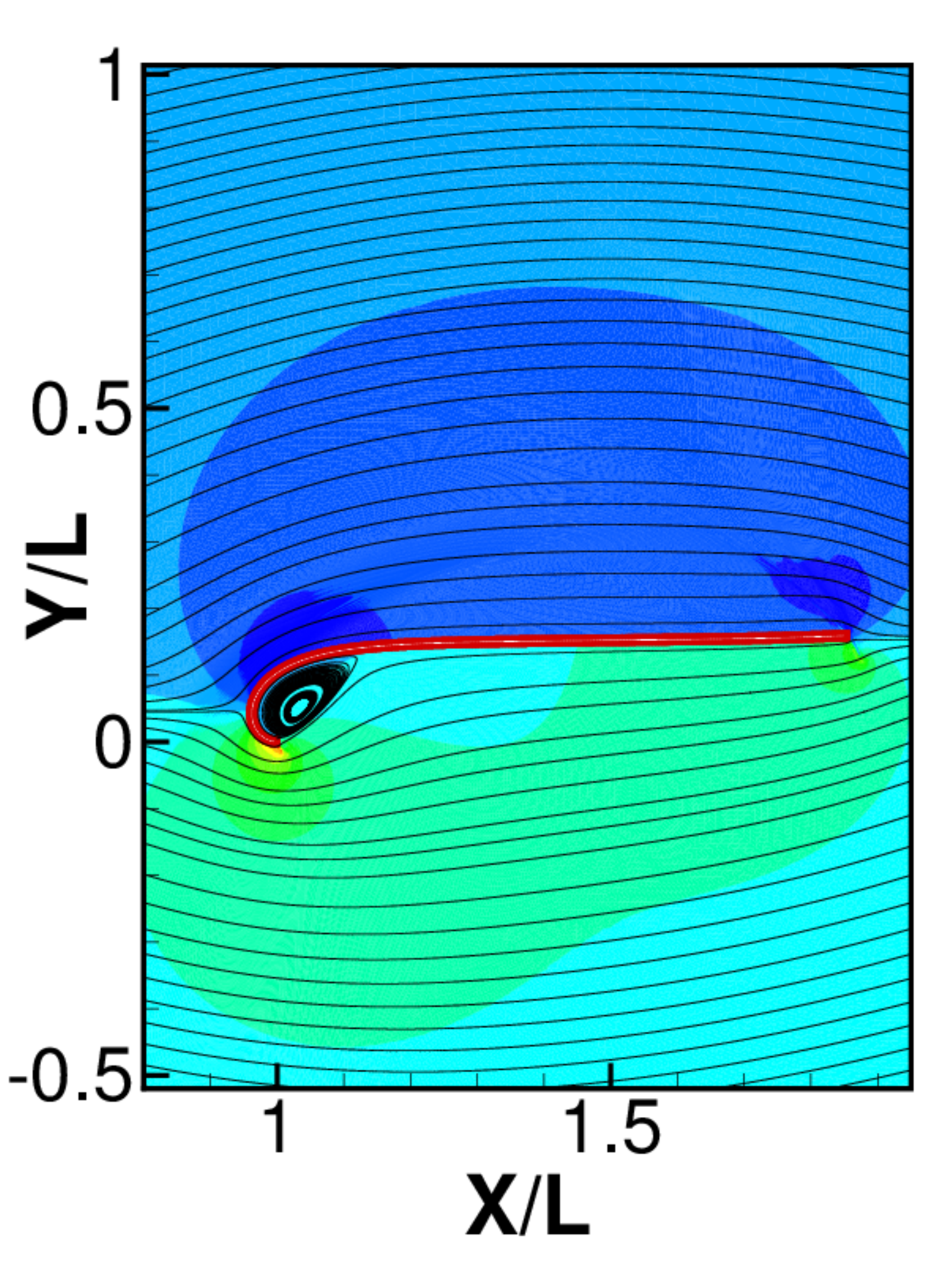}
		\caption{}
	\end{subfigure}
	\begin{subfigure}{0.1\textwidth}
		\vspace{-0.8cm}
		\includegraphics[trim=18mm 0mm 0mm 0mm,clip,width=.99\columnwidth]{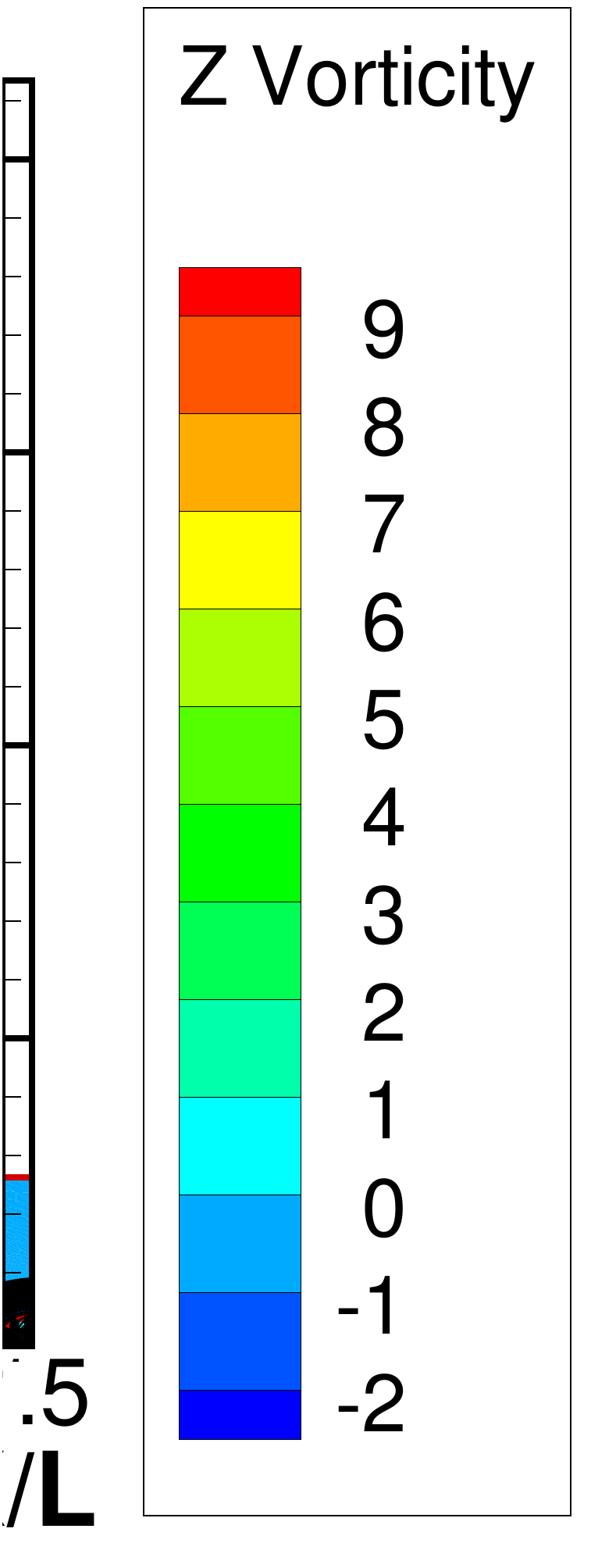}
	\end{subfigure}
	\caption{Streamline topology and spanwise vorticity distributions at $K_B=0.2$ and $m^*=1$ for: $Re=$ (a) 10, (b) 1 and (c) 0.1. }\label{lowReDynamics}
\end{figure}

Figure~\ref{lowReDynamics} shows the topology of the streamlines and spanwise vorticity distribution behind the deformed foil. 
Similar to the vortex shedding phenomenon in  bluff body flows, we no longer observe any LEV shedding for the inverted foil for $Re < 50$ 
 and there exists two counter-rotating steady vortices at $ 10 \le Re < 50$, as shown in figure~\ref{lowReDynamics}a. 
In contrast to the flow around a circular cylinder, the mirror-symmetry concerning the incoming flow is not present in the formation of vortex pair behind the inverted foil. 
%
At the flipped state, the steady vortex at the LE no longer exists, 
as illustrated in figures~\ref{lowReDynamics}b and \ref{lowReDynamics}c. 

%

\subsection{Effect of Aspect Ratio}\label{sec:aspectRatio}
\begin{figure}
	\centering
	\begin{subfigure}{0.49\textwidth}
		\includegraphics[width=0.99\columnwidth,trim=0mm 0mm 17mm 0mm,clip]{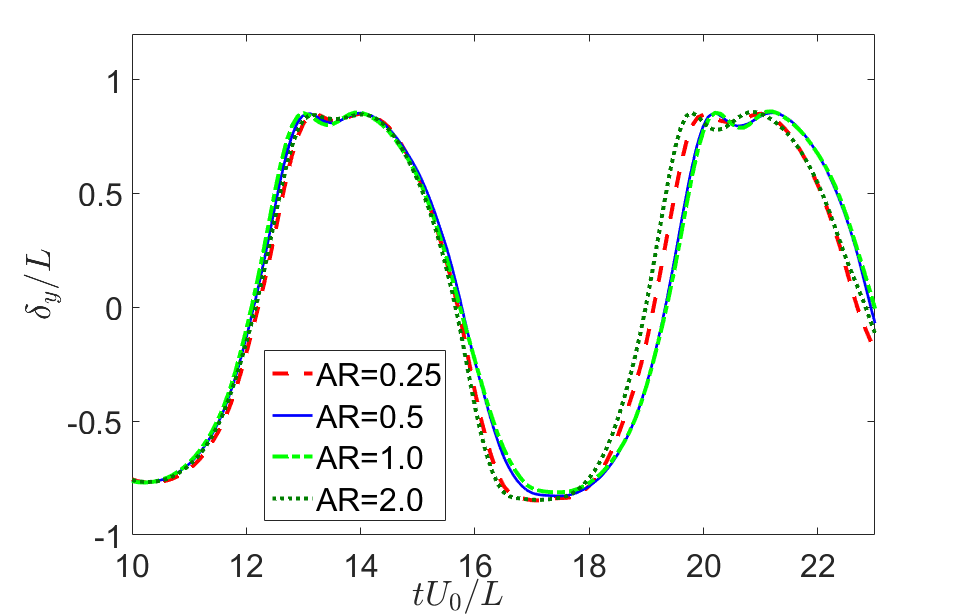}
	\end{subfigure}
	\begin{subfigure}{0.49\textwidth}
		\includegraphics[width=0.99\columnwidth,trim=0mm 0mm 17mm 0mm,clip]{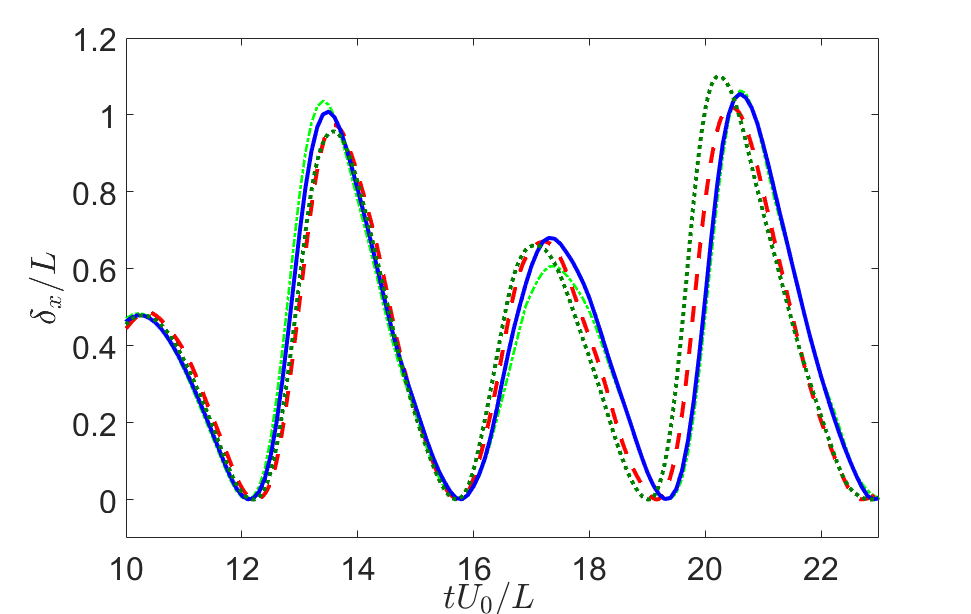}
	\end{subfigure}\\
	\begin{subfigure}{\textwidth}
		\centering
		\caption{}
	\end{subfigure}\\
	\begin{subfigure}{0.49\textwidth}
		\includegraphics[width=0.99\columnwidth,trim=0mm 0mm 17mm 0mm,clip]{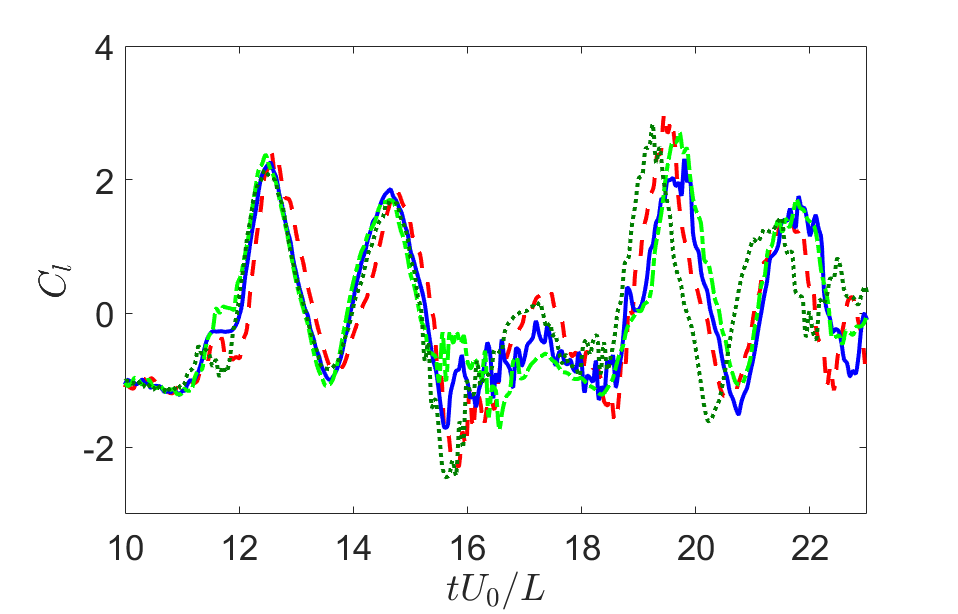}
	\end{subfigure}
	\begin{subfigure}{0.49\textwidth}
		\includegraphics[width=0.99\columnwidth,trim=0mm 0mm 17mm 0mm,clip]{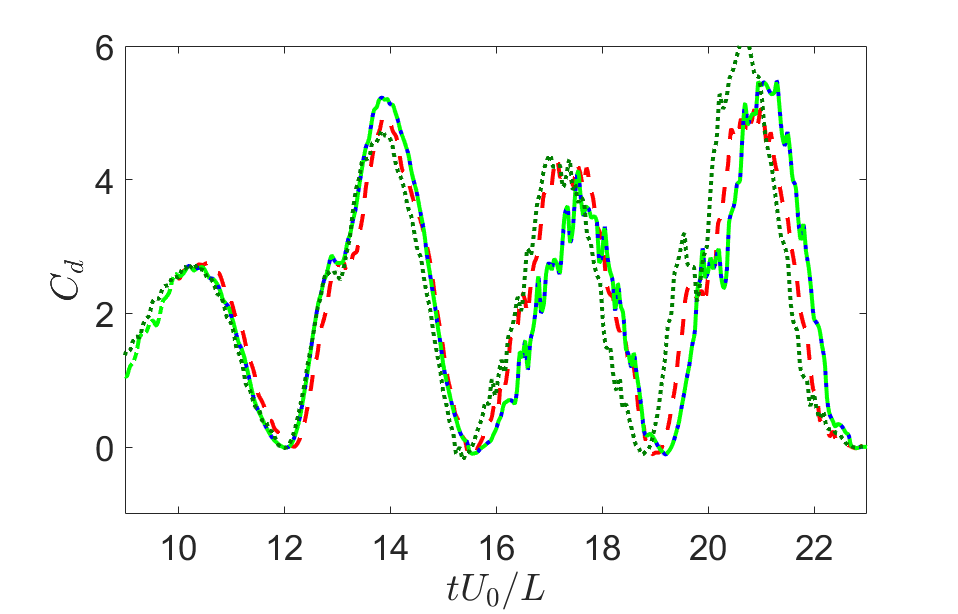}
	\end{subfigure}\\
	\begin{subfigure}{\textwidth}
		\centering
		\caption{}
	\end{subfigure}\\
	\caption{Effect of the inverted foil aspect-ratio ($\mathrm{AR}$) at $Re=30000,\ K_B=0.2, \ m^*=1$  on: (a) LE transverse (left) and streamwise (right) displacements, and (b) transverse lift (left) and streamwise drag (right) forces acting on the foil. }\label{aspect_ratio_disp}
\end{figure}
\begin{figure}
	\begin{subfigure}{0.5\textwidth}
		\centering
		\includegraphics[width=0.98\columnwidth]{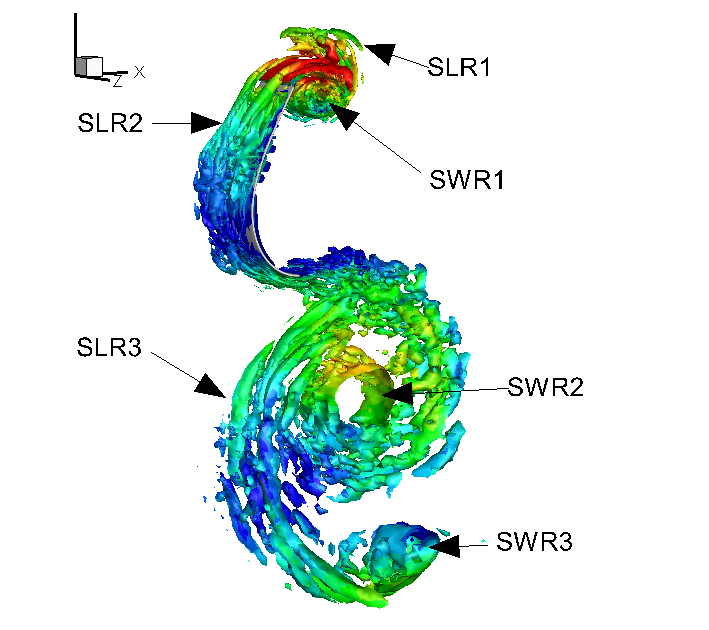}
		\caption{$\mathrm{AR}=0.25$}
	\end{subfigure}
	\begin{subfigure}{0.5\textwidth}
		\centering
		\includegraphics[width=0.98\columnwidth]{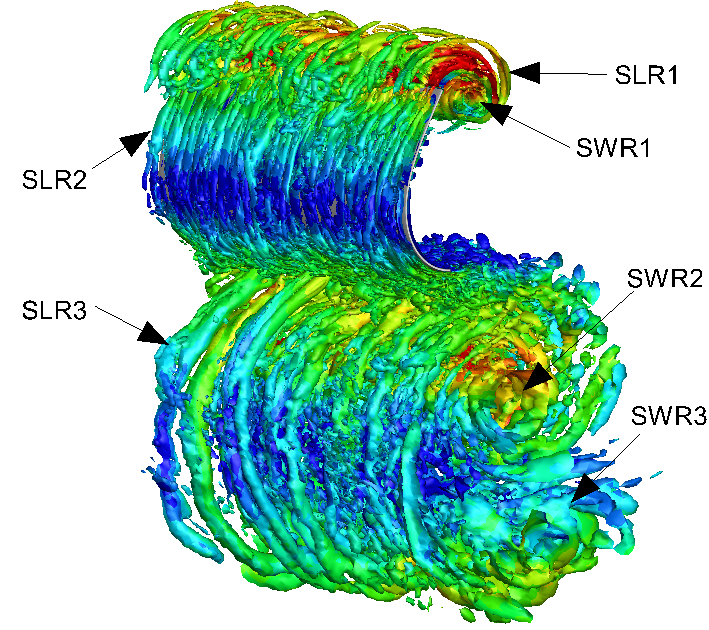}
		\caption{$\mathrm{AR}=2$}
	\end{subfigure}\\
	\begin{subfigure}{\textwidth}
		\centering
		\includegraphics[width=0.65\columnwidth]{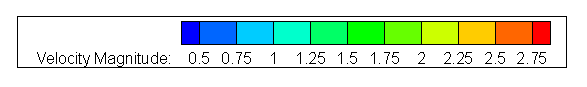}
	\end{subfigure}
	\caption{Iso-surfaces of three-dimensional instantaneous vortical structures at $K_B=0.2,\ Re=30000$ and $m^*=1$ for two representative aspect ratios: $\mathrm{AR}=$ (a) =0.25 and (b) 2.0. Here SLR and SWR denote streamline ribs and spanwise rolls and the iso-surface $\lambda_2=30$  colored by the fluid velocity magnitude is plotted. }\label{aspect_vortex_patterns}
\end{figure}

\revi{Here we will examine the influence of foil AR on the 3D flow structures, the forces and flapping amplitudes for representative AR values at nondimensional parameters $Re=30000,\ K_B=0.2$ and $m^*=1.0$. Figures~\ref{aspect_ratio_disp}a and \ref{aspect_ratio_disp}b compare the time histories of the transverse and streamwise displacements of inverted foil with $\mathrm{AR}=\{0.25, 0.5, 1.0, 2.0\}$. The figures show that the effect of $\mathrm{AR}$  on the transverse and streamwise flapping amplitudes of LAF is marginal. This observation conforms with experimental observations of \cite{kim2013} where the authors did not observe any significant difference in the flapping response for $\mathrm{AR}\in[1.0,1.3]$.
We also compare the time histories of the lift and the drag forces acting on the flexible foil in figure~\ref{aspect_ratio_disp}b. Even the forces acting on the foil as a function of AR show similar time histories barring small variation which can be attributed to the complex 3D turbulent flow structures, since the forces acting on the foil are more sensitive towards the vortex structures than their bulk kinematic response.}

\revi{To investigate the effect of $\mathrm{AR}$ on the large scale 3D flow structures generated by the LAF, we plot the $\lambda_2$ \citep{lambda2} iso-surfaces in figure~\ref{aspect_vortex_patterns} at $\mathrm{AR}=0.25$ and $2.0$ for identical nondimensional parameters $K_B=0.2,\ Re=30000$ and $m^*=1$. We can observe similar flow structures for both the $\mathrm{AR}$ values. The figures show the existence of the streamline ribs SLR1 in the wake behind the deformed foil and SLR2 along the foil surface. The spanwise roll SWR1 represents the LEV vortex formed behind the foil. SWR2 and SWR3 denote the TEV and LEV vortices respectively formed behind the foil during the previous downstroke. SLR3 is the streamline ribs formed behind the deformed foil during the previous downstroke. The streamline ribs SLR3 encloses the LEV+TEV pair of SWR3 and SWR2. It should be noted that in this study, we have ignored the effects of spanwise end effects which can become dominant for AR$\le 0.2$ at $K_B=0.2$, $Re=30000$ and $m^*=1.0$ \citep{sader_2016}.}  



\subsection{Discussion}
\begin{figure}
	\centering
	\begin{subfigure}{0.6\textwidth}
	\includegraphics[width=0.99\columnwidth]{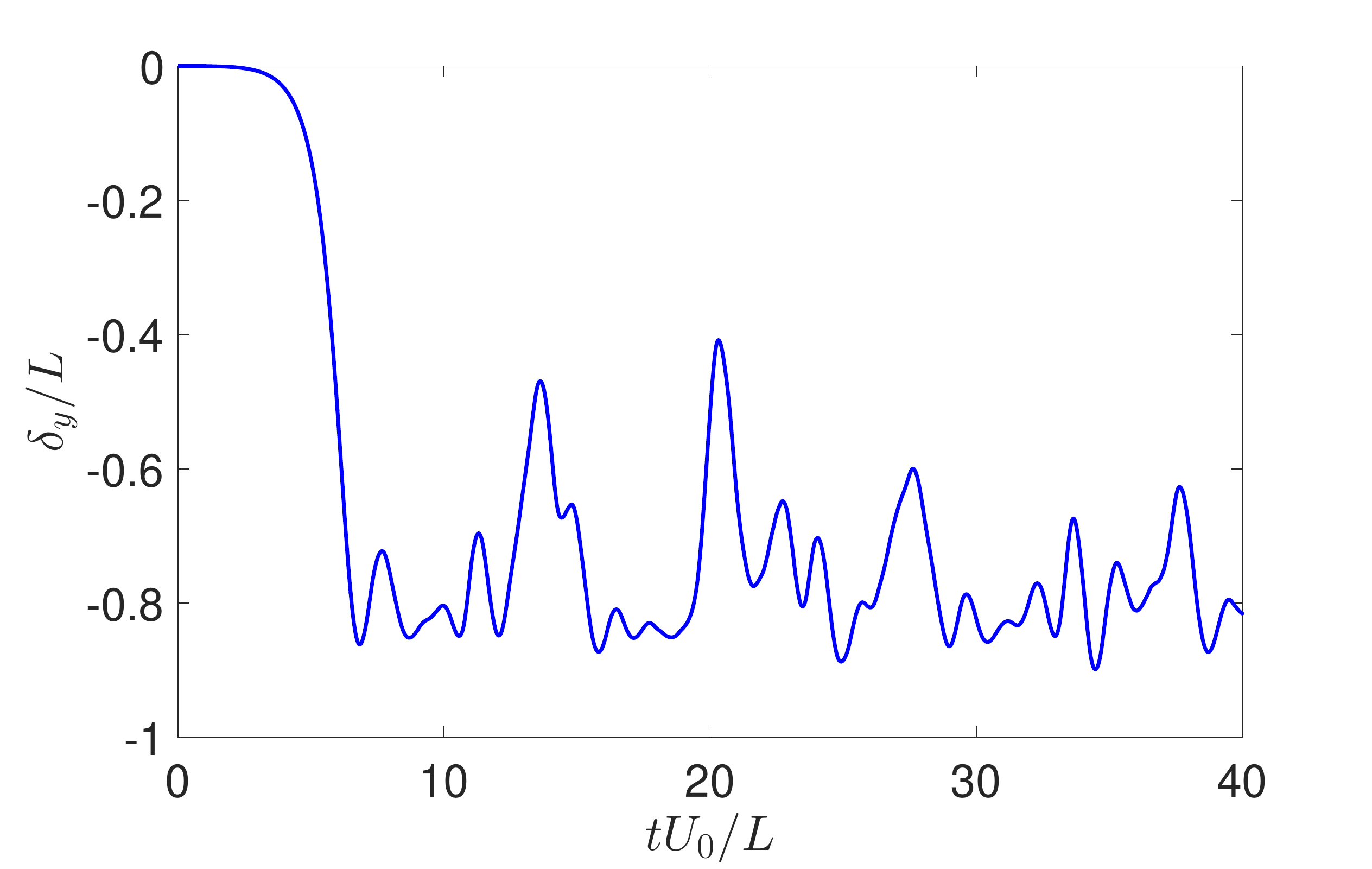}
	\end{subfigure}
	\begin{subfigure}{0.25\textwidth}
		\includegraphics[width=0.99\columnwidth,trim=100mm 20mm 160mm 10mm,clip]{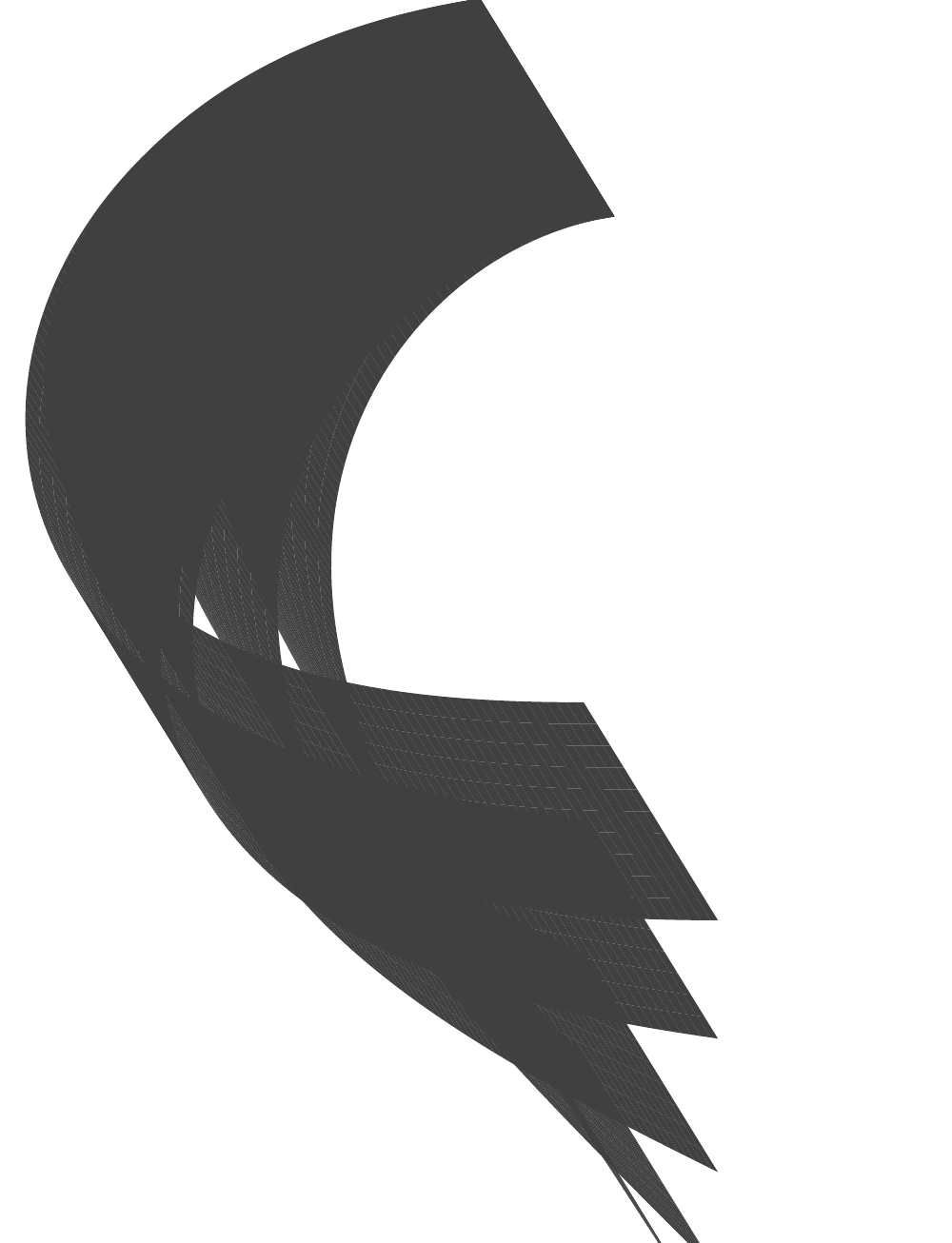}
	\end{subfigure}	
	\caption{Transverse LE response history (left) and foil profiles over a flapping cycle (right) for the inverted foil without splitter at $K_B=0.075,\ Re=30000,\ m^*=1.0$ and AR$=0.5$. Owing to a relatively small elastic recoil at $K_B=0.075$, the inverted foil does not exhibit LAF albeit unsteady vortex shedding.}\label{deformedFlapping}
\end{figure}

\revii{The LAF response of an inverted foil involves large structural deformations with complex interactions between 3D vortical structures shed from the LE and TE and the foil properties such as inertia and elasticity. The underlying physical mechanism behind the LAF phenomenon can be summarized as below:
\begin{itemize}
	\item The onset of LAF is characterized by the breakdown of symmetry through the divergence instability. Once the foil deformation due to the divergence instability becomes large enough, the flow separates at the LE to form a LEV behind the foil.
	\item The formation of LEV results in a low-pressure region behind the foil due to which the lift and drag acting on the foil increase. Both the lift and drag forces play a significant role in the large deformation of the inverted foil. For high-$Re$ turbulent wake flow, initial deformation of the foil from the mean position is dominated by the lift and inertial forces until the drag force on the foil becomes large enough to sustain the deformation. On the other hand for the low-$Re$ laminar flow, the large foil deformation is predominately due to the drag acting on the foil. 
	\item The formation of LEV is not just enough to sustain the LAF response of inverted foil in a uniform flow. 
	This suggests that the periodic flapping motion is due to the complex interplay and the coupled fluid-structure effects of the unsteady shedding of LEV and the structural dynamics of flexible foil due to its elasticity and inertia. 
	Let us first consider a case where there is no shedding. The foil inertial effects will be lessened and an equilibrium between the fluid and structural restoring forces is reached like the LE responses for $Re\le20,\ K_B=0.2$ and $m^*=1.0$ (figure~\ref{lowRe_response}a). The second possibility is where we have unsteady vortex shedding but the combined foil elastic restoring and the inertial effects are not sufficient to overcome the fluid forces acting on it. To demonstrate this case, we present the LE transverse response history at $K_B=0.075,\ Re=30000,\ m^*=1.0$ and AR$=0.5$ in figure~\ref{deformedFlapping}. The figure shows that the inverted foil no longer exhibits LAF, instead it performs  the flapping motion about a deformed state because the combined elastic and inertial effects due to the foil recoil cannot overcome the fluid forces acting on the foil. A similar deformed flapping mode was observed both experimentally \citep{kim2013} and numerically \citep{ryu_2015,gurugubelli_JFM,mittal_2016}.
	\item Mass ratio has a weak influence on the LAF of inverted foil. The 2D simulations of \cite{gurugubelli_JFM} have shown that as the $m^*$ increases the transition from LAF to the deformed flapping can be delayed, i.e. the foil with a greater inertia can overcome the flow-induced forces acting against the foil with more ease 
in comparison to the foil with a lower inertia. 
\end{itemize}}
\changes{To further demonstrate that the physical mechanism behind the LAF is a complex interplay of the foil elastic, the rotational inertia, the LEV and unsteady vortex forces during the periodic LAF motion, we formulate a simplified analogous model based on the elastically mounted plate which is free to undergo single-degree-of-freedom rotation about the TE in a uniform stream. The detail of the analogous formulation is presented in the appendix. 
In this simplified model, 
we estimate the 2D steady force and moment on the plate at a certain rotation angle by means of a simple 
potential theory and the mean drag force acting on the projected length.
We assume that the formation and shedding of vortices do not result in a time-dependent moment on 
the flat plate a given angle of incidence.
The simplified quasi-steady model reasonably predicts the LAF response amplitude and elucidates a direct connection between 
the large-amplitude oscillation and the LAF response amplitude.
It is worthwhile to mention that the inclusion of the improved models for the LEV and TEV in the potential flow model may 
improve the accuracy of the model at large angles of incidence.
We next present the linkage between the dynamics of the LAF with the vortex-induced vibration of a circular cylinder.
}

\changes{With regard to  the underlying fluid-structure interaction and the large amplitude response, 
the LAF holds both similarities and differences with the VIV of an elastically mounted circular cylinder.
Although both configurations are geometrically dissimilar, they exhibit large periodic amplitudes perpendicular to the oncoming flow stream. When an inverted foil 
deforms, the trajectory of LE forms a circular-arc like shape around the TE in a given flapping cycle.  For the oncoming flow stream, 
this time-varying geometrical shape can resemble an effective semi-circular body traversing between the two peak transverse locations and 
emanating two counter-rotating vortices from the leading edge.
 Figure~\ref{VIV_schem} shows the typical schematics of the flow structures exhibited by an elastically mounted circular cylinder, an inverted foil and an inverted foil with a splitter plate. The UVK and LVK represent the upper v{o}n K\'{a}rm\'{a}n and lower von K\'{a}rm\'{a}n vortices formed due to the roll-up of the separated shear layers from the upper and lower cylinder surfaces respectively. The LEV1 and LEV2 in figures~\ref{VIV_schem}b and ~\ref{VIV_schem}c denote the LEV shed during the current and previous half-cycles respectively. Similarly, the TEV1 and TEV2 in figure~\ref{VIV_schem}b denote the TEV shed during the current and previous half-cycles. Based on the results and observations presented in the above subsections, we can deduce the following differences and similarities:
	 \begin{figure}
		\centering
		\begin{subfigure}{0.58\textwidth}
			\includegraphics[width=0.99\columnwidth]{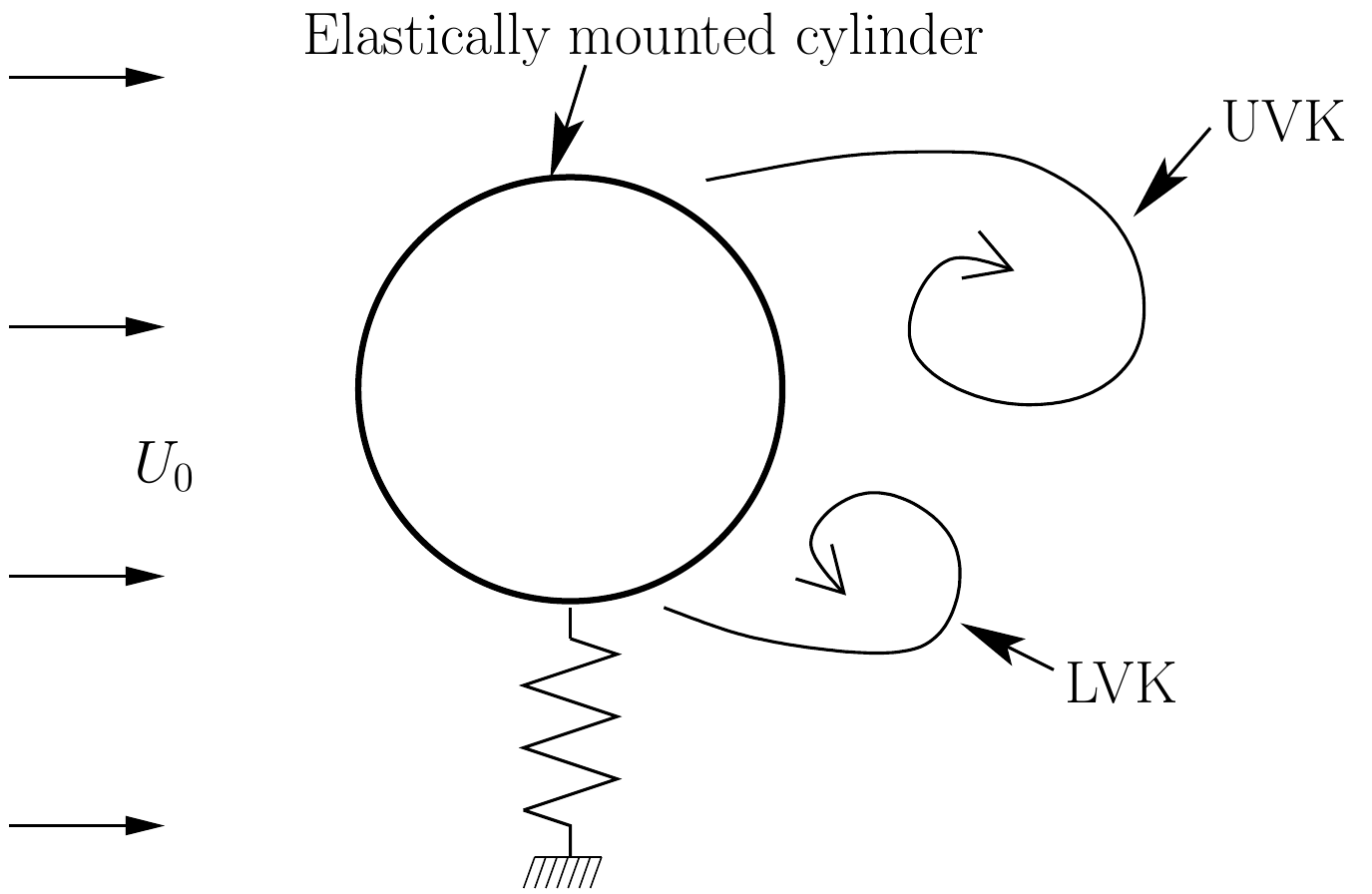}
			\caption{}
		\end{subfigure}
		\begin{subfigure}{0.4\textwidth}
			\includegraphics[width=0.99\columnwidth]{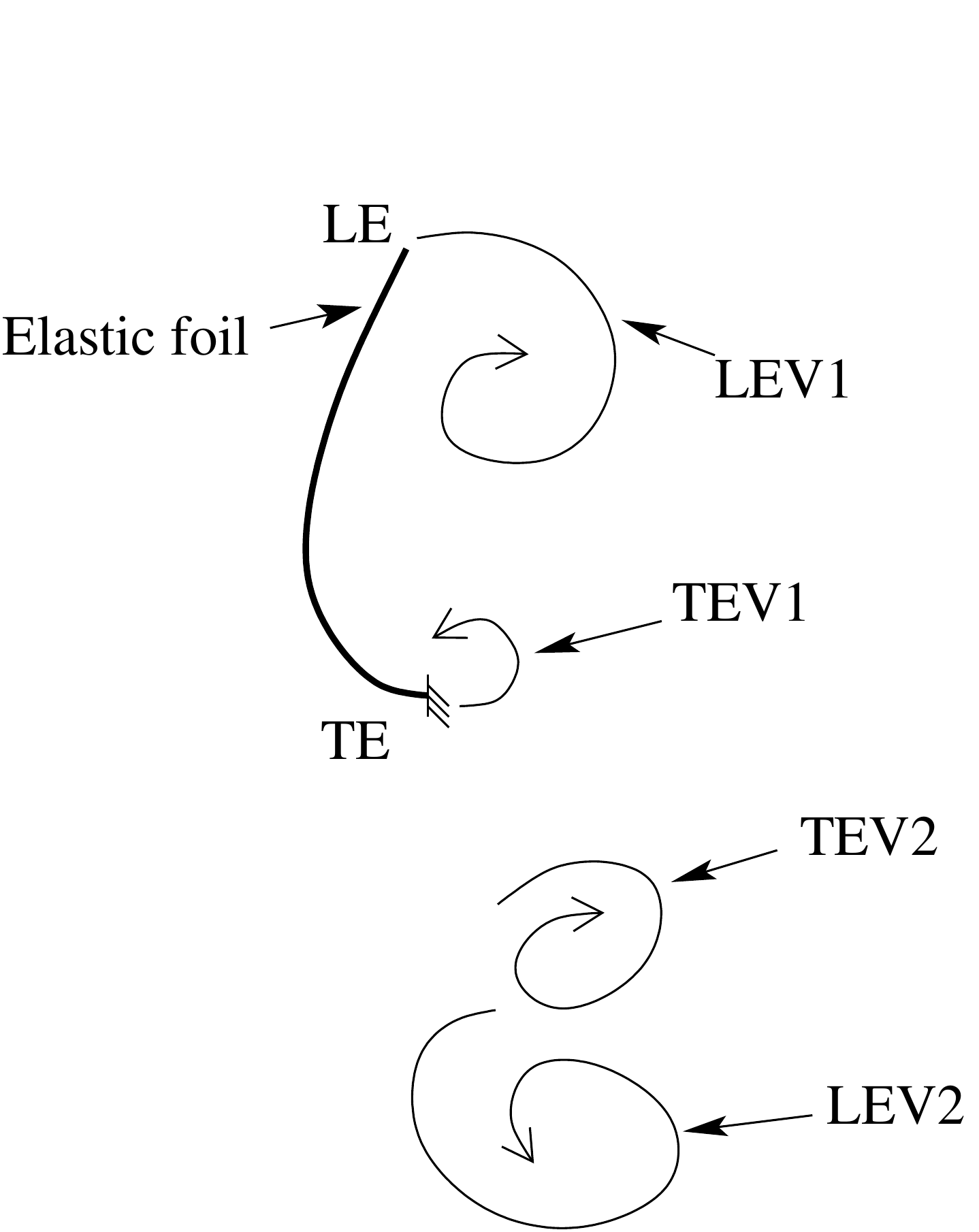}
			\caption{}
		\end{subfigure}\\
		\begin{subfigure}{0.6\textwidth}
			\includegraphics[width=0.99\columnwidth]{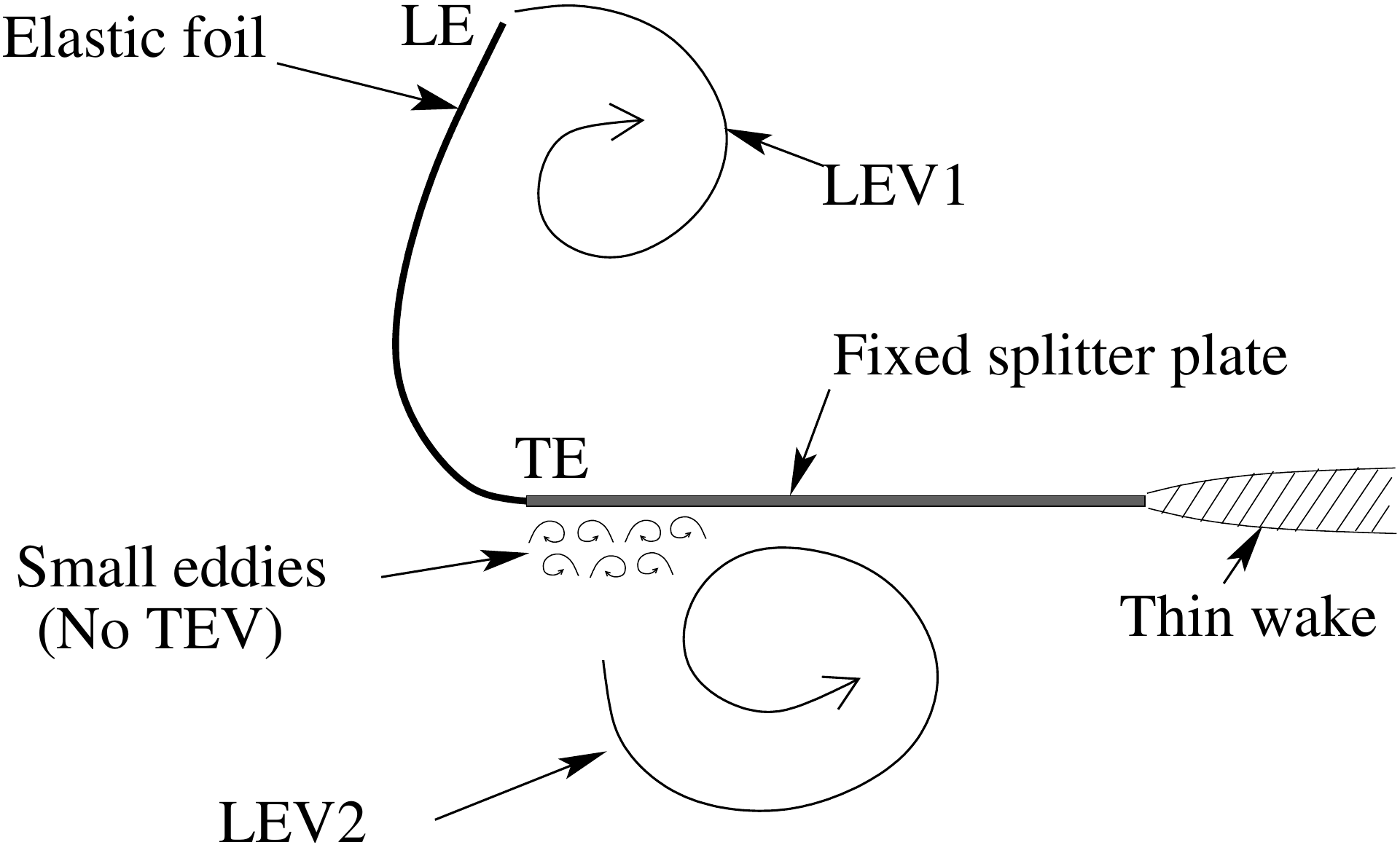}
			\caption{}
		\end{subfigure}
		\caption{ Illustration of typical dominant flow structures during the uniform flow past over: (a)  elastically mounted circular cylinder undergoing VIV with upper von K\'{a}rm\'{a}n (UVK) and lower von K\'{a}rm\'{a}n (LVK) vortices, (b) elastic inverted foil performing LAF with LEV and TEV, and (c) elastic inverted foil clamped at TE to a fixed splitter plate performing LAF with LEV. }\label{VIV_schem}
	\end{figure}
\begin{itemize}
\renewcommand{\labelitemi}{\scriptsize$\blacksquare$}
\item The fundamental difference between the LAF in an inverted foil without splitter and the transverse VIV of an elastically mounted cylinder is that for the LAF frequency may not necessarily be same as that of the vortex-induced forces. However, for the LAF at $K_B=0.2$, $Re=30000$ and $m^*=1$, we have shown that the dominant $C_l$ frequency does not synchronize with the flapping frequency in the transverse direction (see figure~\ref{ComparisonBetweenInvFoilWithAndWithOutSplitter}). On the other hand, the VIV of an elastically mounted cylinder synchronizes with the vortex-induced force frequencies \citep{blevins,naudascher}. 
\item {The strong coupling between the UVK and LVK vortices is critical for the self-sustaining trait of the VIV phenomenon \citep{assi_splitter,yunZhi_2017}. On the other hand, the synchronization between the LEV and TEV vortices has an insignificant influence on the LAF phenomenon. Even the complete suppression of TEV vortex (figure~\ref{VIV_schem}c) has a little influence on the flapping amplitude, which is typically not the case for the VIV of circular cylinders. The LAF dynamics of an inverted foil is a complex interplay of  force dynamics associated with the unsteady LE vortex shedding, the foil inertia and the elastic recoil of the flexible structure. }
\item While both UVK and LVK vortices of similar strengths play an equal role into the VIV of an elastically mounted cylinder, the LAF of inverted foil is predominantly influenced by the LEV1 and LEV2, as illustrated in figures~\ref{VIV_schem}b and \ref{VIV_schem}c. The TEV1 and TEV2 have little effect on the flapping amplitude of inverted foil. However, the TEV1 and TEV2 can modulate the flapping frequency. The suppression of the vortices from the TE reduces the flapping frequency.
\item The wake behind an elastically mounted cylinder vibrating in transverse direction (figure~\ref{VIV_schem}a) and the wake behind a deformable inverted foil with a splitter (figure~\ref{VIV_schem}c) will look similar if the splitter is ignored. Similar to the VIV of an elastically mounted cylinder, which depends on the synchronization between the upper von Karman (UVK) and the lower von Karman (LVK) vortices, the flapping dynamics of an inverted foil depends on the synchronization between the two leading edge vortices. The loss due to the absence of hydrodynamic interaction between the two leading edge vortices is counter-balanced by the foil elastic and inertial effects. Just like during VIV, if we suppress either of the LEV we can suppress the large amplitude flapping. Furthermore, the shedding of LEV is not sufficient to have a large amplitude flapping if the elastic recoil-induced inertia is not sufficient to counteract the fluid dynamic forces and the inverted foil will no longer exhibit the large amplitude flapping (figure~\ref{deformedFlapping}). As far as the interaction between LEV1 and TEV1 is concerned, the large amplitude flapping mechanism does not depend on this interaction. However, if present it will enhance the foil recoil motion from the maximum transverse displacement to the mean position. 
\item Unlike the VIV of an elastically mounted cylinder, the LAF response in inverted foil has its origin in the coupled dynamics of the flexible body and the LEV formed by the roll-up of the separated shear layer (Kelvin-Helmholtz instability) at the LE of the inverted foil. 
In comparison to the strongly coupled asymmetric vortices formed via B\'enard-von K\'arm\'an instability behind circular cylinders, 
the interaction between LEV and TEV in the inverted foil is relatively weak and the placement of a splitter plate has a marginal effect on the overall flapping response amplitudes. 
This implies that the LAF of the inverted foil arises from the intrinsic fluid-body interaction between the elastic foil and the vortex shedding process behind the deformed foil, wherein the vortex forces acting on the foil perform a net work thereby resulting in the energy transfer from the fluid flow in the form of kinetic and strain energies of the foil. 
\item One of key similarity of LAF and VIV is that both the physical phenomenon require the existence of vortex shedding. Suppression of the shedding ceases the flapping and the formation of LEV is necessary for the large amplitude deformation.		
\item The LAF of inverted foil with a splitter closely resemble the VIV of an elastically mounted circular cylinder. In both the cases, the vortex induced  forces synchronize with the flapping response and exhibit `2S' vortex mode. However, LAF with splitter does not involve any interaction between the counter-rotating vortices LEV1 and LEV2. On the other hand, the interaction between UVK and LVK is necessary for the self-sustained large-amplitude oscillations during VIV.
\item Finally, for the VIV of an elastically mounted cylinder, the response amplitude and the synchronization range strongly depend on the mass ratio of immersed body in a flowing stream. However, the experimental \citep{kim2013} and  the numerical \citep{gurugubelli_JFM,mittal_2016} studies have shown that mass ratio has a little impact 
on the LAF amplitude and the synchronization range. 
\end{itemize}}

\section{Concluding Remarks}
The large amplitude limit-cycle flapping of an inverted foil is numerically investigated to elucidate the role of the unsteady counter-rotating vortices shed from the leading and trailing edges of a flexible inverted foil in a uniform flow stream. The coupled 3D fluid-structure interaction formulation relies on the quasi-monolithic formulation with the body-conforming interface
and the variational multiscale turbulence model is employed for the separated wake flow at high Reynolds number.
We have validated the 3D LAF response obtained by the numerical scheme with the published experimental data of \cite{kim2013} at Reynolds number $Re=30,000$.

We first explored the detailed flow field around the inverted foil undergoing the LAF and examined the influence of the vortices shed on the response dynamics. 
In addition to the primary counter-rotating vortex pair (LEV+TEV), 3D spanwise wake structures behind the foil revealed that the LAF can produce a secondary spanwise vortex.
From the evolution of 3D separated flow structures, we find that the vortical structures  are characterized by a large spanwise vortex and small multiple counter-rotating pairs 
of streamline vortices behind the foil. Owing to three-dimensional effects, we have also observed the streamline ribs along the front side of the deformed foil surface.
We have analyzed the relationship of the lift force with the transverse amplitude and the drag force with the streamwise amplitude. 
There exists a critical deformation above which the foil deformation is dominated by the drag until the foil recoils 
due to the elastic restoration forces of the inverted foil. The elastic recoil is followed by the LEV shedding phenomenon.
We also observed that the foil aspect ratio with spanwise periodicity has a minor role on the LAF dynamics.
%

We introduced a novel inverted foil configuration with a fixed splitter plate at TE to suppress the vortex shedding from the TE and thereby to realize the impact of the TEV and the interaction between the TEV-LEV vortex pair. 
We have identified that unlike the self-excited VIV in circular cylinders where synchronized periodic vortex shedding from the top and bottom surfaces is essential for the  self-sustained response, 
the large amplitude periodic flapping of inverted foil does not depend on the synchronized periodic vortex shedding from the leading and trailing edges. 
Notably, the elimination of the TEV via a splitter plate reduces the flapping frequency and modulates the streamwise flapping amplitudes. 
Instead, the foil recoil motion from the maximum transverse displacement depends on the foil inertia attained due to the periodic vortex shedding from LE and the elastic restoring forces.
We also investigated the interaction between an inverted flexible foil without splitter and low-$Re$ flow to generalize the impact of the LEV shedding on the large-amplitude periodic flapping. 
%
Similar to the flow over circular cylinder, the vortex shedding phenomenon ceases for $Re<50$ and the foil no longer exhibits unsteady flapping motion. 
However, the foil exhibits the large static deformation that increases with decrease in $Re$. The foil eventually flips about the leading edge to align its leading edge along flow for $Re\le 1$. For low-$Re$, the large foil deformation is primarily sustained by the large static drag acting on the foil. 
We have shown that the shedding of LEV is necessary but not just enough to sustain the LAF response of inverted foil
in a uniform flow. Instead, the  LAF response is the outcome of strong fluid-structure interaction
associated with the combined effects of unsteady shedding of LEV and the flexible foil's structural elasticity and inertia effects.
\changes{
Based on the aforementioned  investigations,  we also presented a list of similarities and differences between the LAF phenomenon of inverted foil and
the VIV of a circular cylinder.
In addition, we have formulated an analogous elastically mounted rotating plate model with a steady force estimation based on the potential flow theory and the empirical drag force model. The model reasonably reproduced the large-amplitude oscillation similar to the LAF response of a flexible inverted foil in a uniform stream and thus formed an appropriate 
analogous model than the elastically mounted circular cylinder undergoing transverse VIV.
}
The connection between the large-amplitude oscillation of rigid plate and the LAF response 
can help  to develop  an improved understanding of underlying fluid-structure interaction, with a profound impact on energy harvesting and propulsive systems.



\appendix
\setcounter{equation}{0}
\setcounter{figure}{0}
\renewcommand{\theequation}{A.\arabic{equation}}
\renewcommand{\thefigure}{A.\arabic{figure}}
\section*{Appendix: Analogous rotating plate model for the LAF motion of inverted foil }

\begin{figure}
	\centering
	\begin{subfigure}{0.37\textwidth}
		\includegraphics[width=0.99\columnwidth]{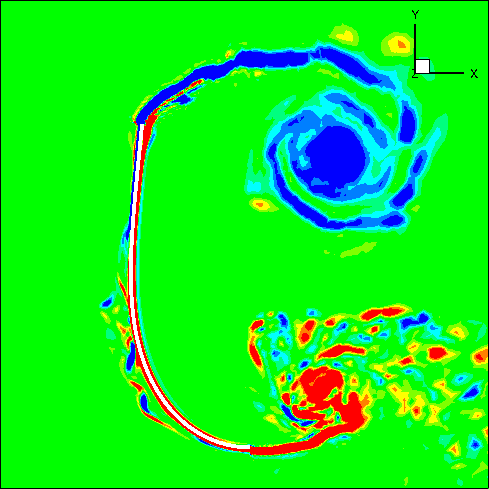}
		\caption{}
	\end{subfigure}
	\begin{subfigure}{0.52\textwidth}
		\includegraphics[width=0.99\columnwidth]{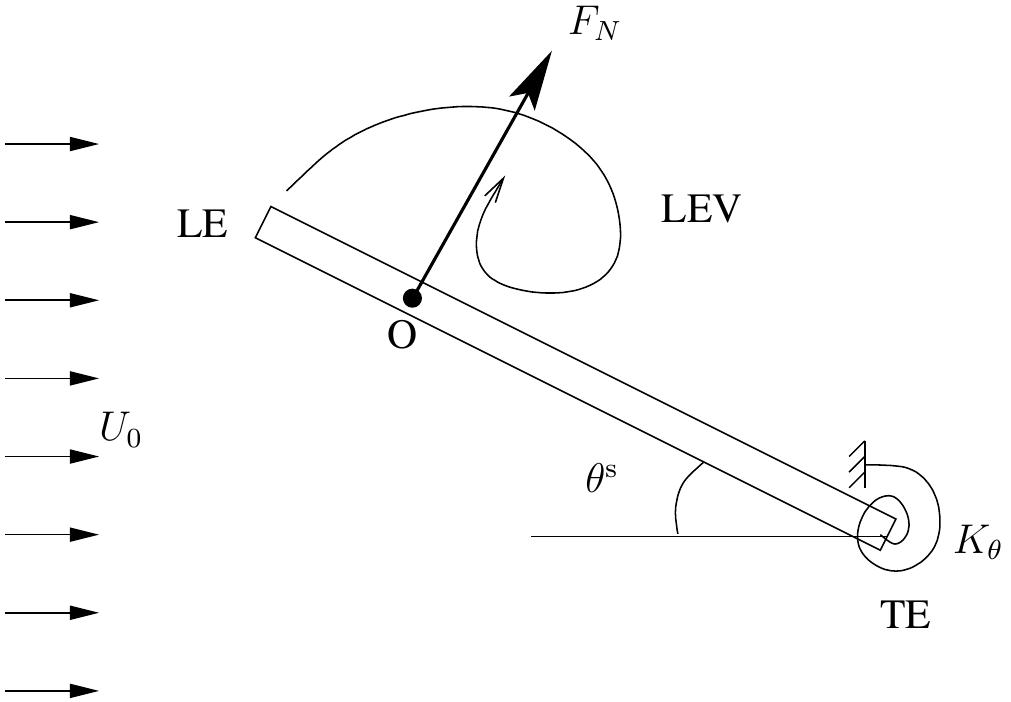}
		\caption{}
	\end{subfigure}
	\caption{(a) Spanwise vorticity contours during LAF motion from 3D FSI simulation of an inverted foil without splitter, (b) sketch of analogous nonlinear pendulum model consisting of a rigid plate mounted on a torsional spring at TE and immersed in a uniform flow stream at freestream velocity $U_0$. $K_\theta$ is the torsion spring constant and $\theta^\mathrm{s}$ is twist angle from the initial equilibrium state.  In (a), LEV and TEV vortical structures during the LAF motion 
can be clearly seen. }\label{nonlinearPendulum}
\end{figure}
Here, we will attempt  to establish a simplified analogous model for the LAF of an inverted flexible foil using 
the elastically mounted flat plate and the 2D force model using a simple potential flow theory. The fluid force represents a potential flow lift and the empirical drag based on the projected plate height. 
As shown in figure~\ref{nonlinearPendulum}a, the LAF response involves complex interactions between the fluid-dynamic forces induced due to the periodic shedding of LEV and TEV vortices from the LE and TE of the foil, the nonlinear structural elastic restoring forces and the foil inertia.
We can idealize the complex nonlinear structural deformations of inverted foil by assuming the flexible foil as a rigid plate mounted on a torsional spring at TE. Figure~\ref{nonlinearPendulum}b illustrates the schematic of the analogous nonlinear pendulum-like model undergoing single-degree-of-freedom pitching motion or rotation in a uniform flow $U_0$ for LAF, where $K_\theta$ is the torsional spring constant.
The equation of motion for the nonlinear pendulum (i.e., elastically mounted rotating plate) for the rotation angle ($\theta^\mathrm{s}$) 
can be written as
\begin{equation}
\begin{aligned}\label{staticEq}
(I+I_\mathrm{a}) \theta_{tt}^\mathrm{s} + C_\theta \theta^\mathrm{s}_t + K_\theta \theta^\mathrm{s} = {M}^\mathrm{f}_z,
\end{aligned}
\end{equation}
where $\theta_{tt}^\mathrm{s}$ and $\theta^\mathrm{s}_t$ represent $\partial^2 \theta^\mathrm{s}/\partial t^2$ and $\partial \theta^\mathrm{s}/\partial t$ respectively, $C_\theta$ is the torsional damping constant, $I=mL^2/3$ and $I_\mathrm{a}=9\pi\rho^\mathrm{f}L^4/128$ \citep{sarpkaya_added_mass,blevins_added_mass} are the moment of inertial and added moment of inertia, respectively for the rigid plate rotating about LE, and $m$ is the mass per unit length. 
${M}^\mathrm{f}_z$  denotes the steady moment acting on the plate due to the fluid loading $F_N$ acting normal to the plate at the aerodynamic center $O$, where the pitching moment does not vary as a function of twist angle. By considering a quasi-static loading on the rotating plate,  the normal force via potential flow model and the force decomposition assumption can be given as
\begin{equation}
\begin{aligned}
F_N &= \frac{1}{2}\rho^\mathrm{f}U_0^2L (C_d \sin\theta^\mathrm{s} \left|\sin\theta^\mathrm{s}\right|+ C_l \cos\theta^\mathrm{s}),\label{eq:nonlinearPendulum}
\end{aligned}
\end{equation} 
where 
$C_l = 2\pi \sin\left(\theta^\mathrm{s}\right)$ is the potential flow lift for an inclined flat plate in a uniform flow
at the angle of incidence $\theta^\mathrm{s}$ 
and $C_d = 1.28$  represents the drag coefficient of a flat plate in a cross flow.
In Eq.~(\ref{eq:nonlinearPendulum}),  the terms $\cos\theta^\mathrm{s}$ and $\sin\theta^\mathrm{s}$ appear due the geometric projection of the lift and drag force normal to the plate. The term $\left|\sin\theta^\mathrm{s}\right|$ projects the inclined plate length in the direction normal to the flow and the absolute operator accounts for the sign of projected length.
In the above analytical form, 
the potential flow is based on the assumption that the flow remains attached even for a large rotation angle.  
Using the Lighthill's force decomposition, we also assume that drag force and the inviscid inertia force can be treated independently. Furthermore, we have ignored 3D effects and the LE suction force associated with the potential flow.

\begin{figure}
	\centering
	\begin{subfigure}{0.49\textwidth}
		\includegraphics[width=0.99\columnwidth,trim=0mm 0mm 20mm 0mm,clip]{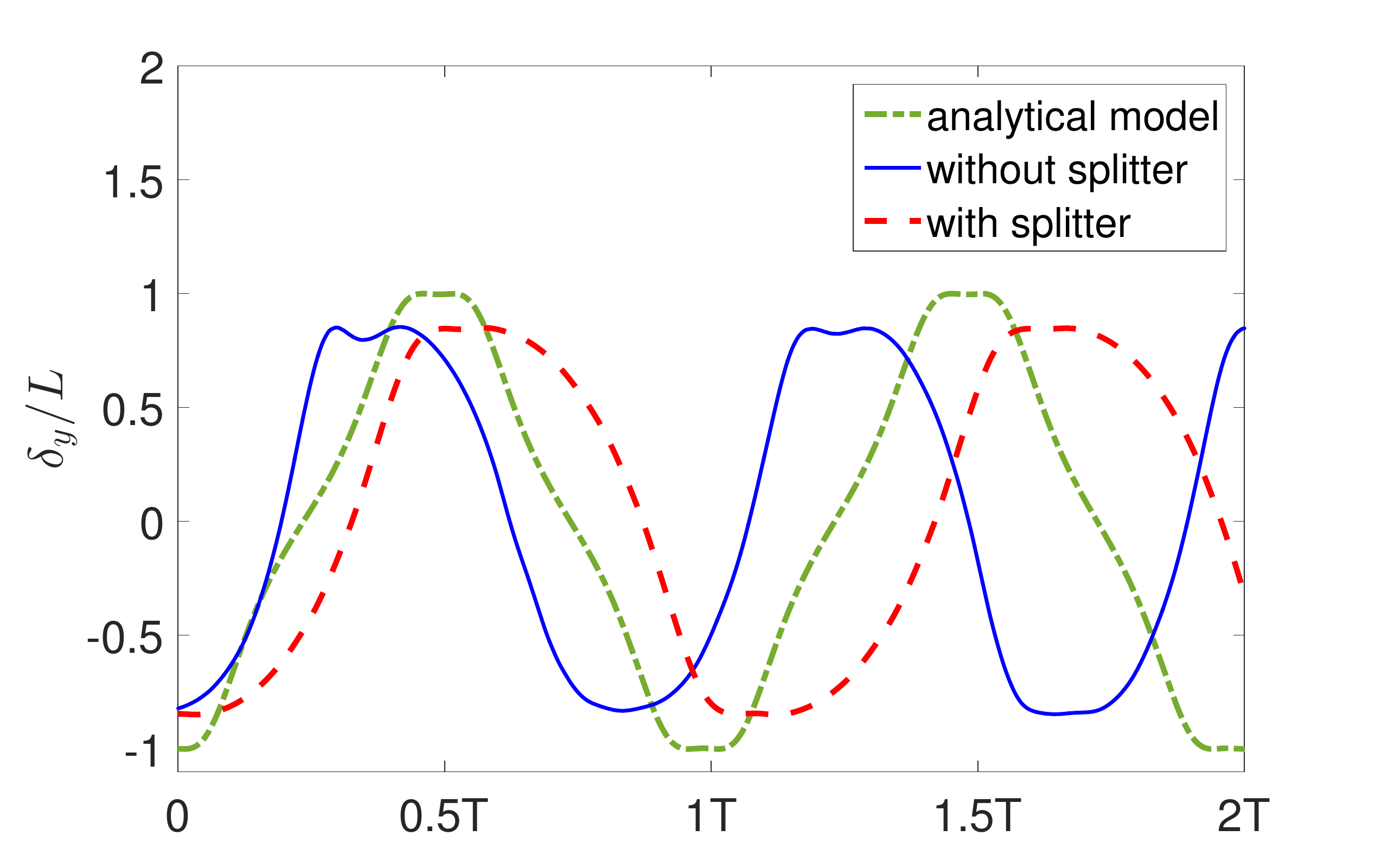}	
		\caption{}
	\end{subfigure}
	\begin{subfigure}{0.49\textwidth}
		\includegraphics[width=0.99\columnwidth,trim=0mm 0mm 20mm 0mm,clip]{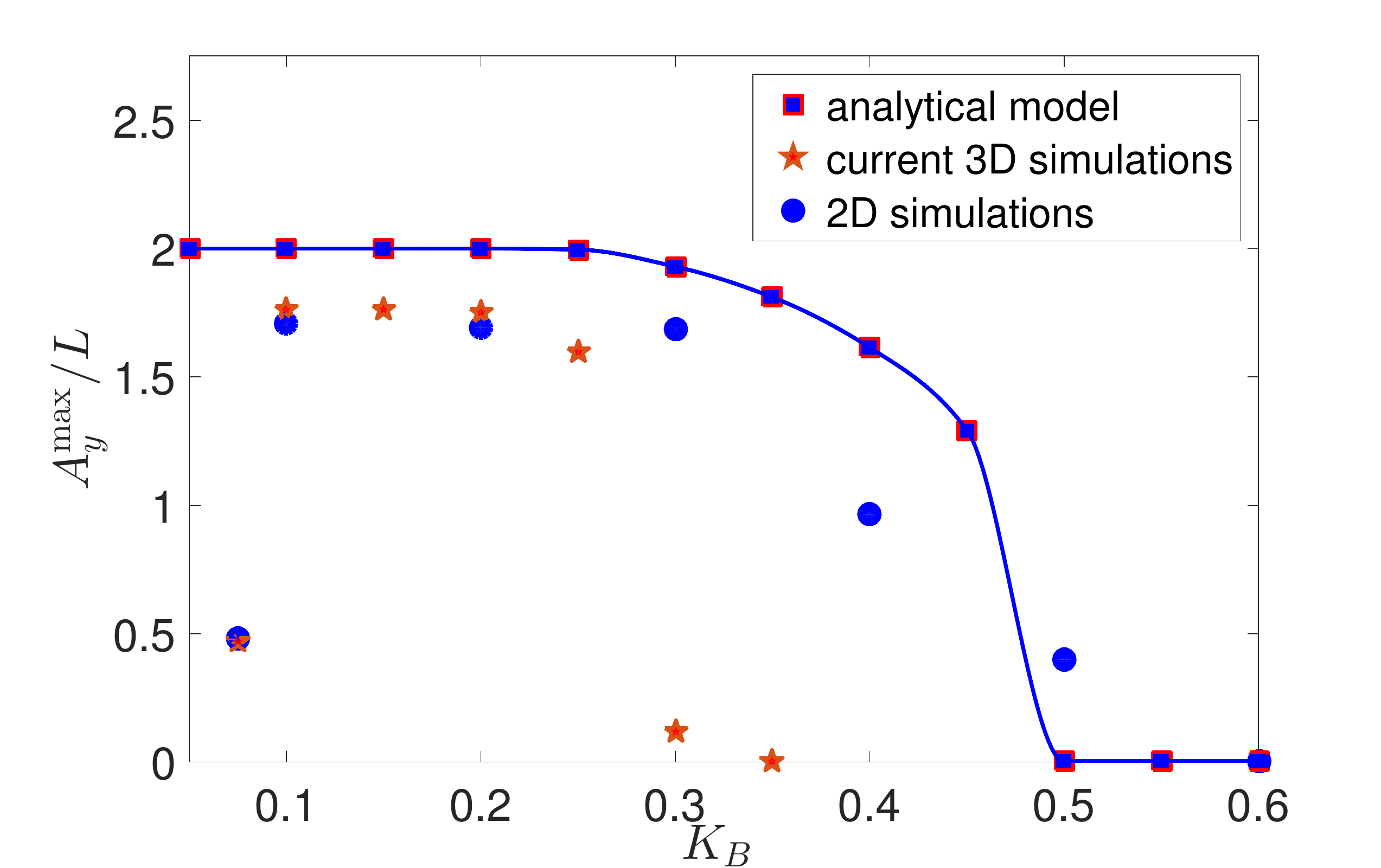}	
		\caption{}
	\end{subfigure}
	\caption{(a) Evolution of LE transverse displacement for nonlinear analytical pendulum model mounted on a torsional spring at $K_\theta=1200 \mathrm{N m^2/ rad}$ and $I_\theta=333.33\ \mathrm{Kg~m^2}$  and  3D FSI simulations of inverted foil with and without splitter at nondimensional parameters $K_B=0.2,\ Re=30000$ and $m^*=0.1$, (b) comparison the maximum transverse amplitude as a function of $K_B$ for the simplified nonlinear pendulum model against 3D FSI simulations at $m^*=1.0$, $Re=30000$ and AR=0.5, and 2D simulations from \cite{gurugubelli_JFM} at $m^*=0.1$ and $Re=1000$. }\label{nonlinearPendulum_timeHistory}
\end{figure}

\begin{figure}
	\centering
	\begin{subfigure}{0.49\textwidth}
		\includegraphics[width=0.99\columnwidth,trim=0mm 0mm 0mm 0mm,clip]{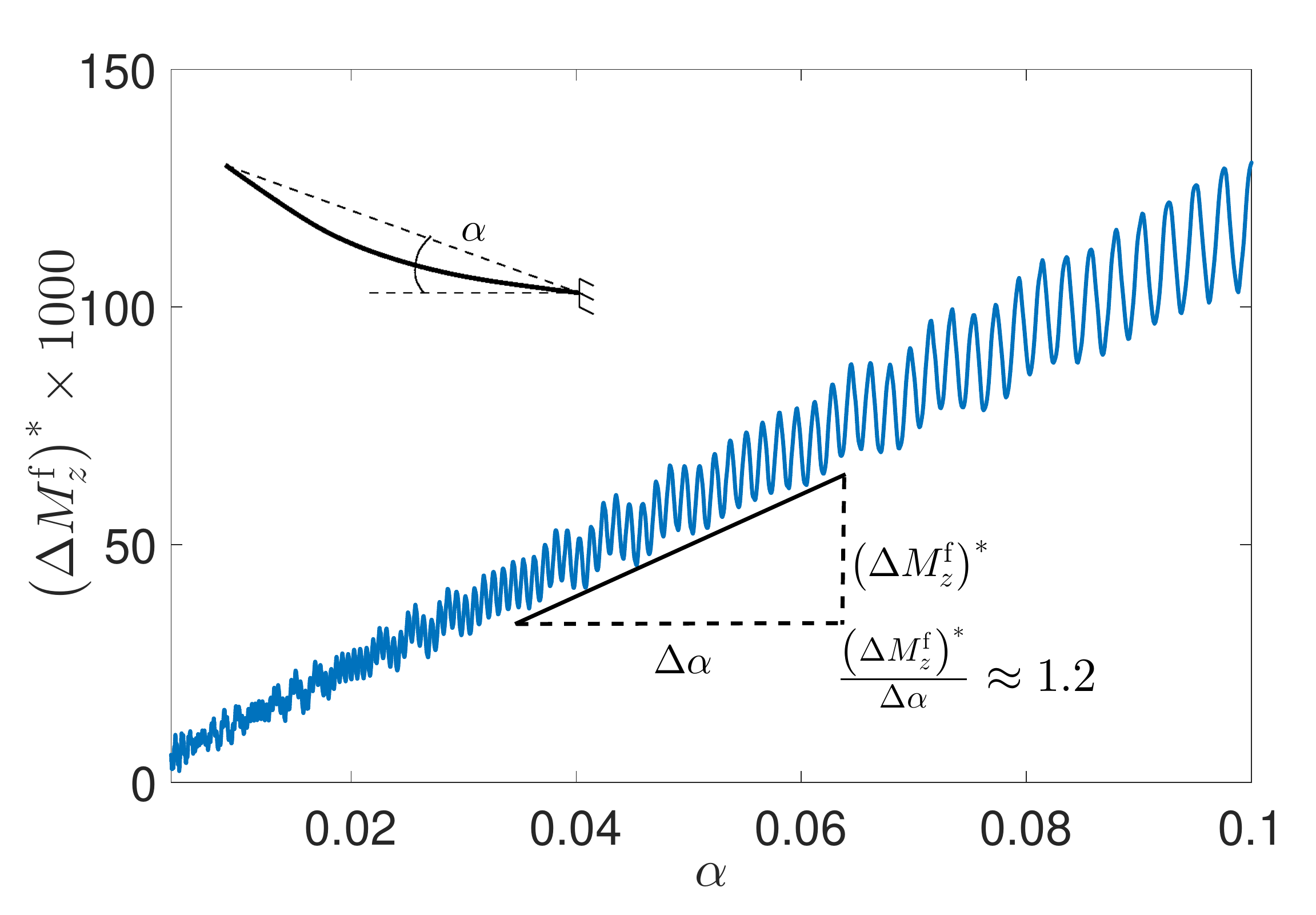}
		\caption{}
	\end{subfigure}
	\begin{subfigure}{0.49\textwidth}
		\includegraphics[width=0.99\columnwidth,trim=0mm 0mm 0mm 0mm,clip]{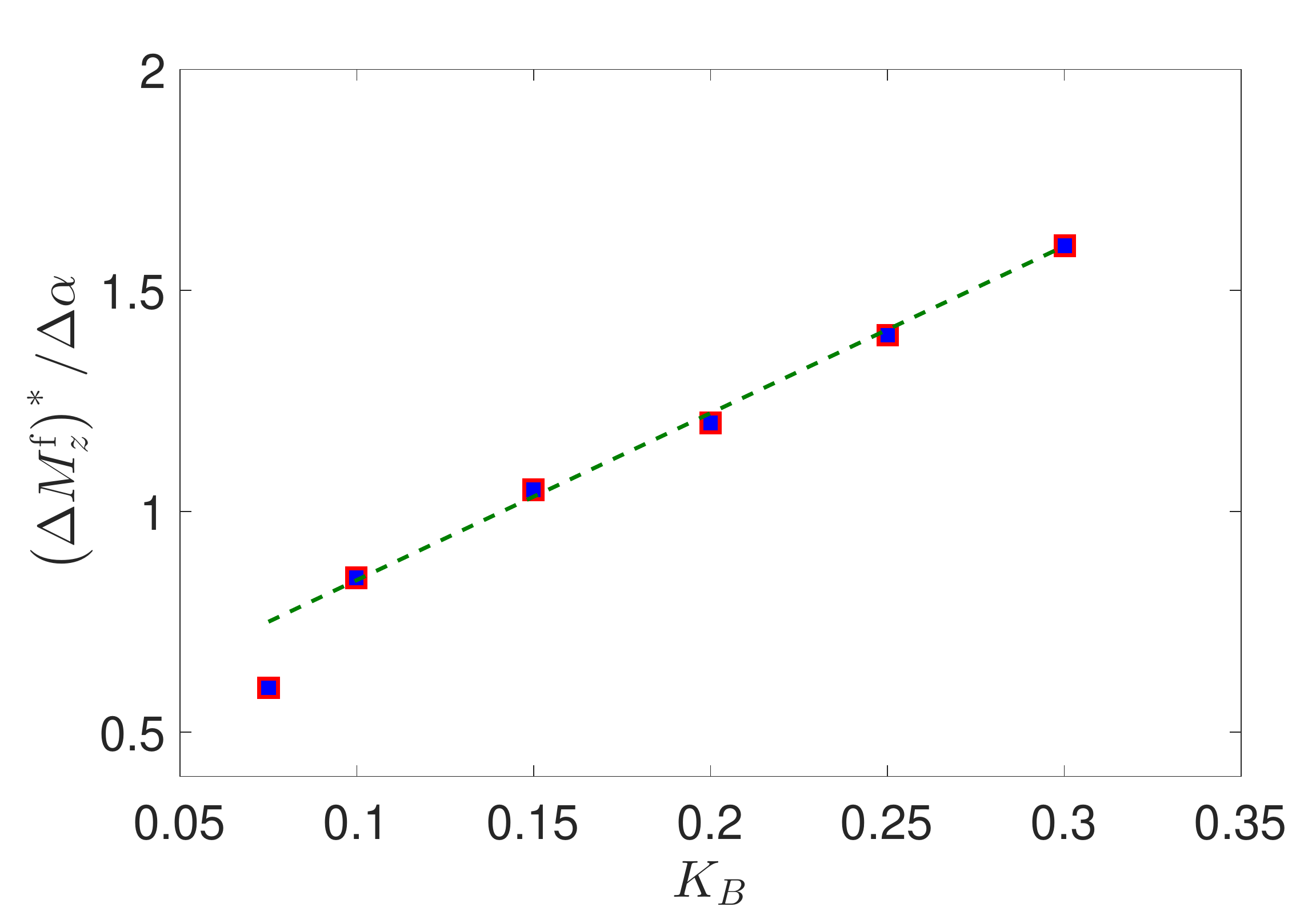}
		\caption{}
	\end{subfigure}
	\caption{(a) Variation of nondimensional moment $\left( M^\mathrm{f}_z\right)^* = M^\mathrm{f}_z/(\rho^\mathrm{f} U_0^2 L^2 W)$ acting on inverted foil with respect to the foil rotation angle $\alpha$ for $tU_0/L\in[0,7]$ at $K_B=0.2,\ Re=3000,\ m^*=1$ and AR=0.5, and (b) relationship between $\Delta\left( M^\mathrm{f}_z\right)^*/\Delta \alpha$ and $K_B$ for $Re=30000,\ m^*=1$ and AR=0.5. The dashed line represents the best fit line given by 
Eq.~(\ref{torsionSpringConst}).}
\label{torsionSpring}
\end{figure}

In figure~\ref{nonlinearPendulum_timeHistory}a, we compare the time traces of the LE transverse displacement by solving the Eq.~\ref{eq:nonlinearPendulum} for $K_\theta=1200 \mathrm{N m^2/rad}$ and $I=333.33\ \mathrm{Kg m^2}$ against the equivalent full-scale 3D simulations at $K_B=0.2$ and $m^*=1.0$. We have considered $\rho^f=1000 \mathrm{Kg/m^3},\ L=1$m and $U_0=1$m/s for the analytical model. For the comparison purpose, we have also included with-splitter case for the identical nondimensional parameters. 
Similar to the LAF response of inverted foil, the elastically mounted plate rotates to a large pitch angle and there is a continuous interplay between the fluid force, the inertia and the restoring torsional spring force.
A small damping coefficient of 0.001 has been considered for the analytical model of the elastically mounted plate. The figure shows the simplified analogous model can qualitatively predict the complex LAF kinematics and the flapping frequency of the simple analogous model lies in between the flapping frequency of the inverted foil with and without splitter. 
For the analogous model of the elastically mounted plate, the equivalent linear torsional constant ($K_\theta$) corresponding to the inverted foil with $K_B=0.2$ is computed from the FSI simulations by plotting the nondimensional torsional moment $\left( M^\mathrm{f}_z\right)^* = M^\mathrm{f}_z/(\rho^\mathrm{f} U_0^2 L^2 W)$ acting on the foil as a function of the foil LE angle of rotation ($\alpha$) over $tU_0/L\in[0,7]$  in figure~\ref{torsionSpring}a. Here, $\alpha$ is defined as $\tan^{-1}{\left(\delta_y/\delta_x\right)}$ and we consider representative values: $K_B=0.2,\ Re=30000,\ m^*=1.0$ and AR=0.5. The slope $\left(\Delta\left( M^\mathrm{f}_z\right)^* / \Delta \alpha\right)\times \rho^\mathrm{f}U_0^2L^2W$ in figure~\ref{torsionSpring}a defines the torsional spring constant $K_\theta$.
For small deformations, the inverted foil behaves as a linear torsional spring and exhibits a constant $K_\theta$ in that range. 
Figure~\ref{torsionSpring}b summarizes the $\Delta \left( M^\mathrm{f}_z\right)^*/\Delta \alpha$ values for different $K_B$ and the dashed line represents the linear correlation between $K_B$ and $\Delta \left( M^\mathrm{f}_z\right)^*/\Delta \alpha$, which is given by
\begin{equation*}
\frac{\Delta \left( M^\mathrm{f}_z\right)^*}{\Delta \alpha} = a\ K_B + b,
\label{torsionSpringConst}
\end{equation*}
where $a=4$ and $b=0.4$. We use this correlation to present the effectiveness of the simplified analytical model presented in Eq.~(\ref{staticEq}). In figure~\ref{nonlinearPendulum_timeHistory}b, we compare the maximum transverse amplitude of the LE as a function of $K_B$ obtained from the analytical model against the 3D FSI simulations at $m^*=1.0$ and $Re=30000$. We have also included 2D simulations at $m^*=0.1$ and $Re=1000$ presented in \cite{gurugubelli_JFM}. 
Overall, the semi-analytical model provides a reasonable agreement with the 3D simulations. Notably, the simplified model exhibits greater maximum transverse amplitudes because a flexible inverted foil needs to bend about the leading edge as shown in figure~\ref{VIV_schem}a,  whereas the rigid plate in the simplified model just rotates about the TE. 
The above analysis provides a direct connection between the large-amplitude oscillation of the elastically mounted plate at the TE with the LAF response of a flexible inverted foil in a uniform flow stream. 
\bibliographystyle{jfm}
\bibliography{refs}

\end{document}